\documentclass[useAMS,referee]{biom}

\usepackage{amsmath}
\usepackage{natbib}
\usepackage[plain,noend]{algorithm2e}
\usepackage{pdflscape}

\usepackage{setspace}
\usepackage{bbm,bm}
\usepackage{soul,color}
\usepackage{amsfonts}
\usepackage{amssymb}
\usepackage{newfloat}
\usepackage{multirow}
\usepackage{arydshln}
\usepackage{mathtools}

\def\bSig\mathbf{\Sigma}

\def\E{\mathbb{E}}

\def\P{\mathbb{P}}
\def\given{\, | \,}

\def\bfp{\mathbf{p}}

\title[A Meta-Learning Method to Assess Time-Varying Moderation]{A Meta-Learning Method for Estimation of Causal Excursion Effects to Assess Time-Varying Moderation}

\author{Jieru Shi$^{*}$\email{herashi@umich.edu} and
Walter Dempsey$^{**}$\email{wdem@umich.edu} \\
Department of Biostatistics, University of Michigan, Ann Arbor, U.S.A.}

\begin{document}


\date{{\it Received October} 2007. {\it Revised February} 2008.  {\it
Accepted March} 2008.}



\pagerange{\pageref{firstpage}--\pageref{lastpage}} 
\volume{64}
\pubyear{2008}
\artmonth{December}


\doi{10.1111/j.1541-0420.2005.00454.x}


\label{firstpage}

\begin{abstract}
\textcolor{black}{Advances in wearable technologies and health interventions delivered by smartphones have greatly increased the accessibility of mobile health (mHealth) interventions.} Micro-randomized trials (MRTs) are designed to assess the effectiveness of the mHealth intervention and introduce a novel class of causal estimands called ``causal excursion effects.'' These estimands enable the evaluation of how intervention effects change over time and are influenced by individual characteristics or context. Existing methods for analyzing causal excursion effects assume known randomization probabilities, complete observations, and a linear nuisance function with prespecified features of the high-dimensional observed history. However, in complex mobile systems, these assumptions often fall short: randomization probabilities can be uncertain, observations may be incomplete, and the granularity of mHealth data makes linear modeling difficult.
To address this issue, we propose a flexible and doubly robust inferential procedure, called ``DR-WCLS,'' for estimating causal excursion effects from a meta-learner perspective.
We present the bidirectional asymptotic properties of the proposed estimators and compare them with existing methods both theoretically and through extensive simulations. The results show a consistent and more efficient estimate, even with missing observations or uncertain treatment randomization probabilities. Finally, the practical utility of the proposed methods is demonstrated by analyzing data from a multiinstitution cohort of first-year medical residents in the United States \citep{necamp2020}.
\end{abstract}



\maketitle

\section{Introduction}
\label{sec:intro}





The use of smart devices, such as smartphones and smartwatches, to deliver mobile health (mHealth) interventions has grown significantly in recent years. These low-cost and accessible interventions can be delivered anywhere, anytime, and in any amount, reaching even reticent or hard-to-reach populations. By personalizing interventions to adapt to the internal and contextual information collected by smart devices, they are hypothesized to lead to meaningful short- and long-term behavior changes.

\textcolor{black}{The evaluation of these time-varying effects led to the development of micro-randomized trials (MRTs), where individuals are randomized to receive notifications at hundreds or thousands of decision points during the study period. The Intern Health Study (IHS) presented in Section~\ref{sec:casestudy} is such an example \citep{necamp2020}. In this study, domain scientists aimed to investigate whether sending targeted notifications to medical interns in stressful work environments can help improve their behavior and mental health. A key goal in scenarios like the IHS is to understand how the effectiveness of targeted notifications varies over time, a concept termed as ``causal excursion effects'' \citep{boruvka2018, qian2020estimating, dempsey2020, shi2022assessing}.} Semiparametric inference of these effects can be performed using the weighted centered least squares (WCLS) criterion \citep{boruvka2018}.

However, in practice, the complexity of mobile systems often violates key assumptions of the WCLS criterion. Treatment randomization probabilities may be unknown or incorrectly recorded due to software errors or deviations from protocol—issues that are especially concerning in MRTs using reinforcement learning algorithms \citep{deliu2024reinforcement}. WCLS also assumes complete outcome data, yet many MRTs rely on self-reports or passive sensing via devices (e.g., Fitbit), leading to missing data. Moreover, the high frequency and dimensionality of mHealth data make it difficult to identify a small set of relevant features for parametric modeling, limiting inferential efficiency.

To address these challenges, we build on the Double Machine Learning (DML) framework \citep{chernozhukov2018}, which enables the use of high-dimensional, data-adaptive models for estimating nuisance components. Although meta-learning methods for Conditional Average Treatment Effect (CATE) estimation are well developed \citep{hill2011bayesian, kunzel2019metalearners, kennedy2020optimal}, their application to longitudinal data is complicated by temporal dependence between observations. Most existing DML approaches in longitudinal settings focus on estimating average treatment effects (ATE) for distal outcomes under fixed or dynamic treatment regimes \citep{lewis2020double, viviano2021dynamic, chernozhukov2022automatic, zhang2021dynamic, bodory2022evaluating}. \textcolor{black}{Related work by \cite{liu2024incorporating} considers meta-learning in MRTs with zero-inflated count outcomes, but assumes correct specification of the causal effect model.}

We present the ``DR-WCLS'', a doubly robust inferential procedure for evaluating time-varying causal effect moderation in MRTs. \textcolor{black}{DR-WCLS flexibly incorporates supervised learning algorithms for valid causal inference, yielding a consistent and asymptotically normal estimator (as $n$ or $T \rightarrow \infty$). We provide theoretical guarantees for its double robustness and improved estimation efficiency compared to the standard WCLS approach. Furthermore, our method addresses common MRT analysis challenges, such as the ``curse of the horizon'' \citep{liu2018breaking} and missing outcomes,} thereby improving scientists' ability to investigate time-varying effect moderation, and find out when, in what context, and what intervention content to deliver to each person to make the intervention more effective \citep{qian2022microrandomized}.

\section{Preliminaries}
\label{sec:existmeth}

\subsection{Micro-Randomized Trials (MRT)}

An MRT consists of a sequence of within-subject decision times $t=1,\ldots, T$ at which treatment options are randomly assigned~\citep{Liaoetal2015}.  Individual-level data can be summarized as~$\{ O_1, A_1, O_2, A_2, \ldots, O_T, A_T, O_{T+1} \}$
where $t$ indexes a sequence of decision points, $O_t$ is the information collected between time $t-1$ and $t$, and $A_t$ is the treatment option provided at time $t$; here we consider binary treatment options, i.e.,~$A_t \in \{ 0, 1\}$.  In an MRT, $A_t$ is randomized with randomization probabilities that may depend on the complete observed history $H_t := \{ O_1, A_1, \ldots, A_{t-1}, O_t \}$, denoted $\bfp = \{ p_t (A_t | H_t) \}_{t=1}^T$. \textcolor{black}{Treatment options are intended to influence a proximal outcome, denoted by $Y_{t+1} \in O_{t+1}$, which depends on the observed history $H_t$ and the most recent treatment $A_t$~\citep{dempsey2020}.}

\subsection{Estimands and Inferential Methods: A Review}

The class of estimands, referred to as ``causal excursion effects'', was developed to assess whether mobile health interventions influence the proximal health outcomes they were designed to impact~\citep{heron2010}. These time-varying effects are a function of the decision point~$t$ and a set of moderators~$S_t$ and marginalize over all other observed and unobserved variables \citep{dempsey2020, qian2020estimating}. We provide formal definitions using potential outcomes~\citep{Rubin, Robins}.

Let~$Y_{t+1} (\bar a_{t-1})$ denote the potential outcome of the proximal response \textcolor{black}{under treatment sequence~$\bar a_{t-1} = \{a_1,...,a_{t-1}\} \in \{0,1\}^{t-1}$}. Let~$O_t (\bar a_{t-1})$ denote the potential information collected between time~$t-1$ and $t$. Let $S_t (\bar a_{t-1})$ denote the potential outcome for a moderator of time-varying effects that is a deterministic function of the potential history up to time $t$, $H_t (\bar a_{t-1})$. \textcolor{black}{We consider the setting in which the potential outcomes are i.i.d. over users according to a distribution $\mathcal{P}$, that is,
    $\left\{ O_{t,j} (\bar a_{t-1,j}) \right\}_{t=1}^T \overset{\text{i.i.d}}{\sim} \mathcal{P}$, $\forall j \in \{1,2,\dots, n\}$.}
The causal excursion effect estimand is:
\begin{equation}
\label{eq:causalexcursion_po} 
\beta_{\bfp}(t;s) = \E_{{\bfp}} \left [ Y_{t+1} \left(\bar A_{t-1}, A_t = 1 \right) - Y_{t+1} \left(\bar A_{t-1}, A_t = 0 \right) | S_t (\bar A_{t-1}) = s\right]. 
\end{equation}
Equation~\eqref{eq:causalexcursion_po} is defined with respect to a reference distribution $\bfp$, i.e., the joint distribution of treatments $\bar A_{t-1}:= \{A_1, A_2, \dots, A_{t-1}\}$. We follow common practice in observational mobile health studies where analyses such as GEEs \citep{liang1986} are conducted marginally over $\bfp$. To express the proximal response in terms of the observed data, we assume positivity, consistency, and sequential ignorability \citep{robins1994,robins1997}:

\begin{assumption}
\label{ass:po}\normalfont
We assume consistency, positivity, and sequential ignorability:
  \begin{enumerate}
  \item Consistency: For each $t \leq T$,
    $\{Y_{t+1} (\bar{A}_{t} ), O_{t} (\bar A_{t-1}), A_{t} (\bar{A}_{t-1} )\}  = \{Y_{t+1}, O_{t}, A_{t}\}$, i.e., observed values equal the corresponding potential outcomes;
  \item Positivity: if the joint density $\{H_t = h_t, A_t = a_t\}$ is greater than zero, there exists a constant \(\epsilon > 0\) such that, 
    $ \epsilon < p_t(A_t | H_t) < 1- \epsilon$ almost surely for all \(t \leq T\);
  \item Sequential ignorability: For each~$t \leq T$, the potential outcomes 
  \newline
  $\{Y_{t+1}(\bar a_t), O_{t+1}(\bar a_t), A_{t+1}(\bar a_t), \dots, Y_{T+1}(\bar a_T)\}$ are independent of $A_{t}$ conditional on the observed history $H_t$.
  \end{enumerate}
\end{assumption}

\noindent Under Assumption~\ref{ass:po}, Equation \eqref{eq:causalexcursion_po} can be re-expressed in terms of observable data:
\begin{equation}
\label{eq:causalexursion}
    \beta_{\bfp}(t;s)= \E \left[ \E_{{\bfp}} \left[  Y_{t+1} \mid A_t = 1, H_t \right] - \E_{{\bfp}} \left[ Y_{t+1} \mid A_t = 0, H_t \right] \mid S_t = s \right]. 
\end{equation}

To evaluate the causal excursion effect, we usually start with a working model assumption on the causal effect. Different choices of effect moderators can be used to address various scientific questions, and our interest lies in making inferences on the corresponding coefficients.

\begin{assumption}
\label{ass:directeffect}
The causal excursion effect takes a known linear form, i.e. $\beta_{{\bfp}} (t;s) = f_t(s)^\top \beta^\star$, where $f_t(s) \in \mathbb{R}^q$ and its Euclidean norm $\| f_t(S_t) \|_2 \leq c_2$ almost surely for some constant $c_2 > 0$ and all $t$.
\end{assumption}

This parametric assumption assumes a correct specification of the causal effect; however, when model misspecification occurs,
we can still interpret the proposed linear form as an $L_2$ projection of the true causal excursion effect onto the space spanned by a $q$-dimensional feature vector $f_t(s)$ that only depends on  $t$ and $s$ \citep{shi2022assessing}. The choice between these interpretations, whether it is a correctly specified causal effect or a projection, reflects a bias-variance trade-off. In practical applications, the projection interpretation ensures a well-defined parameter with practical interest \citep{dempsey2020}. In addition, assuming the causal effect moderators are bounded prevents the excursion effect from diverging and preserves its interpretability.

Previous studies have commonly treat MRTs as experimental studies with prespecified randomization schemes, which leads to the following assumption:
\begin{assumption}
\label{ass:p_correct}
    The randomization probability $p_t(A_t|H_t)$ is known or correctly specified via a parametric model~$p_t(A_t | H_t; \theta)$ for $\theta \in \mathbb{R}^d$.
\end{assumption}
Based on all the assumptions outlined above, a consistent estimator $\hat\beta_n$ can be obtained by minimizing a weighted and centered least squares (WCLS) criterion \citep{boruvka2018}:
\begin{align}
\label{eq:wcls}
     \P_n \Big[\sum_{t=1}^T W_{t} (Y_{t+1}- g_t(H_t)^\top \alpha - (A_{t} - \tilde p_t(1|S_t)) f_t(S_t)^\top \beta)^2 \Big],
\end{align}
where~$\mathbb{P}_n$ is an operator denoting the sample average, $W_t = \tilde p_t (A_t|S_t) / p_t (A_t| H_t)$ is a weight where the numerator is an arbitrary function with range $(0,1)$ that only depends on $S_t$, and $g_t(H_t) \in \mathbb{R}^p$ are $p$ control variables. Important to this paper, the linear term $g_t(H_t)^\top \alpha$ is a working model for $\E[W_t Y_{t+1} | H_t]$, which can be viewed as a nuisance function. A high-quality estimation of the nuisance function can help reduce variance and construct more powerful test statistics. See \citet{boruvka2018} for more details on the estimand formulation and the consistency, asymptotic normality, and robustness properties of this method.

\section{A Meta Learning Approach to Moderation Analysis}
\label{sec:newmethods}




The WCLS criterion presented in display \eqref{eq:wcls} provides a set of estimating equations used to make inferences about the causal parameter $\beta^\star$. 
This approach suggests that the nuisance parameter can be expressed as a sequence of expectations $\mathbf{g} = \{ g_t(H_t) = \E[W_t Y_{t+1} | H_t]\}_{t=1}^T$, with a population value of $\mathbf{g}^\star$. 
To estimate these quantities, the WCLS criterion only considers linear working models $\{g_t(H_t)^\top \alpha\}_{t=1}^T$.

However, prespecifying features from high-dimensional history $H_t$ for linear working models is a significant challenge. To increase flexibility in modeling nuisance functions, we leverage Neyman orthogonality between $\mathbf{g}$ and the causal parameter $\beta$ in Equation \eqref{eq:wcls}. We reformulate the estimating equation into a general form that eliminates parametric assumptions on $g_t (H_t)$ and allows its dimensions to grow with sample size. This leads to our proposed \emph{R-WCLS criterion}, which minimizes:
\begin{equation}
\label{eq:r-wcls}
   \P_n \Big[\sum_{t=1}^T W_t\big(Y_{t+1} - g_t(H_t) -(A_t - \tilde p_t(1|S_t))  f_t (S_t)^\top \beta\big)^2\Big].
\end{equation}

We can recover WCLS by replacing $g_t(H_t)$ with a linear working model with fixed dimension, i.e., $g(H_t)^\top \alpha$ for $\alpha \in \mathbb{R}^p$. Here is the asymptotic property of the proposed estimator:

\begin{theorem}[Asymptotic property of the R-WCLS estimator]
\label{thm:asymptotics_r}
Under Assumptions \ref{ass:po}, \ref{ass:directeffect}, and \ref{ass:p_correct}, given invertibility and moment conditions, the estimator $\hat\beta^{(R)}_n$ minimizes \eqref{eq:r-wcls} is consistent and asymptotically normal: $\sqrt{n}(\hat\beta^{(R)}_n - \beta^\star) \rightarrow \mathcal{N}(0,\Sigma_R)$, where $\Sigma_R$ is defined in Appendix \ref{app:r-wcls}. 
\end{theorem}


\textcolor{black}{Theorem \ref{thm:asymptotics_r} applies to any plug-in estimator $g_t(H_t)$, which can be prespecified or empirically estimated using supervised learning with cross-fitting.
}
A key feature of the R-WCLS criterion is its ability to learn the nuisance function $\mathbf{g}$ without prespecifying features to build a parametric working model. The advantage of ML orthogonalization is that it can estimate more complicated functions with input of high-dimensional data. It can learn interactions and nonlinearities in a way that it is hard to encode into a linear working model. Furthermore, some ML algorithms, especially those based on decision trees, are more flexible and easier to implement compared to linear regression. For further details on implementation, efficiency, connections to meta-learners, and a more efficient R-WCLS version based on the semiparametric efficient influence function \citep{robins1994}, see Appendices \ref{app:R-WCLS} and \ref{app:ar-wcls}.



\subsection{A Doubly-Robust Alternative}

The previous discussion relies on Assumption \ref{ass:p_correct} to be true. In many MRTs, it may not be possible to correctly implement or collect the desired randomization probabilities, leading to unknown randomization probabilities or uncertainty in their recorded values. In such cases, the R-WCLS criterion in \eqref{eq:r-wcls} can only provide consistent estimates of $\beta^\star$ if the outcome regression model for $\E[Y_{t+1} | H_t, A_t]$ has been correctly specified.
This implies that the fully conditional treatment effect depends only on the specified moderators $S_t$ and that the linear model $f_t(S_t)^\top \beta$ is correctly specified. However, in practice, $S_t$ is often a subset of the potential moderators, so this assumption is not expected to hold. Therefore, an estimation procedure that does not rely on a correct model specification will be preferred. In this section, we present an alternative, doubly robust estimator.
We denote $\boldsymbol{\eta}_t(H_t, A_t) = (g(H_t, A_t), p_t(A_t|H_t))$ and define the following estimating equation
\begin{equation}
\label{eq:estimating-eq-psi}
    \psi_t(\beta; \boldsymbol{\eta}_t, A_t, H_t) = \tilde \sigma^2_t(S_t) \Big(\frac{W_t \big(A_t - \tilde p_t (1|S_t)\big)\big(Y_{t+1}- g_t(H_t,A_t)\big)}{\tilde \sigma^2_t(S_t)} + \beta (t; H_t) - f_t(S_t)^\top \beta\Big) f_t(S_t),
\end{equation}
where $\beta (t; H_t) \coloneqq g_t(H_t,1) - g_t(H_t,0) $ is the causal excursion effect under the fully observed history $H_t$, and $\tilde \sigma^2_t(S_t) \coloneqq \tilde p_t (1|S_t) (1- \tilde p_t(1 |S_t)) $ is the projection weight onto the subspace $f_t(S_t)$ \citep{dempsey2020}. Then, the proposed Doubly-Robust Weighted and Centered Least Square \emph{(DR-WCLS)} criterion is given by: 
\begin{align}
\label{eq:drwcls}
     \mathbb{P}_n \Big[\sum_{t=1}^T \psi_t(\beta; \boldsymbol{\eta}_t, A_t, H_t) \Big] = 0 
\end{align}

Theorem \ref{thm:asymptotics_dr} below shows that the estimator $\hat\beta^{(DR)}_n$ obtained from solving \eqref{eq:drwcls} is doubly robust, that is, \eqref{eq:drwcls} will produce a consistent estimator of~$\beta^\star$ if \emph{either} the randomization probability $p_t(A_t|H_t)$ \emph{or} the conditional expectation $g_t(H_t,A_t)$ is correctly specified.

\subsection{Algorithm}
\label{sec:algorithms}


The algorithm exploits the structure of Equation \eqref{eq:drwcls} to characterize the problem as a two-stage weighted regression estimation that regresses the estimated \emph{pseudo-outcomes} on a feature vector. \textcolor{black}{While \cite{chen2022debiased} show that neither sample splitting nor the Donsker property is necessary if the estimator \( \hat{g}(\cdot) \) satisfies leave-one-out stability properties and the moment function meets the weak mean-squared-continuity condition described by \cite{chernozhukov2021simple}, we remain agnostic to the choice of supervised learning algorithms and employ individual-level cross-fitting to establish asymptotic guarantees. The DR-WCLS algorithm is as follows:}

\paragraph{Step I}: Randomly split the $n$ individuals into $K$ equal folds $\{I_k\}_{k=1}^K$, assuming $n$ is a multiple of $K$. Let $I_k^\complement$ denote the complement of fold $k$.

\paragraph{Step II}: For each fold $k$, use data from $I_k^\complement$ to estimate the nuisance functions $\hat g_t^{(k)}(H_t, A_t)$, $\hat p_t^{(k)}(1|H_t)$, and $\hat{\tilde p}_t^{(k)}(1 | S_t)$, \textcolor{black}{and compute the weight $\hat W^{(k)}_{t} = \hat{\tilde p}_t^{(k)}(1|S_t) / \hat p_t^{(k)}(1|H_t)$.} If treatment probabilities are known, set $\hat p_t(A_t|H_t) = p_t(A_t|H_t)$.


\paragraph{Step III}: For each $j \in I_k$ and time $t$, construct the pseudo-outcome $\tilde{Y}^{(DR)}_{t+1}$ as follows, then regress it on $f_t(S_t)^\top \beta$ using weights $\hat{\tilde{p}}_t^{(k)}(1|S_t)(1 - \hat{\tilde{p}}_t^{(k)}(1|S_t))$.
\begin{equation*}
        \tilde Y^{(DR)}_{t+1,j} := \frac{\hat W^{(k)}_{t,j} (A_{t,j} - \hat{\tilde p}^{(k)}_t (1|S_{t,j})) (Y_{t+1,j} - \hat g_t^{(k)} (H_{t,j}, A_{t,j}))}{\hat{\tilde p}^{(k)}_t (1|S_{t,j}) (1-\hat{\tilde p}^{(k)}_t (1|S_{t,j}))} + \left( \hat g_t^{(k)} (H_{t,j}, 1) - \hat g_t^{(k)} (H_{t,j}, 0)\right).
\end{equation*}

\begin{remark}[Connection to the DR-learner]
\sloppy The DR-learner was introduced by \cite{van2006targeted} and later formalized by \cite{kennedy2020optimal} as a two-stage, doubly robust meta-learner for fully conditional causal effects. Our DR-WCLS method extends this to sequential randomization settings, where the causal excursion effect is a time-varying \textit{marginal} effect, obtained by projecting the conditional effect onto moderators and smoothing over time. The projection weight $\tilde \sigma_t^2(S_t)$ in Step III serves this role. See Appendix \ref{app:sec:weights} and \cite{dempsey2020} for more details.
\end{remark}

\vspace{-24pt}
\section{The Asymptotic Properties of the DR-WCLS Estimator}


\subsection{Main Asymptotic Properties}
\label{sec:theoretical}


In this section, we demonstrate the asymptotic theory for the DR-WCLS estimator obtained using the algorithm described in Section \ref{sec:algorithms}. \textcolor{black}{Define the empirical $L_2$ norm of a random variable $X_t$ as $\Vert X_t \Vert = \big(\frac{1}{n}\sum_{j=1}^n  X_{t,j}^\top X_{t,j} \big)^{1/2}$.} To guarantee the consistency of the causal parameter, we require the following assumption.
\begin{assumption}
\label{ass:nuisance-op1}
The data-adaptive plug-ins $ \boldsymbol{\hat\eta}_t$ consistently estimate the true nuisance function $\boldsymbol{\eta}_t$, that is: $\sum_{t=1}^T\sum_{a \in \{0,1 \}}\Vert \boldsymbol{\hat\eta}_t(H_t, a_t) - \boldsymbol{\eta}_t(H_t, a_t) \Vert = o_p(1)$.
\end{assumption}

\begin{theorem}[Asymptotic property of DR-WCLS estimator]
\label{thm:asymptotics_dr}
Assume $T$ and $K$ are both finite and fixed, 
Under Assumption \ref{ass:po}, \ref{ass:directeffect} and \ref{ass:nuisance-op1}, given invertibility and moment conditions, as $n \rightarrow \infty$, the estimator $\hat\beta^{(DR)}_n$ that solves \eqref{eq:drwcls} is subject to an error term which (up to a multiplicative constant) is bounded above by:
    \begin{equation}
    \mathbf{\hat B} = \sum_{t=1}^T \sum_{a \in \{0,1\}}\left\Vert \hat p_t(a|H_t)- p_t(a|H_t)\right\Vert \left\Vert \hat g_t(H_t,a)- g_t(H_t,a)\right\Vert
\end{equation}
If $\mathbf{\hat B} =  o_p(n^{-1/2})$, then $\hat\beta^{(DR)}_n$ is consistent and asymptotically normal such that $\sqrt{n}(\hat\beta^{(DR)}_n - \beta^\star) \rightarrow \mathcal{N}(0,\Sigma_{DR})$, where $\Sigma_{DR}$ is defined in Appendix \ref{app:dr-wlcs}.  In particular, with the algorithm outlined in Section \ref{sec:algorithms}, $\Sigma_{DR}$ can be consistently estimated by
 \begin{equation}
 \label{eq:asymptotic_var}
     \Big[ \frac{1}{K} \sum_{k=1}^K \mathbb{P}_{n,k} \big \{ \dot m (\hat \beta, \hat \eta_k)  \big \} \Big]^{-1} \times 
\Big[ \frac{1}{K} \sum_{k=1}^K \mathbb{P}_{n,k} \big \{ m (\hat \beta, \hat \eta_k) m (\hat \beta, \hat \eta_k)^\top \big \} \Big] \times
\Big[ \frac{1}{K} \sum_{k=1}^K \mathbb{P}_{n,k} \big\{  \dot m (\hat \beta, \hat \eta_k) \big \} \Big]^{-1},
 \end{equation}
 where $m (\hat \beta, \hat \eta_k) = \sum_{t=1}^T \psi_t(\hat\beta, \hat \eta_k;H_t, A_t)$ and $\dot m (\hat \beta, \hat \eta_k)  = \frac{\partial m (\beta, \hat \eta_k)}{\partial \beta} \Big|_{\beta = \hat\beta}$.
\end{theorem}

The bound $\mathbf{\hat B}$ on the DR-WCLS estimator error shows that it can only deviate from $\beta^\star$ by at most a sum of (smoothed) products of errors in the estimation of treatment propensities and conditional expectation of outcomes, thus allowing faster rates for estimating the causal effect even when the nuisance estimates converge at slower rates. This occurs when either $\hat g_t(H_t,a)$ or $\hat p_t(a|H_t)$ are based on correctly specified models, but also achievable for many ML methods under structured assumptions on the nuisance parameters, for example, regularized estimators such as the Lasso and random forest \citep{chernozhukov2018,athey2016efficient}. Importantly, the model-agnostic error bound applies to arbitrary first-stage estimators. For detailed proofs of Theorem \ref{thm:asymptotics_dr}, please refer to Appendix \ref{app:dr-wlcs}.

\subsection{Time Dimension Asymptotic Properties}


In cases where MRTs have a relatively larger time horizon $T$ compared to the sample size $n$, applying a small sample correction in robust variance estimation proves effective in ensuring the robust performance of the estimator (when $n \approx 40$). In such scenarios, the previous algorithm and its asymptotic properties remain applicable. However, in certain extreme cases where the sample size $n$ is quite small and we are interested in ``individual time-averaged effects'', we consider an analogous asymptotic behavior of the estimated causal parameter when $n$ is fixed and $T$ approaches infinity. In this case, the DR-WCLS criterion can be reformulated as follows: 
\begin{equation}
\label{app:eq:bidirectional_asymp}
    \P_n \Big[\frac{1}{T} \sum_{t=1}^T \psi_t(\beta;\boldsymbol{\eta}_t,H_{t},A_{t})\Big] = 0.
\end{equation}

In contrast to the earlier estimating equation in Equation \eqref{eq:drwcls}, which averages over individuals, Equation \eqref{app:eq:bidirectional_asymp} emphasizes averaging over the time horizon. Define the time-averaged norm of a random variable $X_t$ as $\Vert X_t \Vert_T = \big(\frac{1}{T}\sum_{t=1}^T  X_t^\top X_t \big)^{1/2}$. To establish the asymptotic behavior of the estimator $\hat\beta^{(DR)}$, we first introduce the following assumptions. 
\begin{assumption}
\label{ass:T-infinity}
 When $T$ approaches infinity, we require the following conditions to hold:
    \begin{enumerate}
    \item There exists $\beta^\star$, such that $\lim_{T \rightarrow \infty}\frac{1}{T}  \sum_{t=1}^T \E[\psi_t(\beta^\star ;\boldsymbol{\eta}_t, H_{t},A_t)] =0$.
    \item \textcolor{black}{Denote the second-stage residual as \( \xi_t \coloneqq \tilde Y^{(DR)}_{t+1} - f_t(S_t)^\top \beta^\star \). There exists constants \( \delta > 0 \) and \( c_1 > 0 \) such that  $\mathbb{E}[\xi_t^{2+\delta}| H_t, A_t] < c_1$ for all $t$. 
    \item The correlation of residuals decays, i.e., $\lim_{|t - t'| \rightarrow \infty}\text{Corr}\big( \E[\xi_t^2 | H_t, A_t], \E [\xi_{t'}^2 | H_{t'}, A_{t'}] \big) = 0$, and there exists a constant positive definite matrix $\Gamma_{\beta}$, such that }
    \begin{equation*}
        \lim_{T \rightarrow \infty}\frac{1}{T}  \sum_{t=1}^T \E\big[ \psi_{t}(\beta;\boldsymbol{\eta}^{\star}_t, H_{t},A_t)\psi_{t}(\beta;\boldsymbol{\eta}^{\star}_t, H_{t},A_t)^\top\big] = \Gamma_{\beta}.
    \end{equation*}
\item The time-average norm of the nuisance function estimates satisfy $\Vert \boldsymbol{\hat\eta}_t - \boldsymbol{\eta}_t \Vert_T^2 = o_p(1)$
        and
        \begin{equation}
            \sum_{a \in \{0,1\}}\left\Vert \hat p_t(a|H_t)- p_t(a|H_t)\right\Vert_T \left\Vert \hat g_t(H_t,a)- g_t(H_t,a)\right\Vert_T = o_p(T^{-1/2}).
        \end{equation}
    \end{enumerate}
\end{assumption}
The first assumption is the identifiability condition, which is assumed to ensure the consistency of the estimator $\hat\beta^{(DR)}$. 
\textcolor{black}{The second and third assumptions are introduced to establish the asymptotic normality of $\hat\beta^{(DR)}$. In particular, we assume that the residuals have bounded $2 + \delta$ moments to guard against heavy tails—a weaker condition than assuming the outcome is bounded in Euclidean norm \citep{yu2023multiplicative, liu2024incorporating}. Our third assumption aligns with conditions in prior work \citep{bojinov2019time, yu2023multiplicative, liu2024incorporating}. Nevertheless, where these works directly assume the conditional covariance converges to a finite constant matrix for tractability, we offer intuitive, explicit sufficient conditions for this convergence to hold.
}
The fourth assumption outlines the necessary convergence rates for the estimators of the nuisance functions. 
Then we obtain the following theorem on the asymptotic property of the proposed DR-WCLS estimator when the time horizon $T$ goes to infinity. A detailed proof and \textcolor{black}{empirical results can be found in Appendix \ref{app:T-infinity}.}
\begin{theorem}
\label{thm:T-infinity}
Assume that $n$ is finite and fixed 
and \textcolor{black}{the estimated \(\hat{p}_t(A_t | H_t)\) also lies in \((\epsilon, 1 - \epsilon)\) for all $t$}. Under Assumptions \ref{ass:po}, \ref{ass:directeffect} and \ref{ass:T-infinity}, given invertibility and moment conditions, as $T \rightarrow \infty$, the estimator $\hat\beta^{(DR)}$ that solves Equation \eqref{app:eq:bidirectional_asymp} is consistent and asymptotically normal such that $\sqrt{T}(\hat\beta^{(DR)} - \beta^\star)\rightarrow \mathcal{N}(0,B_{\beta}^{-1}\Gamma_{\beta} B_{\beta}^{-1})$, where $\Gamma_{\beta}$ is defined in Assumption \ref{ass:T-infinity} and $B_{\beta}$ is defined in Appendix \ref{app:T-infinity}.
\end{theorem}


For practical implementation, a straightforward approach to train the nuisance function $\boldsymbol{\hat\eta}_t$ is to use only the historical data $H_t$ available at each time point. 
However, this method can result in poor estimates for early time points because of the limited size of the training set.
Sample splitting, on the other hand, is complicated by temporal dependence \citep{gilbert2021causal}, making it challenging to construct independent training and testing sets as done in Section \ref{sec:algorithms}. To address this, we need additional assumptions for time-wise sample splitting.
We provide more details in Appendix \ref{app:time-samplesplit}.

\begin{remark}[Extensions]
Carryover effects are a key challenge in longitudinal analysis but often yield high-variance estimates in long-horizon settings. \cite{shi2022assessing} addressed this ``curse of horizon'' \citep{liu2018breaking}. Building on their work, Appendix \ref{app:lagged} introduces a more efficient doubly robust meta-learning method, with detailed error bounds and asymptotic theory. We also extend the approach to address missing outcomes frequently observed in MRTs due to non-response (Appendix \ref{sec:missingdata}), enabling more efficient data use while preserving valid inference. Lastly, we generalize the framework to log-relative risk estimands (Appendix \ref{app:binaryoutcomes}), following \cite{qian2020estimating}.
\end{remark}

\section{Simulation}
\label{sec:sim}


Motivated by the case study, we extend the simulation setup from \cite{boruvka2018} to empirically verify the performance of our proposed estimators, \textcolor{black}{focusing on their main asymptotic properties as the sample size $n$ approaches infinity.} First, we present a base data generation model. Consider an MRT with a known randomization probability, and $g(H_t)$ in the generative model is a complex function of high-dimensional history information $H_t$. Let $S_{t} \in \{-1,1\}$ denote a single state variable that is an effect moderator, and $S_t \subset H_t$. We have the generative model as follows:
\begin{equation}
\label{eq:generativemodel}
    Y_{t,j} =  g_t(H_t) + \big(A_{t,j} - p_t(1|H_t)\big)(\beta_{10} + \beta_{11} S_{t,j})+  e_{t,j}.
\end{equation}
The randomization probability is $p_t(1|H_{t }) = \text{expit}(\eta_1 A_{t-1,j}+\eta_2 S_{t,j})$ where $\text{expit}(x)=(1+\exp(-x))^{-1}$; the state dynamics are given by $\mathbb{P}(S_{t}=1|A_{t-1 },H_{t-1 })=1/2$ with $A_0 = 0$, and the independent error term satisfies $e_{t,j} \sim \mathcal{N}(0,1)$ with $\text{Corr}(e_{u,j}, e_{t,j'}) = {\bf 1}(j=j') 0.5^{|u-t|/2}$. As in~\cite{boruvka2018}, we set $\eta_1 = -0.8, \eta_2 = 0.8, \beta_{10}=-0.2$, and $\beta_{11} \in \{0.2, 0.5, 0.8\}$, indicating a small, moderate, or large moderation effect. The marginal proximal effect is equal to $\beta_{10} + \beta_{11} \E [S_{t,j} ]=\beta_{10} = -0.2$. The marginal treatment effect is therefore constant over time and is given by $\beta_0^\star = \beta_{10} =  -0.2$.

In the following, we set the complex function $g_t(H_t)$ as a decision tree, and the flow chart Figure \ref{app:fig:generative_tree} in Appendix \ref{app:simulation} visualizes the decision-making process as well as the outcomes. 
We consider the estimation of the fully marginal proximal treatment effect, thus $f_t(S_t)=1$ in Equation (\ref{eq:generativemodel}) (that is, $S_{t} = \emptyset$). The results below report the average point estimate (Est), standard error (SE) and 95\% confidence interval coverage probabilities (CP) in 1000 replicates. \textcolor{black}{Here, we report results with $N= 100$ and $T = 30$} showing the relative advantage of R-WCLS and DR-WCLS over WCLS.

\textbf{WCLS}: Follows \cite{boruvka2018}, modeling control variables linearly as $g(H_t; \alpha) = g(H_t)^\top \alpha$, yielding consistent estimates and valid confidence intervals. Serves as a baseline.

\textbf{R-WCLS}: Estimates $\hat g(H_t,a)$ using supervised learning (random forests) with 5-fold cross-fitting, then combines them as $\hat g(H_t) = \tilde p_t(1|S_t)\hat g(H_t,1) + (1 - \tilde p_t(1|S_t))\hat g(H_t,0)$.

\textbf{DR-WCLS}: Reuses the same plug-in estimates from R-WCLS to compute the contrast $\hat \beta(t;H_t) = \hat g(H_t,1) - \hat g(H_t,0)$.


Table \ref{tab:tabone} reports the simulation results. ``\%RE gain'' indicates the percentage of times we achieve an efficiency gain out of 1000 Monte Carlo replicates. ``mRE'' stands for the average relative efficiency, and ``RSD'' represents the relative standard deviation between two estimates. The proposed R-WCLS and DR-WCLS methods significantly improve the efficiency of the WCLS when estimating the fully marginal causal effect. In addition, we find that mRE varies with $\beta_{11}$. R-WCLS has a higher mRE than DR-WCLS when $\beta_{11}$ is small, and this reverses when $\beta_{11}$ increases. In our simulation, $\beta_{11}$ being large indicates that an important moderator $S_{t,j}$ was not included in the causal effect model (that is, $f_t(S_t)^\top \beta =\beta_{0}$). Therefore, when model misspecification occurs, DR-WCLS shows better performance.

\begin{table}
\caption{Fully marginal causal effect estimation efficiency comparison. \\
The true value of the parameters is $\beta_{0}^\star = -0.2$.}
\label{tab:tabone}
\begin{center}
\begin{tabular}{cccccccc}
\hline
Method & $\beta_{11}$ & Est & SE & CP & \%RE gain  & mRE & RSD  \\\hline
\multirow{3}{*}{WCLS} & 0.2& -0.198 & 0.049 & 0.946 & - & - & -\\
& 0.5&-0.195 & 0.050  & 0.945 & - & - & - \\
&0.8& -0.193 & 0.053  & 0.951 & - & - & - \\
\hline 
\multirow{3}{*}{R-WCLS} & 0.2 &-0.200 & 0.044  & 0.950 & 100\% & 1.231 & 1.260  \\
& 0.5&-0.199 & 0.045  & 0.944 & 100\% & 1.218 & 1.255  \\
&0.8& -0.200 & 0.048  & 0.956 & 99.9\% & 1.203 & 1.236 \\
\hline 
\multirow{3}{*}{DR-WCLS} & 0.2 & -0.200 & 0.045  & 0.954 & 99.7\% & 1.216 & 1.249  \\
& 0.5&-0.199 & 0.045  & 0.947 & 99.9\% & 1.228 & 1.261  \\
&0.8& -0.200 & 0.047  & 0.954 & 99.7\% & 1.254 & 1.282 \\
\hline
\end{tabular}
\end{center}
\end{table}








\section{Intern Health Study: A Worked Example}
\label{sec:casestudy}

The IHS is a 6-month micro-randomized trial on medical interns \citep{necamp2020}, which aimed to investigate when to provide mHealth interventions to individuals in stressful work environments to improve their behavior and mental health. In this section, we evaluate the effectiveness of targeted notifications in improving individuals' moods and step counts. The exploratory and MRT analyses conducted in this paper focus on weekly randomization, thus, an individual was randomized to receive mood, activity, sleep, or no notifications with equal probability ($1/4$ each) every week. 
We choose the outcome $Y_{t+1,j}$ as the self-reported mood score (a Likert scale taking values from 1 to 10) and step count (cubic root) for individual $j$ in study week $t$.

Missing data occurred throughout the trial when interns did not complete the self-reported mood survey or were not wearing their assigned Fitbit wrist-worn device; thus, multiple imputation was originally used to impute missing daily data. 
 See \cite{necamp2020} for further details. The following analysis is based on one of the imputed data sets. The data set used in the analyzes contains 1,562 participants. The average weekly mood score when a notification is delivered is 7.14, and 7.16 when there is no notification; the average weekly step count (cubic root) when a notification is delivered is 19.1, and also 19.1 when there is no notification. In Section \ref{sec:casestudy_1} and \ref{sec:casestudy_2}, we evaluate the targeted notification treatment effect for medical interns using our proposed methods and WCLS. 

\subsection{Comparison of the Marginal Effect Estimation}
\label{sec:casestudy_1}
First, we are interested in assessing the fully marginal excursion effect (that is, $\beta(t) = \beta_0^\star$). For an individual $j$, the study week is coded as a subscript $t$. $Y_{t+1,j}$ is the self-reported mood score or step count (cubic root) of the individual $j$ in study week $t+1$. $A_t$ is defined as the specific type of notification that targets improving the outcome. For example, if the outcome is the self-reported mood score, sending mood notifications would be the action, thus $\P(A_t=1) = 0.25$. We analyze the marginal causal effect $\beta_0$ of the targeted notifications on self-reported mood score and step count using the following model for WCLS: 
$$
    Y_{t+1,j} \sim g_t(H_{t,j})^\top\alpha + (A_{t,j}-\tilde p_t) \beta_0. 
$$ 

The term $g_t(H_t)^\top\alpha$ represents a linear working model of prognostic control variables that includes two baseline characteristics, study week $t$ and the outcome of the previous week $Y_{t,j}$. For the R-WCLS and DR-WCLS methods, we include a total of 12 control variables and use random forests to construct the plug-in estimators $\hat g_t(H_t,A_t)$ as described in Section \ref{sec:algorithms}. For a detailed description of the control variables, see Appendix \ref{app:casestudy}.

\begin{table}
\caption{IHS Study: Fully marginal treatment effect estimation.}
\label{tab:fullymarginal}
\begin{center}
\begin{tabular}{cccccc}
\hline
Outcome & Method & Estimation & Std.err & P-value & RE  \\\hline
\multirow{3}{*}{Mood} & WCLS & -0.016 & $9.03 \times 10^{-3}$  & 0.078 & - \\
& R-WCLS& -0.017 & $8.14 \times 10^{-3}$  & \textbf{0.038} & 1.23  \\
&DR-WCLS & -0.017 & $8.18 \times 10^{-3}$  & \textbf{0.042} & 1.22  \\
\hline 
\multirow{3}{*}{Steps} & WCLS & 0.070 & $2.41 \times 10^{-2}$  & \textbf{0.004} & - \\
& R-WCLS& 0.065 & $2.34 \times 10^{-2}$  & \textbf{0.005} & 1.06 \\
& DR-WCLS & 0.070 & $2.37 \times 10^{-2}$  & \textbf{0.003} & 1.03 \\
\hline
\end{tabular}
\end{center}
\end{table}


Table \ref{tab:fullymarginal} summarizes various estimators, with details in Appendix \ref{app:casestudy}. Compared to WCLS, R-WCLS and DR-WCLS show noticeably smaller standard errors. We conclude that sending activity notifications increases the cubic root of step counts by $0.07$, while mood notifications reduces mood scores by $-0.017$, both significant at 95\%. Unlike WCLS, R-WCLS and DR-WCLS have sufficient power to detect the negative effect on mood scores.

\subsection{Time-varying Treatment Effect Estimation}
\label{sec:casestudy_2}

\begin{figure}
\begin{center}
\includegraphics[width=6in]{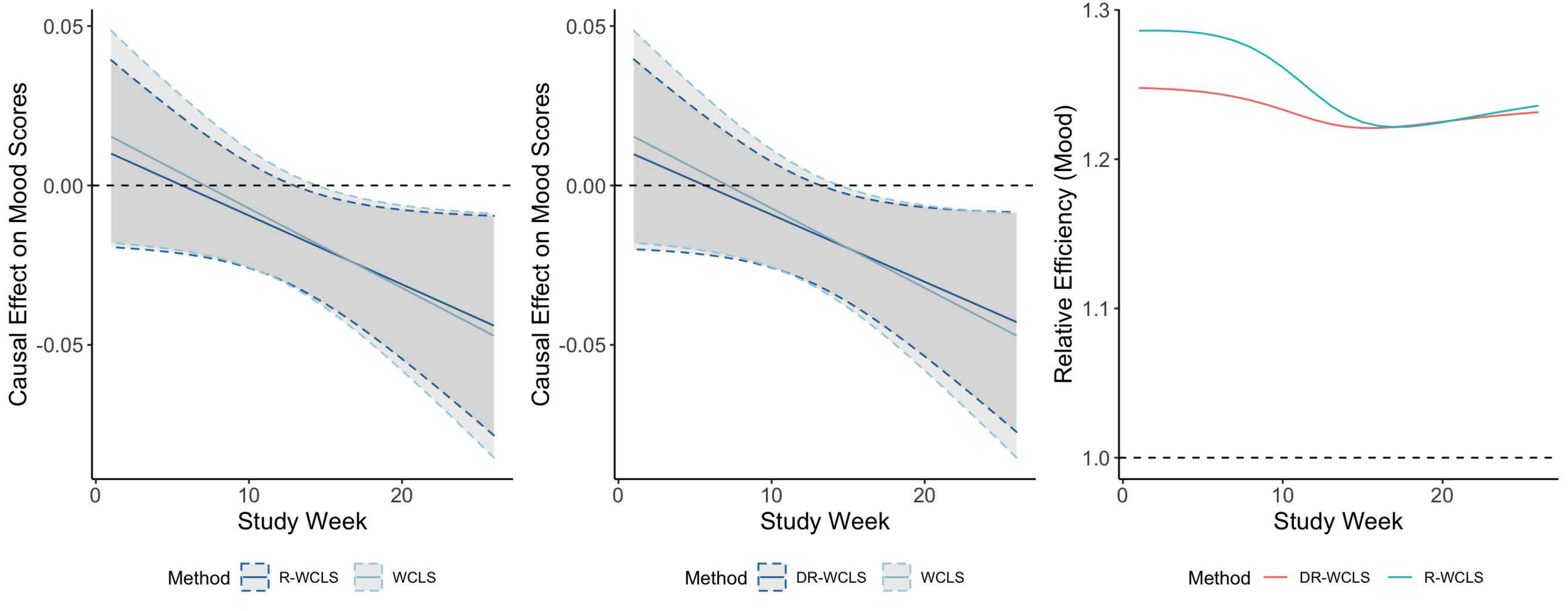}
\end{center}
\caption{Causal effects estimates with confidence intervals of R-WCLS (\textbf{left}) and DR-WCLS  (\textbf{middle}), and their relative efficiency in comparisons with WCLS (\textbf{right}). \label{fig:first}}
\end{figure}

For further analysis, we include study week in the moderated treatment effect model: $\beta(t)  = \beta_0^\star+\beta_1^\star t$, and examine how treatment effect varies over time. Estimated time-varying treatment moderation effects and their relative efficiency are shown in Figure \ref{fig:first}.  The shaded area in Figure \ref{fig:first} represents the 95\% confidence band of the moderation effects as a function of the study week. Narrower confidence bands were observed for estimators constructed using both R-WCLS and DR-WCLS methods. Relative efficiencies between 1.2 and 1.3 were observed over the study week.

Based on these results, sending notifications does not significantly affect mood scores during the first $12$ weeks, but later notifications are less likely to improve mood. Thus, overburdening participants with notifications over time may be unhelpful if there is no therapeutic benefit. We also examined time-varying effects on step counts (see Appendix \ref{app:casestudy}). We applied our methods to raw observed data with 31.3\% and 48.1\% missingness for mood and step count outcomes, respectively. The results in Appendix \ref{sec:casestudy_3} show that mood notifications no longer show a significant overall effect, but activity notifications still positively impact step counts.

\section{Discussion}
\label{sec:discussion}

Scientists wish to take advantage of the large volume of data generated by mobile health systems to better answer scientific questions regarding the time-varying intervention effects. Although machine learning algorithms can effectively handle high-dimensional mobile health data, their black-box nature can sometimes raise concerns about the validity of the results if used without care. In this paper, we introduce two rigorous inferential procedures—(efficient) R-WCLS and DR-WCLS—along with their bidirectional asymptotic properties. These approaches provide flexibility in specifying nuisance models and promise improved estimation efficiency compared to existing methods. In particular, the DR-WCLS criterion is especially powerful when both the treatment randomization probability and the conditional expectation model are correctly specified, resulting in the highest relative asymptotic efficiency. 

\backmatter


\section*{Acknowledgements}

The authors thank Swaraj Bose for running the simulations in Appendices D.2 and D.3 on the Michigan Biostatistics cluster to support the completion of this work.
\vspace*{-8pt}

\section*{Supplementary Materials}

Web Appendices, Tables, and Figures referenced in Sections 3, 4, 5, and 6 are available with this paper at the Biometrics website on Oxford Academic. The R code used to generate the simulation experiments and case study results in this paper can be obtained at \verb"https://github.com/Herashi/Doubly-Robust-WCLS".
\vspace*{-8pt}

\section*{Data Availability}

The dataset used in this paper comes from the Intern Health Sudy (IHS). The dataset is not publicly available, but can be obtained through an internal process of the study, based on a Data and Materials Distribution Agreement (DMDA).
\vspace*{-8pt}

\bibliographystyle{biom} 
\bibliography{paper-ref}

\label{lastpage}

\newpage
\appendix

\section{More on R-WCLS}
\label{app:R-WCLS}

\subsection{Neyman Orthogonality}

To ensure robustness and valid inference for $\beta$, we require Neyman orthogonality for the estimating equation \eqref{eq:r-wcls}~\citep{chernozhukov2015}. The Gateaux derivative operator with respect to $\mathbf{g}$ is:
\begin{equation}
\label{eq:Gateaux_deri}
    G(\mathbf{g}) = \E\Big[\sum_{t=1}^T W_t g_t(H_t)(A_t - \tilde p_t(1|S_t))f_t(S_t)  \Big](\mathbf{g} - \mathbf{g}^\star ) = 0,
\end{equation}
thus Equation \eqref{eq:r-wcls} satisfies Neyman orthogonality. Intuitively, Neyman orthogonality implies that the moment conditions used to identify $\beta^\star$ are sufficiently insensitive to the nuisance parameter estimates, allowing us to directly plug in estimates of $\mathbf{g}^\star$ while still obtaining high-quality inference for $\beta$. 


\subsection{Algorithm}

The algorithm for the R-WCLS criterion follows a routine similar to the DR-WCLS algorithm introduced in Section \ref{sec:algorithms}. The details are outlined below:

\textbf{Step I}: Let $K$ be a fixed integer. Form a $K$-fold random partition of $\{1,2,\dots,n\}$ by dividing it to equal parts, each of size $n/k$, assuming $n$ is a multiple of $k$. From each set $I_k$, let $I_k^\complement$ denote the observation indices that are not in $I_k$.

\textbf{Step II}: Learn the appropriate working models for each fold $I_k$ using the individuals in $I_k^\complement$.  Let~$\hat g_t^{(k)} (H_t, A_t)$, $\hat p_t^{(k)} (1 | H_t)$, and $\hat{\tilde p}_t^{(k)} (1 | S_t)$ denote the estimates for~$\E[Y_{t+1} | H_t, A_t]$, $\E[A_t | H_t]$, and $\E[A_t | S_t]$, respectively, that is, estimates of the nuisance parameters in the $k$th fold.  Note that when randomization probabilities are known, $\hat p_t(A_t | H_t)$ is set equal to $p_t (A_t | H_t)$.

\textbf{Step III}: For individual $j$ at time $t$, define the pseudo-outcome:
    $$\tilde Y^{(R)}_{t+1,j} := Y_{t+1,j} - \hat g^{(k)}_t(H_{t,j}, A_{t,j}) + \left(A_{t,j} - \hat{\tilde p}_t^{(k)} (1 | S_{t,j}) \right) \left( \hat g_t^{(k)} (H_{t,j}, 1) - \hat g_t^{(k)} (H_{t,j}, 0) \right),
    $$
    where $j \in I_k$. Then regress~$\tilde Y^{(R)}_{t+1}$ on $(A_t - \hat{ \tilde p}_t^{(k)} (1|S_t)) f_t (S_t)^\top \beta$ with weights $\hat W_t^{(k)} = \hat{\tilde p}_t^{(k)} (A_t|S_t)/ \hat p_t^{(k)} (A_t | H_t)$ to obtain estimate $\hat\beta_n^{(R)}$. In particular, with the algorithm outlined above, $\Sigma_R$ can be consistently estimated by:
 \begin{equation*}
     \Big[ \frac{1}{K} \sum_{k=1}^K \mathbb{P}_{n,k} \big \{ \dot m (\hat \beta, \hat \eta_k)  \big \} \Big]^{-1} \times 
\Big[ \frac{1}{K} \sum_{k=1}^K \mathbb{P}_{n,k} \big \{ m (\hat \beta, \hat \eta_k) m (\hat \beta, \hat \eta_k)^\top \big \} \Big] \times
\Big[ \frac{1}{K} \sum_{k=1}^K \mathbb{P}_{n,k} \big\{  \dot m (\hat \beta, \hat \eta_k) \big \} \Big]^{-1},
 \end{equation*}
where $\mathbb{P}_{n,k} \{ \bullet \}$ refers to the empirical average within fold $k$, and
\begin{align*}
    m (\hat \beta, \hat \eta_k) &= \sum_{t=1}^T \hat W^{(k)}_t \left( \tilde Y^{(R)}_{t+1}- (A_t - \hat{ \tilde p}_t^{(k)} (1|S_t)) f_t (S_t)^\top \hat\beta^{(R)}_n \right)(A_t - \hat{ \tilde p}_t^{(k)} (1|S_t))f_t (S_t), \\
    \dot m (\hat \beta, \hat \eta_k) &= \frac{\partial m (\beta, \hat \eta_k)}{\partial \beta} \Big|_{\beta = \hat\beta} =  \sum_{t=1}^T \hat{ \tilde p}_t^{(k)} (1|S_t)(1-\hat{ \tilde p}_t^{(k)} (1|S_t)) f_t (S_t)f_t (S_t)^\top.
\end{align*}


\subsection{Connection Between R-WCLS and DR-WCLS}
\label{sec:connect_rdr}

In recent work from \cite{morzywolek2023general}, a unified framework was presented to estimate heterogeneous treatment effects, resulting in a class of weighted loss functions with nuisance parameters. They showed that the R-Learner \citep{nie2021}  and the DR-Learner \citep{kennedy2020optimal} can be seen as special cases resulting from particular weighting choices. Here, we present a complementary viewpoint by showing a simple relationship between the two proposed R-WCLS and DR-WCLS methods. We begin by adding and subtracting $g_t(H_t,A_t) = A_t g_t (H_t,1) + (1-A_t) g_t (H_t, 0)$ from Equation \eqref{eq:r-wcls}:
$$
\mathbb{P}_n \Big[ \sum_{t=1}^T W_t \left( Y_{t+1}
- g_t(H_t, A_t) + \left(A_t - \tilde p_t (1 | S_t) \right) \left( \beta (t; H_t) - f_t (S_t)^\top \beta \right) \right)^2 \Big].
$$
One can then obtain an estimate of $\beta^\star$ by solving the following estimating equation:
\begin{align}
 &0 = \mathbb{P}_n \Big[ \sum_{t=1}^T W_t \left( Y_{t+1}
- g_t(H_t, A_t)  \right)\left(A_t - \tilde p_t (1 | S_t) \right)f_t (S_t) \Big] + \label{eq:firstterm}\\
&~~~~~~~~~~~~~~~~~~~~~ \mathbb{P}_n \Big[ \sum_{t=1}^T W_t   \left(A_t - \tilde p_t (1 | S_t) \right)^2 \left( \beta (t; H_t) - f_t (S_t)^\top \beta \right) f_t (S_t) \Big]. \label{eq:secondterm}
\end{align}

Under the correct specification of the randomization probabilities, the Gateaux derivative with respect to $\mathbf{g}$ of both terms \eqref{eq:firstterm} and \eqref{eq:secondterm} will be 0. However, if the randomization probabilities are not specified correctly, term \eqref{eq:secondterm} may not have a Gateaux derivative of 0.
To address this, we replace the stochastic term $W_t (A_t - \tilde p_t(1|S_t))^2$ in \eqref{eq:secondterm} with its expectation under the correct randomization probability:
\begin{align*}
    \E\Big[ \sum_{t=1}^T \tilde \sigma^2_t(S_t) (\beta (t; H_t) - f_t (S_t)^\top \beta^\star) f_t (S_t)\Big].
\end{align*}
After this substitution, we recover \eqref{eq:drwcls}. And by doing so, the Gateaux derivative with respect to $\mathbf{g}$ of both terms will no longer be affected by the randomization probability specification. The above derivation links the R-WCLS and DR-WCLS, showing that the doubly-robust estimators can be constructed from R-learner methods. Finally, \eqref{eq:r-wcls} and \eqref{eq:drwcls} yield estimation procedures that are presented in Section~\ref{sec:algorithms}.

\begin{remark}[Connection to the WCLS criterion]
The R-WCLS criterion replaces $g_t (H_t)^\top \alpha$ in the WCLS criterion, which was a linear working model for $\E[ W_{t} Y_{t+1} | H_t ]$, with a general choice of working models.  
Setting $g_t(H_t,A_t)$ to be the linear working model~$g_t(H_t)^\top \alpha + (A_t - \tilde p_t (1|S_t)) f_t(S_t)^\top \beta$, the R-WCLS criterion recovers the original WCLS criterion.  Thus,~\eqref{eq:r-wcls} is a strict generalization of~\eqref{eq:wcls}. 
\end{remark}

\begin{remark}[Connection to the R-learner]
In traditional causal inference with a single treatment~$A$, fully-observed set of confounders $X$, and outcome~$Y$, a two-stage estimator, referred to as the R-Learner, was previously proposed by \citet{nie2021}.  Beyond our extension to the time-varying setting, there are two key distinguishing features of R-WCLS in \eqref{eq:r-wcls} compared to R-Learner. 
First, we focus on estimating a low-dimensional target parameter, whereas R-learner seeks to estimate the conditional average treatment effect and allows it to be a complex function of baseline covariates.
Second, the weight $W_t$ in R-WCLS criterion implicitly depends on the propensity $\tilde p_t (1|S_t)$, we thereby replace the R-learner data-adaptive model for $\E[W_t Y_{t+1} | H_t]$ with one for each $\E[Y_{t+1} | H_t, a]$, $a \in \{0,1\}$, which is invariant to different choices of moderators $S_t$.

In particular, R-WCLS suggests incorporating data-adaptive plug-ins $\hat g_t(H_t,A_t)$ into the estimating equation instead of directly forming a projection estimation using $\hat g_t(H_t,1) - \hat g_t(H_t,0)$. This consideration is driven by the potential risk of introducing causal bias using the latter approach, unless the ML predictions converge fast enough (at a rate of $O_p(n^{-1/2})$).

\end{remark}

\subsection{Proof of Theorem \ref{thm:asymptotics_r}}
\label{app:r-wcls}

Assume $\hat\beta^{(R)}_n$ minimizes the R-WCLS criterion:
\begin{equation}
\P_n \Big[ \sum_{t=1}^T W_t \left( Y_{t+1} - \left( \tilde p_t (1 | S_t) g_t (1, H_t) + (1-\tilde p_t (1 | S_t)) g_t (0, H_t)  \right) - (A_t - \tilde p_t (1 | S_t)) ) f_t (S_t)^\top \beta \right)^2 \Big].
\end{equation}

Denote $g^\star_t(H_t) = \tilde p_t (1 | S_t) g_t (H_t,1 ) + (1-\tilde p_t (1 | S_t)) g_t ( H_t,0) = \E[W_tY_{t+1}|H_t]$, where we applied a supervised learning algorithm and obtain an estimator $\hat g_t(H_t)$. The asymptotic properties of the R-WCLS estimator follow from the expansion.
\begin{align}
\label{app:eq:estimatingeq_R}
    0 &= \P_n \Big[\sum_{t=1}^T W_t \left( Y_{t+1}- \hat g(H_t) - (A_t - \tilde p_t (1 | S_t)) f_t (S_t)^\top \hat\beta^{(R)}_n \right)(A_t - \tilde p_t (1 | S_t))f_t (S_t) \Big] \nonumber \\
    & = \P_n \Big[\sum_{t=1}^T W_t \left( Y_{t+1}- g^\star_t(H_t) -  (A_t - \tilde p_t (1 | S_t)) f_t (S_t)^\top \beta^{\star} \right)(A_t - \tilde p_t (1 | S_t))f_t (S_t) \Big] \nonumber \\
    & ~~~~~ - \P_n\Big[\sum_{t=1}^T W_t (A_t - \tilde p_t (1 | S_t))^2 f_t (S_t)f_t (S_t)^\top  \Big](\hat\beta^{(R)}_n-\beta^{\star}) \nonumber \\
    & ~~~~~ + \P_n\Big[\sum_{t=1}^T W_t (A_t - \tilde p_t (1 | S_t)) (g^\star_t(H_t) - \hat g(H_t)) f_t (S_t)  \Big].
\end{align}
By the Weak Law of Large Number (WLLN), we have the following:
\begin{align*}
    &\P_n\Big[\sum_{t=1}^T W_t (A_t - \tilde p_t (1 | S_t))^2 f_t (S_t)f_t (S_t)^\top  \Big]  \overset{P}{\to} \E\Big[\sum_{t=1}^T \tilde p_t (1 | S_t)(1-\tilde p_t (1 | S_t)) f_t (S_t)f_t (S_t)^\top  \Big], \\
    &\P_n \Big[\sum_{t=1}^T W_t (A_t - \tilde p_t (1 | S_t)) f_t (S_t) (g^\star(H_t)-\hat g(H_t)) \Big] \overset{P}{\to} 0 ~~~~\text{(by design).}
\end{align*}
The second convergence result holds true for any $\hat g(H_t)$ by design because:
\begin{align*}
    &\E\Big[\sum_{t=1}^TW_t (A_t - \tilde p_t (1 | S_t)) f_t (S_t) (g^\star(H_t)-\hat g(H_t)) \Big] \\
    =& \E\Big[\sum_{t=1}^T \E[ \tilde p_t (1 | S_t)(1 - \tilde p_t (1 | S_t)) f_t (S_t) (g^\star(H_t)-\hat g(H_t))|A_t = 1] \\
    & ~~~~~~~~~~ +\E[ (1-\tilde p_t (1 | S_t))(0 - \tilde p_t (1 | S_t)) f_t (S_t) (g^\star(H_t)-\hat g(H_t))|A_t = 0] \Big] \\
    =& 0.
\end{align*}

Therefore, under regularity conditions, the estimator $\hat\beta^{(R)}_n \overset{P}{\to} \beta^{\star}$; that is, $\hat\beta^{(R)}_n$ is a consistent estimator of $\beta^{\star}$. Denote $\tilde Y^{(R)}_{t+1} =Y_{t+1}- \hat g_t(H_t) $. Therefore, when $n \rightarrow \infty$, after solving \eqref{app:eq:estimatingeq_R}, we obtain the following:
\begin{align*}
    n^{1/2}(\hat\beta^{(R)}_n-\beta^{\star}) & = n^{1/2}~ \P_n \Big\{\sum_{t=1}^T ~ \E\Big[\sum_{t=1}^T \tilde p_t (1 | S_t)(1-\tilde p_t (1 | S_t)) f_t (S_t)f_t (S_t)^\top  \Big]^{-1} \times \\
    &~~~~~~~~~~~~~~~~~ W_t \left( \tilde Y^{(R)}_{t+1}- (A_t - \tilde p_t (1 | S_t)) f_t (S_t)^\top \beta^{\star} \right)(A_t - \tilde p_t (1 | S_t))f_t (S_t) \Big\} + o_p(1).
\end{align*}
By the definition of $\beta^\star$:
\begin{equation*}
    \E \left[\sum_{t=1}^T W_t \left( \tilde Y^{(R)}_{t+1}- (A_t - \tilde p_t (1 | S_t)) f_t (S_t)^\top \beta^{\star} \right)(A_t - \tilde p_t (1 | S_t))f_t (S_t) \right] =0
\end{equation*}
The influence function for $\hat\beta^{(R)}_n$ is:
\begin{align}
        &\sum_{t=1}^T ~  \E\Big[\sum_{t=1}^T \tilde p_t (1 | S_t)(1-\tilde p_t (1 | S_t)) f_t (S_t)f_t (S_t)^\top  \Big]^{-1} \times \nonumber \\
        & ~~~~~~~~~~~~~~~~~~~~~~~~~W_t \left( \tilde Y^{(R)}_{t+1}- (A_t - \tilde p_t (1 | S_t)) f_t (S_t)^\top \beta^{\star} \right)(A_t - \tilde p_t (1 | S_t))f_t (S_t).
\end{align}
Under moment conditions, we have asymptotic normality with variance given by $\Sigma_R=Q^{-1} W Q^{-1}$, where
\begin{align*}
    Q & = \E\Big[\sum_{t=1}^T \tilde p_t (1 | S_t)(1-\tilde p_t (1 | S_t)) f_t (S_t)f_t (S_t)^\top  \Big], \\
    W &= \E \Big[ \Big(\sum_{t=1}^T W_t \left( \tilde Y^{(R)}_{t+1}- (A_t - \tilde p_t (1 | S_t)) f_t (S_t)^\top \beta^{\star} \right)(A_t - \tilde p_t (1 | S_t))f_t (S_t) \Big)^2\Big],
\end{align*}
due to space constraints, we use $\E[X^2]$ to denote $\E[X X^\top]$. In conclusion, we establish that the estimator minimizing the R-WCLS criterion $\hat\beta^{(R)}_n$ is consistent and asymptotically normal: 
\begin{equation*}
    n^{1/2}(\hat\beta^{(R)}_n-\beta^{\star}) \sim \mathcal{N} (0, \Sigma_R).
\end{equation*}



We further prove that the variance is consistently estimated by Equation \eqref{eq:asymptotic_var}. Using sample splitting, the estimating equation can be written as:

\begin{align}
    0 &= \frac{1}{K}\sum_{k=1}^K \mathbb{P}_{n,k}  \Big[\sum_{t=1}^T W_t \left( Y_{t+1}- \hat g^{(k)}(H_t) - (A_t - \tilde p^{(k)}_t (1 | S_t)) f_t (S_t)^\top \hat\beta^{(R)}_n \right)(A_t - \tilde p^{(k)}_t (1 | S_t))f_t (S_t) \Big] \nonumber \\
    & = \frac{1}{K}\sum_{k=1}^K \mathbb{P}_{n,k} \Big[\sum_{t=1}^T W_t \left( Y_{t+1}- g^\star_t(H_t) -  (A_t - \tilde p^{(k)}_t (1 | S_t)) f_t (S_t)^\top \beta^{\star} \right)(A_t - \tilde p^{(k)}_t (1 | S_t))f_t (S_t) \Big] \nonumber \\
    & ~~~~~ - \frac{1}{K}\sum_{k=1}^K \mathbb{P}_{n,k} \Big[\sum_{t=1}^T W_t (A_t - \tilde p^{(k)}_t (1 | S_t))^2 f_t (S_t)f_t (S_t)^\top  \Big](\hat\beta^{(R)}_n-\beta^{\star}) \nonumber \\
    & ~~~~~ + \frac{1}{K}\sum_{k=1}^K \mathbb{P}_{n,k} \Big[\sum_{t=1}^T W_t (A_t - \tilde p_t (1 | S_t)) (g^\star_t(H_t) - \hat g^{(k)}(H_t)) f_t (S_t)  \Big]
\end{align}

Assume $K$ is finite and fixed, and we have the same reasoning as above:
\begin{align*}
    &\mathbb{P}_{n,k} \Big[\sum_{t=1}^T W_t (A_t - \tilde p^{(k)}_t (1 | S_t))^2 f_t (S_t)f_t (S_t)^\top  \Big]  \overset{P}{\to} \E\Big[\sum_{t=1}^T \tilde p^{(k)}_t (1 | S_t)(1-\tilde p^{(k)}_t (1 | S_t)) f_t (S_t)f_t (S_t)^\top  \Big], \\
    &\mathbb{P}_{n,k} \Big[\sum_{t=1}^T W_t (A_t - \tilde p^{(k)}_t (1 | S_t)) f_t (S_t) (g^\star(H_t)-\hat g^{(k)}(H_t)) \Big] \overset{P}{\to} 0 ~~~~\text{(by design)}
\end{align*}

Then we obtain the following:

\begin{align*}
    n^{1/2}(\hat\beta^{(R)}_n-\beta^{\star}) & = n^{1/2}~ \frac{1}{K}\sum_{k=1}^K \P_{n,k} \bigg[\sum_{t=1}^T ~ \bigg\{\frac{1}{K}\sum_{k=1}^K \E\Big[\sum_{t=1}^T \tilde p^{(k)}_t (1 | S_t)(1-\tilde p^{(k)}_t (1 | S_t)) f_t (S_t)f_t (S_t)^\top  \Big]\bigg\}^{-1} \times \\
    &~~~~~~~~~~~~~~~~~ W_t \left( \tilde Y^{(R)}_{t+1}- (A_t - \tilde p^{(k)}_t (1 | S_t)) f_t (S_t)^\top \beta^{\star} \right)(A_t - \tilde p^{(k)}_t (1 | S_t))f_t (S_t) \bigg] + o_p(1).
\end{align*}
By the definition of $\beta^\star$:
\begin{equation*}
    \E \left[\sum_{t=1}^T W_t \left( \tilde Y^{(R)}_{t+1}- (A_t - \tilde p^{(k)}_t (1 | S_t)) f_t (S_t)^\top \beta^{\star} \right)(A_t - \tilde p^{(k)}_t (1 | S_t))f_t (S_t) \right] =0
\end{equation*}
Consequently, under regularity conditions, the estimator $\hat\beta^{(R)}_n \overset{P}{\to} \beta^{\star}$; that is, $\hat\beta^{(R)}_n$ is consistent. The influence function for $\hat\beta^{(R)}_n$ is:
\begin{align}
        &\frac{1}{K}\sum_{k=1}^K \sum_{t=1}^T ~ \bigg\{ \frac{1}{K}\sum_{k=1}^K \E\Big[\sum_{t=1}^T \tilde p^{(k)}_t (1 | S_t)(1-\tilde p^{(k)}_t (1 | S_t)) f_t (S_t)f_t (S_t)^\top  \Big]\bigg\}^{-1} \times \nonumber \\
        & ~~~~~~~~~~~~~~~~~~W_t \left( \tilde Y^{(R)}_{t+1}- (A_t - \tilde p^{(k)}_t (1 | S_t)) f_t (S_t)^\top \beta^{\star} \right)(A_t - \tilde p^{(k)}_t (1 | S_t))f_t (S_t).
\end{align}

Recall
\begin{align*}
    m &= \sum_{t=1}^T \hat W^{(k)}_t \left( \tilde Y^{(R)}_{t+1}- (A_t - \hat{ \tilde p}_t^{(k)} (1|S_t)) f_t (S_t)^\top \hat\beta^{(R)}_n \right)(A_t - \hat{ \tilde p}_t^{(k)} (1|S_t))f_t (S_t) ,\\
    \dot m & = \frac{\partial m(\beta, \eta)}{\partial \beta}  =  \sum_{t=1}^T \hat{ \tilde p}_t^{(k)} (1|S_t)(1-\hat{ \tilde p}_t^{(k)} (1|S_t)) f_t (S_t)f_t (S_t)^\top .
\end{align*}
Then the variance can be consistently estimated by:
\begin{equation*}
     \Big[ \frac{1}{K} \sum_{k=1}^K \mathbb{P}_{n,k} \big \{ \dot m (\hat \beta, \hat \eta_k)  \big \} \Big]^{-1} \times 
\Big[ \frac{1}{K} \sum_{k=1}^K \mathbb{P}_{n,k} \big \{ m (\hat \beta, \hat \eta_k) m (\hat \beta, \hat \eta_k)^\top \big \} \Big] \times
\Big[ \frac{1}{K} \sum_{k=1}^K \mathbb{P}_{n,k} \big\{  \dot m (\hat \beta, \hat \eta_k) \big \} \Big]^{-1}.
 \end{equation*}

\section{An efficient R-WCLS estimator}
\label{app:ar-wcls}

As presented in the R-WCLS criterion in Equation \eqref{eq:r-wcls}, $ g_t(H_t,A_t)$ is only used to construct the plug-in estimator $g_t(H_t)$, but the difference between $ g_t(H_t,1)$ and $g_t(H_t,0)$, that is, the causal excursion effect under fully observed history, is not incorporated into the estimating equation \eqref{eq:r-wcls}. Here we introduce a more efficient R-WCLS criterion as follows:
\begin{equation}
\label{eq:ar-wcls}
    0 = \mathbb{P}_n \Big[ \sum_{t=1}^T W_t \big( Y_{t+1}
- g_t(H_t) - \left(A_t - \tilde p_t (1 | S_t) \right) \left( f_t (S_t)^\top \beta + \Lambda_t^\perp  \right) \big)\left(A_t - \tilde p_t (1 | S_t) \right)f_t (S_t) \Big],
\end{equation}
where $\Lambda_t^\perp$ denotes the projection of $g_t(H_t,1)-g_t(H_t,0)$ onto the orthogonal complement of $f_t(S_t)$. The definition of the orthogonal complement is provided in Appendix \ref{app:lemma:ar-wcls-implement} \eqref{ass:ortho-condition}, along with details on constructing a plug-in estimator of $\Lambda_t^\perp$. 


\subsection{Implementation of the efficient R-WCLS criterion}
\label{app:lemma:ar-wcls-implement}

Let $f_t(S_t)^\perp$ denote the orthogonal complement of $f_t(S_t)$ in $H_t$, which refers to the set of random variables that are uncorrelated with $f_t(S_t)$ smoothing over time. Here, a rigorous definition of the orthogonal complement of $f_t(S_t)$ is given below \citep{Shia2-wcls}:
\begin{equation}
    \label{ass:ortho-condition}
   f_t(S_t)^\perp \coloneqq \Big\{X_t \in H_t : \E\Big[\sum_{t=1}^T \tilde p_t (1- \tilde p_t)X_t f_t(S_t)Big] = 0 \Big\}.
\end{equation}

To construct $\Lambda_t^\perp$, i.e., the projection of $\beta(t;H_t)$ onto $f_t(S_t)^\perp$, we can apply a linear working model as follows:
\begin{equation*}
    \beta(t;H_t) \sim (f_t(S_t)^\perp)^\top\eta + f_t(S_t)^\top \beta.
\end{equation*}

Therefore, $\Lambda_t^\perp = (f_t(S_t)^\perp)^\top\eta$. This approach allows us to effectively leverage the information from the nuisance functions $\beta(t;H_t) = g_t(H_t,1) - g_t(H_t,0)$, which can be decomposed into $\Lambda_t^\perp$ and $f_t(S_t)^\top \beta$. Most importantly, the inclusion of $\Lambda_t^\perp$ in the estimating equation does not compromise the consistency of the estimator $\hat\beta_n^{(ER)}$.

\begin{lemma}
\label{lemma:ar-wcls}
    Let $\hat\beta^{(ER)}_n$ denote the efficient R-WCLS estimator obtained from solving Equation \eqref{eq:ar-wcls} in Appendix \ref{app:ar-wcls}. 
    Under Assumptions \ref{ass:po}, \ref{ass:directeffect}, and \ref{ass:p_correct}, given invertibility and moment conditions, $\hat\beta^{(ER)}_n$ is consistent and asymptotically normal such that $\sqrt{n}(\hat\beta^{(ER)}_n - \beta^\star) \rightarrow \mathcal{N}(0, \Sigma_{ER})$, where $\Sigma_{ER}$ is defined in Appendix \ref{app:lemma:ar-wcls}.
\end{lemma}

\subsection{Asymptotic properties}
\label{app:lemma:ar-wcls}

The asymptotic properties of the efficient R-WCLS estimator follow from the expansion:
\begin{align*}
    0 &= \P_n \Big[\sum_{t=1}^T W_t \left( Y_{t+1}- \hat g(H_t) - (A_t - \tilde p_t (1 | S_t)) \left( f_t (S_t)^\top \hat\beta^{(ER)}_n + \hat \Lambda_t^\perp  \right) \right)(A_t - \tilde p_t (1 | S_t))f_t (S_t) \Big] \nonumber \\
    & = \P_n \Big[\sum_{t=1}^T W_t \left( Y_{t+1}- g^\star_t(H_t) -  (A_t - \tilde p_t (1 | S_t)) \left(f_t (S_t)^\top \beta^{\star} +\Lambda_t^\perp \right) \right)(A_t - \tilde p_t (1 | S_t))f_t (S_t) \Big] \nonumber \\
    & ~~~~~ + \P_n\Big[\sum_{t=1}^T W_t (A_t - \tilde p_t (1 | S_t)) (g^\star_t(H_t) - \hat g(H_t)) f_t (S_t)  \Big]\\
    & ~~~~~ - \P_n\Big[\sum_{t=1}^T W_t (A_t - \tilde p_t (1 | S_t))^2  (\hat \Lambda_t^\perp-\Lambda_t^\perp) f_t (S_t)  \Big]\\
    & ~~~~~ - \P_n\Big[\sum_{t=1}^T W_t (A_t - \tilde p_t (1 | S_t))^2 f_t (S_t)f_t (S_t)^\top  \Big](\hat\beta^{(ER)}_n-\beta^{\star}) \nonumber \\
\end{align*}
By the WLLN, we have:
\begin{align*}
    &\P_n\Big[\sum_{t=1}^T W_t (A_t - \tilde p_t (1 | S_t))^2 f_t (S_t)f_t (S_t)^\top  \Big]  \overset{P}{\to} \E\Big[\sum_{t=1}^T \tilde p_t (1 | S_t)(1-\tilde p_t (1 | S_t)) f_t (S_t)f_t (S_t)^\top  \Big], \\
    &\P_n \Big[\sum_{t=1}^T W_t (A_t - \tilde p_t (1 | S_t)) f_t (S_t) (g^\star(H_t)-\hat g(H_t)) \Big] \overset{P}{\to} 0 ~~~~\text{(by design),} \\
    & \P_n\Big[\sum_{t=1}^T W_t (A_t - \tilde p_t (1 | S_t))^2  (\hat \Lambda_t^\perp -\Lambda_t^\perp )f_t (S_t)  \Big] \overset{P}{\to} 0 ~~~~\text{(orthogonal projection).}
\end{align*}
Consequently, under regularity conditions, the estimator $\hat\beta^{(ER)}_n \overset{P}{\to} \beta^{\star}$; that is, $\hat\beta^{(ER)}_n$ is consistent. Recall $\tilde Y^{(R)}_{t+1} =Y_{t+1}- \hat g_t(H_t) $. Setting $n \rightarrow \infty$, we obtain 
\begin{align*}
    n^{1/2}(\hat\beta^{(ER)}_n-\beta^{\star}) & = n^{1/2}~ \P_n \bigg\{\sum_{t=1}^T ~ \E\Big[\sum_{t=1}^T \tilde p_t (1 | S_t)(1-\tilde p_t (1 | S_t)) f_t (S_t)f_t (S_t)^\top  \Big]^{-1} \times \\
    &~~~~~~~ W_t \left( \tilde Y^{(R)}_{t+1}- (A_t - \tilde p_t (1 | S_t)) \left( f_t (S_t)^\top \beta^{\star} +\Lambda_t^\perp \right)  \right)(A_t - \tilde p_t (1 | S_t))f_t (S_t) \bigg\} + o_p(1).
\end{align*}

By definition of $\beta^\star$:
\begin{equation*}
    \E \Big[\sum_{t=1}^T W_t \left( \tilde Y^{(R)}_{t+1}- (A_t - \tilde p_t (1 | S_t)) \left( f_t (S_t)^\top \beta^{\star} +\Lambda_t^\perp \right)  \right)(A_t - \tilde p_t (1 | S_t))f_t (S_t) \Big] =0
\end{equation*}

The influence function for $\hat\beta^{(ER)}_n$ is:
\begin{align}
        &\sum_{t=1}^T ~  \E\Big[\sum_{t=1}^T \tilde p_t (1 | S_t)(1-\tilde p_t (1 | S_t)) f_t (S_t)f_t (S_t)^\top  \Big]^{-1} \times \nonumber \\
        & ~~~~~~~~~~~~~W_t \left( \tilde Y^{(R)}_{t+1}- (A_t - \tilde p_t (1 | S_t)) \left( f_t (S_t)^\top \beta^{\star} +\Lambda_t^\perp \right)  \right)(A_t - \tilde p_t (1 | S_t))f_t (S_t).
\end{align}

Then under moment conditions, we have asymptotic normality with variance given by $\Sigma_R=Q^{-1} W Q^{-1}$, where
\begin{align*}
    Q & = \E\Big[\sum_{t=1}^T \tilde p_t (1 | S_t)(1-\tilde p_t (1 | S_t)) f_t (S_t)f_t (S_t)^\top  \Big], \\
    W &= \E \Big[ \Big(\sum_{t=1}^T W_t \left( \tilde Y^{(R)}_{t+1}- (A_t - \tilde p_t (1 | S_t)) \left( f_t (S_t)^\top \beta^{\star} +\Lambda_t^\perp \right)  \right)(A_t - \tilde p_t (1 | S_t))f_t (S_t) \Big)^2\Big].
\end{align*}

In conclusion, we establish that the estimator minimizing the efficient R-WCLS criterion $\hat\beta^{(ER)}_n$ is consistent and asymptotically normal. Under sample splitting, the asymptotic variance can be estimated by Equation \eqref{eq:asymptotic_var} with:
\begin{align*}
    m &= \sum_{t=1}^T \hat W^{(k)}_t \left( \tilde Y^{(R)}_{t+1}- (A_t - \hat{ \tilde p}_t^{(k)} (1|S_t)) \left( f_t (S_t)^\top \hat\beta^{(ER)}_n + \hat \Lambda_t^{\perp (k)} \right) \right)(A_t - \hat{ \tilde p}_t^{(k)} (1|S_t))f_t (S_t) ,\\
    \dot m & = \frac{\partial m(\beta, \eta)}{\partial \beta}  =  \sum_{t=1}^T \hat{ \tilde p}_t^{(k)} (1|S_t)(1-\hat{ \tilde p}_t^{(k)} (1|S_t)) f_t (S_t)f_t (S_t)^\top .
\end{align*}

\subsection{Efficiency gain over the WCLS estimator}
\label{app:dif_rnwcls}

To reconcile the notations, we write the estimating equation in a general form, from which can obtain a consistent estimate of $\beta^\star$ by solving:
$$
\E \Big[ \sum_{t=1}^T W_t \left( Y_{t+1}
- g_t(H_t) - \left(A_t - \tilde p_t (1 | S_t) \right) \left(  f_t (S_t)^\top \beta + \Lambda_t^\perp \right) \right)\left(A_t - \tilde p_t (1 | S_t) \right)f_t (S_t) \Big] = 0.
$$



For WCLS, denote the linear working model for $\E[Y_{t+1}|H_t,A_t]$ as $\tilde g_t(H_t, A_t)$. We can then write the estimating equation as:
\begin{align*}
    \E \Big[ \sum_{t=1}^T W_t \left( Y_{t+1}
- \tilde g_t(H_t) - \left(A_t - \tilde p_t (1 | S_t) \right)  (f_t (S_t)^\top \beta + \tilde \Lambda_t^\perp)\right)\left(A_t - \tilde p_t (1 | S_t) \right)f_t (S_t) \Big] =0.
\end{align*}

For an efficient R-WCLS estimator, recall $ \tilde Y^{(R)}_{t+1} = Y_{t+1}- g(H_t)$, the estimating equation can be written as:
\begin{align*}
   0 &= \E \Big[ \sum_{t=1}^T W_t \left( Y_{t+1}
- g_t(H_t) - \left(A_t - \tilde p_t (1 | S_t) \right) \left( f_t (S_t)^\top \beta + \Lambda_t^\perp  \right) \right)\left(A_t - \tilde p_t (1 | S_t) \right)f_t (S_t) \Big] \\
&= \E \Big[ \sum_{t=1}^T W_t \left( \tilde Y^{(R)}_{t+1} - \left(A_t - \tilde p_t (1 | S_t) \right) \left( f_t (S_t)^\top \beta + \Lambda_t^\perp  \right) \right)\left(A_t - \tilde p_t (1 | S_t) \right)f_t (S_t) \Big].
\end{align*}

Since both methods yield consistent estimates, we now compare their asymptotic variances. To demonstrate that the efficient R-WCLS estimator has a smaller asymptotic variance than the WCLS estimator (i.e., $\Sigma^{(ER)} - \Sigma$ is negative semidefinite), we require the following assumption:
\begin{assumption}
\label{ass:conditional-ind}
    The residual $e_t \coloneqq Y_{t+1} - g(H_t, A_t)$ is  uncorrelated with future states given history~$H_t$ and treatment~$A_t$, i.e., $\E[e_t f_{t'}(S_{t'})\Lambda^\perp_{t'}|H_t, A_t] =0, ~ \forall t < t'$.
\end{assumption}
For the WCLS estimator, the asymptotic variance can be calculated as:
\begin{align*}
    \Sigma = & \E\Big[\sum_{t=1}^T \tilde p_t (1 | S_t)(1-\tilde p_t (1 | S_t)) f_t (S_t)f_t (S_t)^\top  \Big]^{-1} \times \\
    &~~~~~ \E \Big[\Big(\sum_{t=1}^T W_t \left( Y_{t+1} -\tilde g_t(H_t) - (A_t - \tilde p_t (1 | S_t)) (f_t (S_t)^\top \beta^{\star}+\tilde \Lambda_t^\perp) \right)(A_t - \tilde p_t (1 | S_t)) f_t (S_t)\Big)^2 \Big] \times\\
    &~~~~~ \E\Big[\sum_{t=1}^T \tilde p_t (1 | S_t)(1-\tilde p_t (1 | S_t)) f_t (S_t)f_t (S_t)^\top  \Big]^{-1},
\end{align*}
and for the efficient R-WCLS estimator, the asymptotic variance can be calculated as:
\begin{align*}
    \Sigma^{(ER)} = & \E\Big[\sum_{t=1}^T \tilde p_t (1 | S_t)(1-\tilde p_t (1 | S_t)) f_t (S_t)f_t (S_t)^\top  \Big]^{-1} \times \\
    &~~~~~~~~~~~~ \E \Big[\Big(\sum_{t=1}^T W_t \left(\tilde Y^{(R)}_{t+1}  - (A_t - \tilde p_t (1 | S_t)) (f_t (S_t)^\top \beta^{\star} + \Lambda_t^\perp )\right)(A_t - \tilde p_t (1 | S_t)) f_t (S_t)\Big)^2 \Big] \times\\
    &~~~~~~~~~~~~  \E\Big[\sum_{t=1}^T \tilde p_t (1 | S_t)(1-\tilde p_t (1 | S_t)) f_t (S_t)f_t (S_t)^\top  \Big]^{-1}
\end{align*}

Denote $\epsilon(H_t) = g(H_t)  -\tilde g(H_t)$, we have the following derivation: 
\begin{align*}
    & \E \Big[\Big(\sum_{t=1}^T W_t \left( Y_{t+1} -\tilde g_t(H_t) - (A_t - \tilde p_t (1 | S_t)) (f_t (S_t)^\top \beta^{\star}+\tilde \Lambda_t^\perp) \right)(A_t - \tilde p_t (1 | S_t)) f_t (S_t)\Big)^2 \Big] \\
    =& \E \Big[\Big(\sum_{t=1}^T W_t \left(\tilde Y^{(R)}_{t+1} + \epsilon(H_t) - (A_t - \tilde p_t (1 | S_t)) (f_t (S_t)^\top \beta^{\star}+\tilde \Lambda_t^\perp + \Lambda_t^\perp -\Lambda_t^\perp ) \right)(A_t - \tilde p_t (1 | S_t)) f_t (S_t)\Big)^2 \Big] \\
    =& \E \Big[\Big(\sum_{t=1}^T W_t \left(\tilde Y^{(R)}_{t+1}  - (A_t - \tilde p_t (1 | S_t))(f_t (S_t)^\top \beta^{\star} + \Lambda_t^\perp )\right)(A_t - \tilde p_t (1 | S_t)) f_t (S_t)\Big)^2 \Big]\\
    & ~~~~~~ +\E \Big[\Big(\sum_{t=1}^T W_t \epsilon(H_t,A_t)(A_t - \tilde p_t (1 | S_t)) f_t (S_t)\Big)^2 \Big]\\
    \geq & \E \Big[\Big(\sum_{t=1}^T W_t \left(\tilde Y^{(R)}_{t+1}  - (A_t - \tilde p_t (1 | S_t)) (f_t (S_t)^\top \beta^{\star} + \Lambda_t^\perp )\right)(A_t - \tilde p_t (1 | S_t)) f_t (S_t)\Big)^2 \Big],
\end{align*}
where $\epsilon(H_t,A_t) = \epsilon(H_t) + (A_t -\tilde p_t (1 | S_t) )(\Lambda_t^\perp-\tilde \Lambda_t^\perp)$. The interaction term is 0 because:
\begin{align*}
    &\E\Big[\sum_{t, t'}^T W_t \left( \tilde Y^{(R)}_{t+1}  - (A_t - \tilde p_t ) (f_t (S_t)^\top \beta^{\star} + \Lambda_t^\perp )\right)(A_t - \tilde p_t)f_t (S_t)W_{t'}\epsilon(H_{t'},A_{t'})(A_{t'} - \tilde p_{t'})f_{t'} (S_{t'})^\top \Big]\\
    =& \E\Big[\sum_{t, t'}^T W_t \left(Y_{t+1}  - g^\star_t(H_t,A_t) \right) (A_t - \tilde p_t)f_t (S_t)W_{t'}\epsilon(H_{t'},A_{t'})(A_{t'} - \tilde p_{t'})f_{t'} (S_{t'})^\top \Big]
\end{align*}
Here the first to second line uses the fact that $f_t(S_t)^\top \beta^\star + \Lambda_t^\perp = g_t(H_t,1) - g_t (H_t, 0)$ so we can then get $\tilde Y^{(R)}_{t+1}  - (A_t - \tilde p_t ) (f_t (S_t)^\top \beta^{\star} + \Lambda_t^\perp ) = Y_{t+1}  - g^\star_t(H_t,A_t) $. For $t \geq t'$, by iterated expectation, we have:
\begin{align*}
    & \E\Big[\sum_{t, t'}^T W_t \underbrace{\E\left[\left(Y_{t+1}  - g^\star_t(H_t,A_t) \right) |H_t,A_t\right] }_{=0}W_{t'}\epsilon(H_{t'},A_{t'})(A_{t'} - \tilde p_{t'})f_{t'} (S_{t'})^\top(A_t - \tilde p_t)f_t (S_t)\Big]\\
    & = 0.
\end{align*}

For $t<t'$, by iterated expectation, we have:

\begin{align*}
    & \E\Big[\sum_{t, t'}^T W_t \left(Y_{t+1}  - g^\star_t(H_t,A_t) \right) \E\left[W_{t'}\epsilon(H_{t'},A_{t'})(A_{t'} - \tilde p_{t'})f_{t'} (S_{t'})^\top |H_{t'}\right](A_t - \tilde p_t)f_t (S_t)\Big] \\
   = & \E\Big[\sum_{t, t'}^T W_t \left(Y_{t+1}  - g^\star_t(H_t,A_t) \right) \E\left[\tilde p_{t'}(1 - \tilde p_{t'})\Lambda_{t'}^\perp f_{t'} (S_{t'})^\top |H_{t'}\right](A_t - \tilde p_t)f_t (S_t)\Big] \\
   = & \E\Big[\sum_{t, t'}^T W_t \E \big[\underbrace{\left(Y_{t+1}  - g^\star_t(H_t,A_t) \right) \tilde p_{t'}(1 - \tilde p_{t'})\Lambda_{t'}^\perp f_{t'} (S_{t'})^\top  |A_t,H_t }_{\text{conditionally independent by Assumption \ref{ass:conditional-ind}}} \big] (A_t - \tilde p_t)f_t (S_t)\Big] \\
    = & \E\Big[\sum_{t, t'}^T W_t \underbrace{\E \left[Y_{t+1}  - g^\star_t(H_t,A_t)|A_t,H_t\right] }_{=0}\E\left[\tilde p_{t'}(1 - \tilde p_{t'})\Lambda_{t'}^\perp f_{t'} (S_{t'})^\top |A_t,H_t \right] (A_t - \tilde p_t)f_t (S_t)\Big] \\
   = & 0.
\end{align*}


Therefore, the above derivation shows that $\Sigma^{(ER)}-\Sigma$ is negative semidefinite. This indicates that using the efficient R-WCLS to estimate treatment effect $\beta^\star$ is more efficient than WCLS. In the case when we estimate $\beta_t^\star$ nonparametrically rather than smoothing over time, the interaction terms for $t \neq t'$ do not exist, therefore the conclusion holds without the conditional independence assumption.

\section{Proof of Theorem \ref{thm:asymptotics_dr}}
\label{app:dr-wlcs}

\subsection{Double robustness property}

The following is proof of the double robustness of the DR-WCLS estimator. Assume $\hat \beta^{(DR)}_n$ minimizes the \emph{DR-WCLS} criterion:

\begin{equation*}
    \mathbb{P}_n \Big[\sum_{t=1}^T \tilde \sigma^2_t(S_t) \Big(\frac{W_t (A_t - \tilde p_t (1|S_t))(Y_{t+1}- g_t(H_t,A_t))}{\tilde \sigma^2_t(S_t)} + \beta (t; H_t) - f_t(S_t)^\top \beta\Big)^2\Big].  
\end{equation*}

Here the true randomization probability is $p_t(A_t|H_t)$, and the outcome conditional expectation (also known as the outcome regression):
\begin{equation*}
    \E[Y_{t+1}|H_t,A_t] = g_t^\star(H_t,A_t)
\end{equation*}
Denote $\beta (t; H_t) = g_t^\star(H_t,1)-g_t^\star(H_t,0) $. The corresponding ML estimators are denoted as $\hat g_t(H_t,A_t)$ and $\hat \beta (t; H_t)$. We consider the estimating equation of the objective function above:
\begin{align*}
    0 &= \E \Big[\sum_{t=1}^T \tilde \sigma^2_t(S_t) \Big(\frac{W_t (A_t - \tilde p_t (1|S_t))(Y_{t+1}-  g_t(H_t,A_t))}{\tilde \sigma^2_t(S_t)} +  \beta (t; H_t) - f_t(S_t)^\top \beta^{(DR)}\Big) f_t(S_t) \Big]\\
    & = \E \Big[ \sum_{t=1}^T W_t (A_t - \tilde p_t (1|S_t))(Y_{t+1}-  g_t(H_t,A_t))f_t(S_t) \Big]+ \E \Big[\sum_{t=1}^T \tilde \sigma^2_t(S_t) \left( \beta (t; H_t) - f_t(S_t)^\top \beta^{(DR)} \right)f_t(S_t) \Big] 
\end{align*}

If the conditional expectation $g_t(H_t,A_t)$ is correctly specified, the first term above boils down to: 
\begin{align*}
    & \E \Big[ \sum_{t=1}^T W_t (A_t - \tilde p_t (1|S_t))(Y_{t+1}-  g_t(H_t,A_t))f_t(S_t) \Big] \\
    = & \E \Big[ \sum_{t=1}^T W_t (A_t - \tilde p_t (1|S_t))\E[Y_{t+1}-  g_t(H_t,A_t)|H_t,A_t]f_t(S_t) \Big]\\
    =& 0,
\end{align*}
and only the second term remains. To estimate $\hat\beta_n^{(DR)}$, we then solve the following equation:
\begin{align*}
    \E \Big[\sum_{t=1}^T \tilde \sigma^2_t(S_t) \left( \beta (t; H_t) - f_t(S_t)^\top \beta \right)f_t(S_t) \Big] =0 
\end{align*}
By the definition of $\beta^\star$, $\E \big[\sum_{t=1}^T \tilde \sigma^2_t(S_t) \left( \beta (t; H_t) - f_t(S_t)^\top \beta^\star \right)f_t(S_t) \big] =0$ holds.
Under regularity conditions, the estimator $\hat\beta^{(DR)}_n \overset{P}{\to} \beta^{\star}$; that is, $\hat\beta^{(DR)}_n$ is consistent. Another case is when the treatment randomization probability is correctly specified. Then we have:
\begin{align*}
    &\E \Big[ \sum_{t=1}^T W_t (A_t - \tilde p_t (1|S_t))(Y_{t+1}-  \hat g_t(H_t,A_t))f_t(S_t) \Big]+ \E \Big[\sum_{t=1}^T \tilde \sigma^2_t(S_t) \left( \hat\beta (t; H_t) - f_t(S_t)^\top \beta  \right)f_t(S_t) \Big] \\
    =& \E \Big[ \sum_{t=1}^T \tilde p_t (1|S_t)(1 - \tilde p_t (1|S_t))\big(\E[Y_{t+1}|H_t, A_t =1]-  \E[Y_{t+1}|H_t, A_t =0] - \hat\beta(H_t,A_t)\big)f_t(S_t) \Big] \\
     &+\E \Big[\sum_{t=1}^T \tilde \sigma^2_t(S_t) \big( \hat\beta (t; H_t) - f_t(S_t)^\top \beta  \big)f_t(S_t) \Big]\\
     =&  \E \Big[\sum_{t=1}^T \tilde \sigma^2_t(S_t) \left( \E[Y_{t+1}|H_t, A_t =1]-  \E[Y_{t+1}|H_t, A_t =0] - f_t(S_t)^\top \beta  \right)f_t(S_t) \Big]
\end{align*}
Similar argument as above, under regularity conditions, the estimator $\hat\beta^{(DR)}_n \overset{P}{\to} \beta^{\star}$; that is, $\hat\beta^{(DR)}_n$ is consistent.

\subsection{More on the second stage weights} 
\label{app:sec:weights}

    The second-stage regression weight in Step III of Algorithm 3.2 can facilitate a consistent estimation of the causal parameter of interest. We regress the constructed pseudo-outcome, $\tilde Y^{(DR)}_{t+1}$, onto the marginal treatment effect subspace $f_t(S_t)$ using the weight $\tilde\sigma_t^2(S_t) = \tilde p_t(1|S_t) \big(1 - \tilde p_t(1|S_t)\big)$ to obtain a consistent estimate of the causal parameter $\hat\beta$. To see this:
    
    \begin{align*}
        0 & = \E\big[\tilde\sigma_t^2(S_t) \big(\tilde Y^{(DR)}_{t+1} - f_t(S_t)^\top \beta \big)f_t(S_t)\big] \\
        &= \E\Big[\tilde \sigma^2_t(S_t) \Big(\frac{W_t \big(A_t - \tilde p_t (1|S_t)\big)\big(Y_{t+1}- g_t(H_t,A_t)\big)}{\tilde \sigma^2_t(S_t)} + \beta (t; H_t) - f_t(S_t)^\top \beta\Big) f_t(S_t)\Big]\\
        &= \E\big[W_t \big(A_t - \tilde p_t (1|S_t)\big)\big(Y_{t+1}- g_t(H_t,A_t)\big)f_t(S_t)\big] \\
        & ~~~~~~~~~~~~~~~~~~~~~~~~~~~~~~~~~~~~~~~~~~~~~~~~~~~~~~~~~~+ \E\big[\tilde\sigma_t^2(S_t)\big(\beta (t; H_t) - f_t(S_t)^\top \beta\big) f_t(S_t) \big]\\
        &= \E\Big[ \E\big[W_t \big(A_t - \tilde p_t (1|S_t)\big)\big(Y_{t+1}- g_t(H_t,A_t)\big)f_t(S_t) | H_t\big]\Big] \\
        & ~~~~~~~~~~~~~~~~~~~~~~~~~~~~~~~~~~~~~~~~~~~~~~~~~~~~~~~~~~+ \E\big[\tilde\sigma_t^2(S_t)\big(\beta (t; H_t) - f_t(S_t)^\top \beta\big) f_t(S_t) \big]\\
        &\overset{\star}{=} \E\Big[ \underbrace{\tilde\sigma_t^2(S_t) \Big(\big(\E\big[Y_{t+1}| 1, H_t\big] - \E\big[Y_{t+1}| 1, H_t\big]\big)- \beta(t;H_t)\Big)f_t(S_t)}_{\text{term I}} \Big] \\
        & ~~~~~~~~~~~~~~~~~~~~~~~~~~~~~~~~~~~~~~~~~~~~~~~~~~~~~~~~~~+ \E\big[\underbrace{\tilde\sigma_t^2(S_t)\big(\beta (t; H_t) - f_t(S_t)^\top \beta\big) f_t(S_t)}_{\text{term II}} \big]\\
        &= \E\Big[ \tilde\sigma_t^2(S_t) \Big(\big(\E\big[Y_{t+1}| 1, H_t\big] - \E\big[Y_{t+1}| 1, H_t\big]\big)- f_t(S_t)^\top \beta\Big)f_t(S_t) \Big]
    \end{align*}
    \sloppy where $\beta(t;H_t) = g_t(H_t,1)- g_t(H_t,0) $, and $g_t(H_t,A_t)$ is a working model for $\E \left[  Y_{t+1} \mid A_t, H_t \right]$. The highlighted “$\overset{\star}{=}$” step shows that $\tilde \sigma_t^2(S_t)$ in term I arises from the expansion of the following expectation. 
    \begin{align*}
        \E[W_t(A_t - \tilde p_t(1|S_t))|H_t, A_t =1] & = \tilde p_t(1|S_t)(1- \tilde p_t(1|S_t)) = \tilde \sigma_t^2(S_t), \\
        \E[W_t(A_t - \tilde p_t(1|S_t))|H_t, A_t =0] & = -\tilde p_t(1|S_t)(1- \tilde p_t(1|S_t)) = - \tilde \sigma_t^2(S_t), 
    \end{align*}
    while the $\tilde \sigma_t^2(S_t)$ in term II is inherited from our definition of the second-stage regression weight. Intuitively, this can be viewed as a projection weight, where we project the outcome \( Y_{t+1} \) onto the treatment to estimate the causal effect of interest. In the case of a potentially misspecified conditional mean working model $g_t(H_t, A_t)$, it is crucial to ensure that the coefficients for $\beta(t;H_t)$ in terms I and II have the same magnitude but opposite signs so that they cancel out. To achieve this, the proposed second-stage regression weight, $\tilde \sigma_t^2(S_t)$, in term II aligns with the expectation derived from term I, ensuring the consistency of the causal parameter estimate.
    
\subsection{Asymptotic properties for DR-WCLS estimators }

Assume $\hat \beta^{(DR)}_n$ minimizes the \emph{DR-WCLS} criterion:
\begin{equation*}
    \mathbb{P}_n \Big[\sum_{t=1}^T \tilde \sigma^2_t(S_t) \Big(\frac{\hat W_t (A_t - \tilde p_t (1|S_t))(Y_{t+1}- \hat g_t(H_t,A_t))}{\tilde \sigma^2_t(S_t)} + \hat \beta (t; H_t) - f_t(S_t)^\top \beta^{(DR)}\Big)^2\Big].
\end{equation*}

The estimated treatment randomization probability is denoted as $\hat p_t = \hat p(A_t|H_t)$, thus we have the weight $W_t$ estimated by $ \hat W_t = \tilde p_t(A_t|S_t)/ \hat p_t(A_t|H_t)$. And the estimating equation is:
\begin{align}
    0 &= \P_n \Big[\sum_{t=1}^T \tilde \sigma^2_t(S_t) \Big(\frac{ \tilde p_t (A_t|S_t) (A_t - \tilde p_t (1|S_t))(Y_{t+1}-  \hat g_t(H_t,A_t))}{\hat p_t(A_t|H_t) \tilde \sigma^2_t(S_t)} +  \hat \beta (t; H_t) - f_t(S_t)^\top \beta^{(DR)}\Big) f_t(S_t) \Big].
\end{align}

Expand the right-hand side, we have:
\begin{align*}
    &\P_n \Big[\sum_{t=1}^T \tilde \sigma^2_t(S_t) \Big(\frac{ \tilde p_t (A_t|S_t) (A_t - \tilde p_t (1|S_t))\big(Y_{t+1}-  \hat g_t(H_t,A_t)\big)}{\hat p_t(A_t|H_t) \tilde \sigma^2_t(S_t)} +  \hat \beta (t; H_t) - f_t(S_t)^\top \beta \Big) f_t(S_t) \Big] \\
    =&\P_n \Big[\sum_{t=1}^T \tilde \sigma^2_t(S_t) \Big(\frac{ \tilde p_t (A_t - \tilde p_t(1|S_t) )\big(Y_{t+1}- g^\star_t(H_t,A_t)+g^\star_t(H_t,A_t)- \hat g_t(H_t,A_t)\big)}{\tilde \sigma^2_t(S_t)}\Big(\frac{1}{\hat p_t }-\frac{1}{ p_t }+\frac{1}{ p_t }\Big) + \\
    & \beta (t; H_t)-f_t(S_t)^\top \beta^\star  +(\hat \beta (t; H_t)-\beta (t; H_t)) - f_t(S_t)^\top (\beta -\beta^\star)\Big) f_t(S_t) \Big]\\
    =& \P_n \Big[\sum_{t=1}^T \tilde \sigma^2_t(S_t)\Big( \frac{ W_t (A_t - \tilde p_t )\big(Y_{t+1}- g^\star_t(H_t,A_t)\big)}{\tilde \sigma^2_t(S_t)}+\beta (t; H_t)-f_t(S_t)^\top \beta^\star\Big) f_t(S_t)\Big]\\
    &   + \P_n \Big[\sum_{t=1}^T \tilde p_t (A_t - \tilde p_t )\big(Y_{t+1}- g^\star_t(H_t,A_t)\big)\Big(\frac{1}{\hat p_t }-\frac{1}{ p_t }\Big)f_t(S_t)\Big]\\
    & + \P_n \Big[\sum_{t=1}^T\tilde p_t (A_t - \tilde p_t )(g^\star_t(H_t,A_t)- \hat g_t(H_t,A_t))\Big(\frac{1}{\hat p_t }-\frac{1}{ p_t }\Big) f_t(S_t)\Big]\\
    & + \P_n \Big[\sum_{t=1}^T W_t (A_t - \tilde p_t (1|S_t))(g^\star(H_t,A_t)-\hat g_t(H_t,A_t))f_t(S_t)\Big] \\
    & + \P_n \Big[\sum_{t=1}^T  \tilde \sigma^2_t(S_t) (\hat \beta (t; H_t)-\beta (t; H_t))f_t(S_t)\Big]\\
    & -\P_n \Big[\sum_{t=1}^T \tilde \sigma^2_t(S_t) f_t(S_t)f^\top_t(S_t)\Big](\beta -\beta^\star).
\end{align*}
By WLLN, the following convergence result holds:
\begin{align*}
 \P_n \Big[\sum_{t=1}^T \tilde p_t (A_t - \tilde p_t )(Y_{t+1}- g^\star_t(H_t,A_t))\Big(\frac{1}{\hat p_t }-\frac{1}{ p_t }\Big)f_t(S_t)\Big]&\overset{P}{\to} 0 ~~~~\text{(correct model specification),} \\
    \P_n \Big[\sum_{t=1}^T \tilde \sigma^2_t(S_t) f_t(S_t)f^\top_t(S_t)\Big] &\overset{P}{\to} \E\Big[\sum_{t=1}^T \tilde \sigma^2_t(S_t) f_t(S_t)f^\top_t(S_t)\Big]. 
\end{align*}
and
\begin{align*}
    & \P_n \Big[\sum_{t=1}^T W_t (A_t - \tilde p_t (1|S_t))(g^\star(H_t,A_t)-\hat g_t(H_t,A_t))f_t(S_t)\Big] +\\
    &~~~~~~~~~~~~~~\P_n \Big[\sum_{t=1}^T  \tilde \sigma^2_t(S_t) (\hat \beta (t; H_t)-\beta (t; H_t))f_t(S_t)\Big]\overset{P}{\to} 0 ~~~~\text{(terms cancellation).}
\end{align*}
To see this, we have:
\begin{align*}
    & \E \big[  W_t (A_t - \tilde p_t (1|S_t))\big(g^\star(H_t,A_t)-\hat g_t(H_t,A_t)\big)f_t(S_t) +\tilde \sigma^2_t(S_t) \big(\hat \beta (t; H_t)-\beta (t; H_t)\big)f_t(S_t)\big]\\
    = &  \E\Big[\E\big[W_t (A_t - \tilde p_t (1|S_t))\big(g^\star(H_t,A_t)-\hat g_t(H_t,A_t)\big) | H_t\big]f_t(S_t) +\tilde \sigma^2_t(S_t) \big(\hat \beta (t; H_t)-\beta (t; H_t)\big)f_t(S_t) \Big]\\
    = & \E\big[\tilde \sigma^2_t(S_t) \big(\beta (t; H_t)-\hat \beta (t; H_t)\big)f_t(S_t) +\tilde \sigma^2_t(S_t) \big(\hat \beta (t; H_t)-\beta (t; H_t)\big)f_t(S_t) \big]\\
    = & 0
\end{align*}
Apart from the nicely-behaved term above, the only term that might be problematic and causes bias is:
$$
\P_n \Big[\sum_{t=1}^T\tilde p_t (A_t - \tilde p_t )(g^\star_t(H_t,A_t)- \hat g_t(H_t,A_t))\Big(\frac{1}{\hat p_t }-\frac{1}{ p_t }\Big) f_t(S_t)\Big],
$$
which could be further written as:
\begin{align*}
    \sum_{a \in \{0,1 \}} \sum_{t=1}^T \P_n\Big[\Big(\frac{ \tilde \sigma^2_t(S_t)}{a\hat p_t(1|H_t)+ (1-a)(1-\hat p_t(1|H_t))}(p_t(1|H_t)-\hat p_t(1|H_t))(g(H_t,a)-\hat g(H_t,a))\Big) f_t(S_t)\Big].
\end{align*}

In our context, $T$ is finite and fixed. Therefore, by the fact that $\hat p_t(1|H_t))$ is bounded away from zero and one, along with the Cauchy–Schwarz inequality, we have that (up to a multiplicative constant) the term above is bounded above by:
\begin{align}
\label{app:eq-error}
\mathbf{\hat B} & = \sum_{a \in \{0,1 \}} \sum_{t=1}^T \sqrt{\P_n \Big[ \big(p_t(1|H_t)-\hat p_t(1|H_t)\big)^2\Big]} \sqrt{\P_n \Big[\big( g(H_t,a)-\hat g(H_t,a)\big)^2 \Big]}  \nonumber \\ 
    & = \sum_{t=1}^T \sum_{a \in \{0,1\}}\left\Vert \hat p_t(a|H_t)- p_t(a|H_t)\right\Vert \left\Vert \hat g_t(H_t,a)- g_t(H_t,a)\right\Vert 
\end{align}
Assuming we have nuisance estimates that can make $\mathbf{\hat B}$ asymptotically negligible, and along side with other terms converge at a $o_p(n^{-1/2})$ rate:
\begin{align*}
    &\text{Var}\Big(\P_n \Big[\sum_{t=1}^T \tilde p_t (A_t - \tilde p_t)\big(Y_{t+1}- g^\star_t(H_t,A_t)\big)\Big(\frac{1}{\hat p_t }-\frac{1}{ p_t }\Big)f_t(S_t)\Big]\Big)^2\\
    & \lesssim  \frac{1}{n}\P_{n} \Big[\sum_{t=1}^T \E\big[\big(\tilde p_t (A_t - \tilde p_t)\big(Y_{t+1}- g^\star_t(H_t,A_t)\big)\Big(\frac{1}{\hat p_t }-\frac{1}{ p_t }\Big)\big)^2 f_t(S_t) f_t(S_t)^\top  \big] \Big] \\
    & = \frac{1}{n}\P_{n} \Big[\sum_{t=1}^T \E\big[\tilde p^2_t (A_t - \tilde p_t )^2 \E[(Y_{t+1}- g^\star_t(H_t,A_t))^2|H_t, A_t]\Big(\frac{1}{\hat p_t }-\frac{1}{ p_t }\Big)^2 f_t(S_t) f_t(S_t)^\top  \big] \Big] \\
    & \lesssim  \frac{1}{n}\P_{n} \Big[\sum_{t=1}^T \E\big[(p_t - \hat p_t)^2 \big] \Big]\times \pmb{1}_{q\times q}\\
    & = \sum_{t=1}^T  \E\Big[\P_{n}  \big[(p_t - \hat p_t)^2 \big] \Big]\times \frac{1}{n} \pmb{1}_{q\times q}\\
     & = o_p(1) \times \frac{1}{n}\pmb{1}_{q\times q} \\
    & = o_p(n^{-1}) \times \pmb{1}_{q\times q},
\end{align*}
where we use the notation $\lesssim$ to represent the left-hand side is bounded by a finite constant times the right-hand side. The first inequality follows from the assumption that $T$ is finite and fixed, while the second inequality holds because of the moment conditions and $f_t(S_t)$, $\tilde p_t$, $\hat p_t$ and $p_t$ are all bounded. The last equality holds due to the Dominated Convergence Theorem and the assumption that the nuisance functions $\hat p_t$ should at least estimate the true randomization probability $p_t$ consistently. By Chebyshev inequality, this term should converge at a $o_p(n^{-1/2})$ rate. Similarly,
\begin{align*}
    & \Big(\P_n \Big[\sum_{t=1}^T 
     W_t (A_t - \tilde p_t (1|S_t))(g^\star(H_t,A_t)-\hat g_t(H_t,A_t))f_t(S_t) \Big]\Big)^2\\
    &\lesssim \frac{1}{n}\P_{n} \Big[\sum_{t=1}^T \big( W_t (A_t - \tilde p_t (1|S_t))(g^\star(H_t,A_t)-\hat g_t(H_t,A_t))\big)^2 f_t(S_t)f_t(S_t)^\top \Big]\\
    & \lesssim \P_{n}\Big[ \sum_{t=1}^T (g^\star(H_t,A_t)-\hat g_t(H_t,A_t))^2\Big]\times \frac{1}{n}\pmb{1}_{q\times q}\\
    & = o_p(1) \times \frac{1}{n}\pmb{1}_{q\times q} \\
    & = o_p(n^{-1}) \times \pmb{1}_{q\times q},
\end{align*}
The first inequality follows from the assumption that $T$ is finite and fixed, while the second inequality holds because $W_t$ and $f_t(S_t)$ are bounded. The $o_p(1)$ comes from the assumption that the nuisance functions should at least estimate the true outcome conditional expectation consistently. Thus, this term should converge to $0$ at a $o_p(n^{-1/2})$ rate. Last but not the least, we have:
\begin{align*}
&\Big(\P_n \Big[\sum_{t=1}^T  \tilde \sigma^2_t(S_t) (\hat \beta (t; H_t)-\beta (t; H_t))f_t(S_t)\Big] \Big)^2 \\
    & \lesssim \frac{1}{n}\P_{n} \Big[\sum_{t=1}^T  \tilde \sigma^4_t(S_t) (\hat \beta (t; H_t)-\beta (t; H_t))^2 f_t(S_t)f_t(S_t)^\top\Big] \\
    & \lesssim \frac{1}{n}\P_{n} \Big[\sum_{t=1}^T  (\hat \beta (t; H_t)-\beta (t; H_t))^2 \Big]  \times \pmb{1}_{q\times q}\\
    & = o_p(1) \times \frac{1}{n}\pmb{1}_{q\times q} \\
    & = o_p(n^{-1}) \times \pmb{1}_{q\times q}.
\end{align*}
The first inequality follows from the assumption that $T$ is finite and fixed, while the second inequality holds because $\sigma^4_t(S_t)$ and $f_t(S_t)$ are bounded. The $o_p(1)$ comes from the assumption that the nuisance functions should at least estimate the true outcome conditional expectation consistently. Thus, this term should converge to $0$ at a $o_p(n^{-1/2})$ rate.
Denote the pseudo-outcomes with true nuisance parameters as $\tilde Y^{(DR)}_{t+1} = \frac{W_t (A_t - \tilde p_t (1|S_t))(Y_{t+1}- g^\star(H_t,A_t))}{\tilde\sigma_t^{2}}+ \beta(t;H_t)$, the DR-WCLS estimator satisfies:
\begin{align*}
    n^{1/2}(\hat\beta^{(DR)}_n-\beta^{\star}) & = n^{1/2}~ \P_n \bigg[\sum_{t=1}^T ~ \E\Big[\sum_{t=1}^T \tilde \sigma^2_t(S_t) f_t(S_t)f^\top_t(S_t)\Big]^{-1} \tilde \sigma^2_t(S_t) (\tilde Y^{(DR)}_{t+1} - f_t(S_t)\beta^\star)f_t(S_t) \bigg] + o_p(1),
\end{align*}
and it is efficient with influence function:
\begin{align*}
        &\sum_{t=1}^T ~  \E\Big[\sum_{t=1}^T \tilde \sigma^2_t(S_t) f_t (S_t)f_t (S_t)^\top  \Big]^{-1} \tilde \sigma^2_t(S_t) (\tilde Y^{(DR)}_{t+1} - f_t(S_t)\beta^\star)f_t(S_t).
\end{align*}

Under moment conditions, we have asymptotic normality with variance given by $\Sigma_{DR}=Q^{-1} W Q^{-1}$, where
\begin{align*}
    Q & = \E\Big[\sum_{t=1}^T \tilde \sigma^2_t(S_t) f_t (S_t)f_t (S_t)^\top  \Big], \\
    W &= \E \Big[ \Big(\sum_{t=1}^T \tilde \sigma^2_t(S_t) (\tilde Y^{(DR)}_{t+1} - f_t(S_t)\beta^\star)f_t(S_t)\Big)^2\Big].
\end{align*}

\subsection{Asymptotic variance using sample splitting}

Built on the previous doubly robust property, we know that if either the conditional expectation model $g_t(H_t,A_t)$ or the treatment randomization probability $p_t(A_t|H_t)$ is correctly specified, we can obtain a consistent estimator of $\beta^\star$. In this section, we provide the asymptotic variance estimation under sample splitting. Without loss of generality, we assume that the treatment randomization probability $p_t(A_t|H_t)$ is correctly specified. For simplicity, we use $\tilde\sigma_t^{2(k)}$ to denote $\tilde \sigma^2_t(S_t)^{(k)}$ The asymptotic properties of the DR-WCLS estimator follow from the expansion:
\begin{align*}
    0 &= \frac{1}{K}\sum_{k=1}^K \mathbb{P}_{n,k} \Big[\sum_{t=1}^T \tilde\sigma_t^{2(k)} \Big(\frac{W_t (A_t - \tilde p^{(k)}_t (1|S_t))(Y_{t+1}- \hat g^{(k)}_t(H_t,A_t))}{\tilde\sigma_t^{2(k)}} + \hat \beta^{(k)} (t; H_t) - f_t(S_t)^\top \hat\beta^{(DR)}_n\Big) f_t(S_t) \Big]\\
    & =  \frac{1}{K}\sum_{k=1}^K \mathbb{P}_{n,k} \bigg[\sum_{t=1}^T \tilde\sigma_t^{2(k)} \Big(\frac{W_t (A_t - \tilde p_t (1|S_t))(Y_{t+1}- g^\star(H_t,A_t)+g^\star(H_t,A_t)-\hat g^{(k)}_t(H_t,A_t))}{\tilde\sigma_t^{2(k)}} + \\
    & ~~~~~~~\beta (t; H_t) +(\hat \beta^{(k)} (t; H_t)-\beta (t; H_t)) - f_t(S_t)^\top \beta^\star - f_t(S_t)^\top (\hat\beta^{(DR)}_n-\beta^\star) \Big) f_t(S_t) \bigg]\\
    & = \frac{1}{K}\sum_{k=1}^K \mathbb{P}_{n,k}  \Big[\sum_{t=1}^T\tilde\sigma_t^{2(k)} \Big( \frac{W_t (A_t - \tilde p^{(k)}_t (1|S_t))(Y_{t+1}- g^\star(H_t,A_t))}{\tilde\sigma_t^{2(k)}}+ \beta(t;H_t)-f_t(S_t)^\top \beta^\star \Big)f_t(S_t)\Big]\\
    &+\frac{1}{K}\sum_{k=1}^K \mathbb{P}_{n,k}  \Big[\sum_{t=1}^T W_t (A_t - \tilde p^{(k)}_t (1|S_t))(g^\star(H_t,A_t)-\hat g^{(k)}_t(H_t,A_t))f_t(S_t)\Big]\\
    & + \frac{1}{K}\sum_{k=1}^K \mathbb{P}_{n,k}  \Big[\sum_{t=1}^T  \tilde\sigma_t^{2(k)} (\hat \beta^{(k)} (t; H_t)-\beta (t; H_t))f_t(S_t)\Big]\\
    & -\frac{1}{K}\sum_{k=1}^K \mathbb{P}_{n,k}  \Big[\sum_{t=1}^T \tilde\sigma_t^{2(k)} f_t(S_t)f^\top_t(S_t)\Big](\hat\beta^{(DR)}_n-\beta^\star)
\end{align*}
By the WLLN, we have the term cancellation as follows:
\begin{align*}
    & \frac{1}{K}\sum_{k=1}^K \mathbb{P}_{n,k} \Big[\sum_{t=1}^T W_t (A_t - \tilde p^{(k)}_t (1|S_t))(g^\star(H_t,A_t)-\hat g^{(k)}_t(H_t,A_t))f_t(S_t)\Big] +\\
    &~~~~~~~~~~~~~~~~~~~~~~~~~~~~~~\frac{1}{K}\sum_{k=1}^K \mathbb{P}_{n,k} \Big[\sum_{t=1}^T  \tilde\sigma_t^{2(k)} (\hat \beta^{(k)} (t; H_t)-\beta (t; H_t))f_t(S_t)\Big]\overset{P}{\to} 0, 
\end{align*}
and
\begin{align*}
    \mathbb{P}_{n,k} \Big[\sum_{t=1}^T \tilde\sigma_t^{2(k)} f_t(S_t)f^\top_t(S_t)\Big] &\overset{P}{\to} \E\Big[\sum_{t=1}^T \tilde\sigma_t^{2(k)} f_t(S_t)f^\top_t(S_t)\Big]. 
\end{align*}

With an abuse of notation, we denote $\tilde Y^{(DR)}_{t+1} = \frac{W_t (A_t - \tilde p^{(k)}_t (1|S_t))(Y_{t+1}- g^\star(H_t,A_t))}{\tilde\sigma_t^{2(k)}}+ \beta(t;H_t),$ thus we obtain:
\begin{align*}
    n^{1/2}(\hat\beta^{(DR)}_n-\beta^{\star}) & = n^{1/2}~ \frac{1}{K}\sum_{k=1}^K \mathbb{P}_{n,k} \bigg[\sum_{t=1}^T ~ \bigg\{\frac{1}{K}\sum_{k=1}^K \E\Big[\sum_{t=1}^T \tilde \sigma^2_t(S_t) f_t(S_t)f^\top_t(S_t)\Big]\bigg\}^{-1} \times \\
    & ~~~~~~~~~~~~~~~~~~~~~~~~~~~~~~~~~~~~~~~~~\tilde\sigma_t^{2(k)} (\tilde Y^{(DR)}_{t+1} - f_t(S_t)\beta^\star)f_t(S_t) \bigg] + o_p(1).
\end{align*}
By definition of $\beta^\star$:
\begin{equation*}
    \frac{1}{K}\sum_{k=1}^K \E \Big[\sum_{t=1}^T \tilde\sigma_t^{2(k)} \left( \tilde Y^{(DR)}_{t+1}- f_t (S_t)^\top \beta^{\star} \right)f_t (S_t) \Big] =0.
\end{equation*}

Consequently, the influence function for $\hat\beta^{(DR)}_n$ is:
\begin{align}
        &\frac{1}{K}\sum_{k=1}^K \sum_{t=1}^T ~  \bigg\{\frac{1}{K}\sum_{k=1}^K \E\Big[\sum_{t=1}^T \tilde\sigma_t^{2(k)} f_t (S_t)f_t (S_t)^\top  \Big]\bigg\}^{-1} \tilde\sigma_t^{2(k)} (\tilde Y^{(DR)}_{t+1} - f_t(S_t)\beta^\star)f_t(S_t).
\end{align}

Then, under moment conditions, we have asymptotic normality with variance given by Equation \eqref{eq:asymptotic_var} where:
  \begin{align*}
     m (\hat \beta, \hat \eta_k) &= \sum_{t=1}^T \psi_t(\hat\beta, \hat \eta_k;H_t, A_t)=  \sum_{t=1}^T \hat{ \tilde p}_t^{(k)} (1|S_t)(1-\hat{ \tilde p}_t^{(k)} (1|S_t)) (\tilde Y^{(DR)}_{t+1} - f_t(S_t)\hat\beta)f_t(S_t), \\
     \dot m (\hat \beta, \hat \eta_k)  &= \frac{\partial m (\beta, \hat \eta_k)}{\partial \beta} \Big|_{\beta = \hat\beta} =   \sum_{t=1}^T \hat{ \tilde p}_t^{(k)} (1|S_t)(1-\hat{ \tilde p}_t^{(k)} (1|S_t)) f_t (S_t)f_t (S_t)^\top. 
 \end{align*}
In conclusion, we establish that the estimator that minimizes the DR-WCLS criterion $\hat\beta^{(DR)}_n$ is consistent and asymptotically normal. 

\section{Proof of Theorem \ref{thm:T-infinity}}
\label{app:T-infinity}

The asymptotic property when $T \rightarrow \infty$ is comparatively more challenging, as the dependence between time points is not negligible, and we may expect that the convergence rate can be impacted by the number of other time points on which each time point depends. Intuitively speaking, this means that adding more dependent time points does not necessarily translate to including more information compared to including more independent participants. A similar argument can be found in \cite{ogburn2022causal,van2014causal}. Define the operator $\P_{n,T} = \frac{1}{T}  \sum_{t=1}^T \P_n$, and in this section, when we discuss the convergence rate of certain vectors or matrices, we specifically refer to \emph{coordinate-wise convergence}. To prove consistency, we start with the following expansion:
\begin{align}
    &\P_{n,T} \Big[  \tilde \sigma^2_t(S_t) \big(\frac{ \tilde p_t (A_t|S_t) (A_t - \tilde p_t (1|S_t))(Y_{t+1}-  \hat g_t(H_t,A_t))}{\hat p_t(A_t|H_t) \tilde \sigma^2_t(S_t)} +  \hat \beta (t; H_t) - f_t(S_t)^\top \beta\big) f_t(S_t) \Big] \nonumber \\
    =&\P_{n,T} \Big[  \tilde \sigma^2_t(S_t) \Big(\frac{ \tilde p_t (A_t - \tilde p_t(1|S_t) )\big(Y_{t+1}- g^\star_t(H_t,A_t)+g^\star_t(H_t,A_t)- \hat g_t(H_t,A_t)\big)}{\tilde \sigma^2_t(S_t)}\Big(\frac{1}{\hat p_t }-\frac{1}{ p_t }+\frac{1}{ p_t }\Big) + \nonumber \\
    & \beta (t; H_t)-f_t(S_t)^\top \beta^\star  +\big(\hat \beta (t; H_t)-\beta (t; H_t)\big) - f_t(S_t)^\top (\beta-\beta^\star)\Big) f_t(S_t) \Big] \nonumber\\
    =& \P_{n,T} \Big[  \tilde \sigma^2_t(S_t)\Big( \frac{ W_t (A_t - \tilde p_t(1|S_t) )(Y_{t+1}- g^\star_t(H_t,A_t))}{\tilde \sigma^2_t(S_t)}+\beta (t; H_t)-f_t(S_t)^\top \beta^\star\Big) f_t(S_t)\Big] \label{term_1}\\
    &  + \P_{n,T} \Big[  \tilde p_t (A_t - \tilde p_t(1|S_t) )(Y_{t+1}- g^\star_t(H_t,A_t))\Big(\frac{1}{\hat p_t }-\frac{1}{ p_t }\Big)f_t(S_t)\Big] \label{term_2}\\
    & + \P_{n,T} \Big[ \tilde p_t (A_t - \tilde p_t(1|S_t) )(g^\star_t(H_t,A_t)- \hat g_t(H_t,A_t))\Big(\frac{1}{\hat p_t }-\frac{1}{ p_t }\Big) f_t(S_t)\Big] \label{term_3}\\
    &+ \P_{n,T} \big[  W_t (A_t - \tilde p_t (1|S_t))\big(g^\star(H_t,A_t)-\hat g_t(H_t,A_t)\big)f_t(S_t)\big] \label{term_4}\\
    & + \P_{n,T} \big[   \tilde \sigma^2_t(S_t) \big(\hat \beta (t; H_t)-\beta (t; H_t)\big)f_t(S_t)\big] \label{term_5}\\
    & -\P_{n,T} \big[  \tilde \sigma^2_t(S_t) f_t(S_t)f^\top_t(S_t)\big](\beta-\beta^\star) \label{term_6}
\end{align}

Based on Assumption \ref{ass:T-infinity} (1), we can conclude that Term \eqref{term_1} $\overset{p}{\rightarrow} 0$ when $T \rightarrow \infty$. Term \eqref{term_2} is a Martingale Difference Sequence (MDS) with respect to the filtration $\mathcal{F}(H_t, A_t)$, where $\mathcal{F}(H_t, A_t)$ represents the $\sigma$-algebra generated by $\{H_t, A_t\}$. To see this, we show:
\begin{align*}
    &\E\Big[  \tilde p_t (A_t - \tilde p_t(1|S_t) )(Y_{t+1}- g^\star_t(H_t,A_t))\Big(\frac{1}{\hat p_t }-\frac{1}{ p_t }\Big)f_t(S_t)|H_t,A_t\Big]\\
    = &  \tilde p_t (A_t - \tilde p_t(1|S_t) )\E\Big[ (Y_{t+1}- g^\star_t(H_t,A_t))\Big(\frac{1}{\hat p_t }-\frac{1}{ p_t }\Big)|H_t,A_t\Big]f_t(S_t)\\
    = &  \tilde p_t (A_t - \tilde p_t(1|S_t) )\E\Big[ \E\big[Y_{t+1}- g^\star_t(H_t,A_t)|H_t,A_t, \tilde H_{t+r}\big]\Big(\frac{1}{\hat p_t }-\frac{1}{ p_t }\Big)|H_t,A_t\Big]f_t(S_t)\\
    = &  \tilde p_t (A_t - \tilde p_t(1|S_t) )\E\Big[ \underbrace{\E\big[Y_{t+1}- g^\star_t(H_t,A_t)|H_t,A_t\big]}_{=0}\Big(\frac{1}{\hat p_t }-\frac{1}{ p_t }\Big)|H_t,A_t\Big]f_t(S_t)\\
    = & 0.
\end{align*}
Thus when $T \rightarrow \infty$, Term \eqref{term_2} $\overset{p}{\rightarrow} 0$. We can also demonstrate that this holds true when the nuisance model is trained on a subset of $\{H_t, A_t, \tilde H_{t+r}\}$.
In addition, the sum of Term \eqref{term_4} and \eqref{term_5} also forms an MDS with respect to the filtration $\mathcal{F}(H_t)$. 
\begin{align*}
    & \E \big[  W_t (A_t - \tilde p_t (1|S_t))\big(g^\star(H_t,A_t)-\hat g_t(H_t,A_t)\big)f_t(S_t) +\tilde \sigma^2_t(S_t) \big(\hat \beta (t; H_t)-\beta (t; H_t)\big)f_t(S_t) | H_t\big]\\
    = &  \E\big[W_t (A_t - \tilde p_t (1|S_t))\big(g^\star(H_t,A_t)-\hat g_t(H_t,A_t)\big) | H_t\big]f_t(S_t) +\tilde \sigma^2_t(S_t) \big(\hat \beta (t; H_t)-\beta (t; H_t)\big)f_t(S_t)\\
    = & \tilde \sigma^2_t(S_t) \big(\beta (t; H_t)-\hat \beta (t; H_t)\big)f_t(S_t) +\tilde \sigma^2_t(S_t) \big(\hat \beta (t; H_t)-\beta (t; H_t)\big)f_t(S_t) \\
    = & 0
\end{align*}
Thus, when $T \rightarrow \infty$, Term \eqref{term_4} + \eqref{term_5} $\overset{p}{\rightarrow} 0$. Based on Assumption \ref{ass:T-infinity} (4), we have the following inequality for Term \eqref{term_3}:
\begin{align*}
    & \frac{1}{T}\sum_{t=1}^T\tilde p_t (A_t - \tilde p_t(1|S_t) )(g^\star_t(H_t,A_t)- \hat g_t(H_t,A_t))\left(\frac{1}{\hat p_t }-\frac{1}{ p_t }\right)\\
    \lesssim &\left( \frac{1}{T}\sum_{t=1}^T (p_t^\star(1|H_t)-\hat p_t(1|H_t))^2 \right)^{1/2}  \times \left(\frac{1}{T}\sum_{t=1}^T \big(g_t^\star(H_t,A_t)-\hat g_t(H_t,A_t)\big)^2\right)^{1/2} \\
    = & \Vert p_t^\star(1|H_t)-\hat p_t(1|H_t) \Vert_T \Vert g_t^\star(H_t,A_t)-\hat g_t(H_t,A_t) \Vert_T\\
    =&  o_p(T^{-1/2} ).
\end{align*}
 In summary, the sum of Terms (16) to (19) is asymptotically negligible, meaning their total converges at the \(o_p(1)\) rate. Assumption \ref{ass:T-infinity} guarantees $\P_{n,T} \big[  \tilde \sigma^2_t(S_t) f_t(S_t)f^\top_t(S_t)\big]$ in Term \eqref{term_6} exists and converges to $B_{\beta}$ when $T \rightarrow \infty$, where:
\begin{align*}
    B_{\beta} &= \lim_{T \rightarrow \infty} \frac{1}{T} \sum_{t=1}^T \E[\dot \psi_{t}(\beta;H_{t},A_t)].
\end{align*}
In conclusion, $\hat\beta^{(DR)} \overset{p}{\rightarrow} \beta^\star$ when $T \rightarrow \infty$, that is, $\hat\beta^{(DR)}$ is a consistent estimator of the true causal parameter $\beta^\star$. Now we consider the asymptotic normality. First, define $\dot \psi_{t}(\beta;H_{t},A_t)  = \partial \psi_{t}(\beta;H_{t},A_t )/\partial \beta = \tilde \sigma_t^2 (S_t)f_t(S_t)f_t(S_t)^\top$. 
Next, by the martingale central limit theorem presented in \cite{dvoretzky1972asymptotic}, we need to verify the following two conditions:
\begin{enumerate}
    \item (\textbf{Conditional Variance}) There exists a constant positive definite matrix $\Gamma_\beta$ that: $$\frac{1}{T}\sum_{t=1}^T \E[\psi_t(\beta^\star,\boldsymbol{\eta}^\star,H_{t},A_{t})\psi_t(\beta^\star,\boldsymbol{\eta}^\star,H_{t},A_{t})^\top| \mathcal{F}(H_t, A_t) ] \overset{p}{\rightarrow} \Gamma_\beta ;$$ 
    \item (\textbf{Conditional Lindeberg}) For any $e > 0$, and any fixed $\mathbf{c} \in \mathbb{R}^q$ with $\Vert \mathbf{c} \Vert =1$:
    $$\frac{1}{T}\sum_{t=1}^T \E[\Vert\mathbf{c}^\top \Gamma_\beta^{-1/2} \psi_t(\beta^\star,\boldsymbol{\eta}^\star,H_{t},A_{t}) \Vert^2 \mathbbm{1}_{\Vert \mathbf{c}^\top\Gamma_\beta^{-1/2} \psi_t\Vert>eT}|\mathcal{F}(H_{t},A_{t})] \overset{p}{\rightarrow} 0.$$
\end{enumerate}

To verify Condition 1, recall $\tilde Y^{(DR)}_{t+1} = \frac{W_t (A_t - \tilde p_t (1|S_t))(Y_{t+1}- g^\star(H_t,A_t))}{\tilde\sigma^2_t(S_t)}+ \beta(t;H_t),$
\begin{align*}
    \psi_t(\beta^\star,\boldsymbol{\eta}^\star,H_{t},A_{t}) &=\tilde \sigma^2_t(S_t) \Big(\frac{W_t \big(A_t - \tilde p_t (1|S_t)\big)\big(Y_{t+1}- g^\star_t(H_t,A_t)\big)}{\tilde \sigma^2_t(S_t)} + \beta (t; H_t) - f_t(S_t)^\top \beta^\star\Big) f_t(S_t) \\
    &= \tilde \sigma^2_t(S_t)\big(\tilde Y^{(DR)}_{t+1} - f_t(S_t)^\top \beta^\star \big)f_t(S_t)
\end{align*}
Then we have the conditional variance as:
\begin{align*}
    \E[\psi_t\psi_t^\top| \mathcal{F}(H_t, A_t) ] = & 
    \tilde \sigma^4_t(S_t) \E\big[\big(\tilde Y^{(DR)}_{t+1} - f_t(S_t)^\top \beta^\star \big)^2 | \mathcal{F}(H_t, A_t) \big] f_t(S_t)f_t(S_t)^\top \\
    = & \tilde \sigma^4_t(S_t) \E[\xi_t^2 | \mathcal{F}(H_t, A_t) ]  f_t(S_t)f_t(S_t)^\top
\end{align*}
By Assumption \ref{ass:T-infinity}, each summand is bounded and we have:
\begin{equation*}
    \lim_{T \rightarrow \infty}\frac{1}{T}\sum_{t=1}^T\E[\psi_t\psi_t^\top] = \lim_{T \rightarrow \infty}\frac{1}{T}\sum_{t=1}^T \E\big[\tilde \sigma^4_t(S_t) \E[\xi_t^2 | \mathcal{F}(H_t, A_t) ] f_t(S_t)f_t(S_t)^\top\big] = \Gamma_\beta
\end{equation*}
To show that Var$(\frac{1}{T}\sum_{t=1}^T\E[\psi_t\psi_t^\top| \mathcal{F}(H_t, A_t) ]) = o(\pmb{1}_{q\times q})$, we first check the covariance across time, for a pair of time points $t \neq t'$: 
\begin{align*}
        & \text{Cov}\big(\E[\psi_t\psi_t^\top| \mathcal{F}(H_t, A_t) ],\E[\psi_{t'}\psi_{t'}^\top| \mathcal{F}(H_{t'}, A_{t'}) ]\big)\\
        & = \text{Cov}\Big(\tilde \sigma^4_t(S_t) \E[\xi_t^2 | \mathcal{F}(H_t, A_t) ] f_t(S_t)f_t(S_t)^\top, \tilde\sigma^4_{t'}(S_{t'}) \E[\xi_{t'}^2 | \mathcal{F}(H_{t'}, A_{t'}) ] f_{t'}(S_{t'})f_t(S_{t'})^\top\Big)\\
        & \leq c^2_2 \cdot \pmb{1}_{q\times q} \cdot \text{Cov}\big(\E[\xi_t^2 | \mathcal{F}(H_t, A_t) ],\E[\xi_{t'}^2 | \mathcal{F}(H_{t'}, A_{t'}) ]\big)\\
        & \leq c_1^2 c^2_2 \cdot \rho(t,t') \cdot \pmb{1}_{q\times q} 
\end{align*}
where $\pmb{1}_{q\times q}$ is a $q\times q$ matrix of $1$. Thus, we have:
\begin{align*}
    \text{Var}&(\frac{1}{T}\sum_{t=1}^T\E[\psi_t\psi_t^\top| \mathcal{F}(H_t, A_t) ])\\
    &= \frac{1}{T^2} \sum_{t=1}^T \text{Var}\big(\E[\psi_t\psi_t^\top| \mathcal{F}(H_t, A_t) ]\big) +  \frac{1}{T^2} \sum_{t\neq t'} \text{Cov}\big(\E[\psi_t\psi_t^\top| \mathcal{F}(H_t, A_t) ],\E[\psi_{t'}\psi_{t'}^\top| \mathcal{F}(H_{t'}, A_{t'}) ]\big)\\
    & \leq \Big(o(1) + c_1^2 c^2_2 \cdot \frac{1}{T^2} \sum_{t\neq t'}\rho(t,t')\Big)\times \pmb{1}_{q\times q}  \\
    & = \big(o(1) +o(1) \big)\times \pmb{1}_{q\times q}  \\
    &= o(\pmb{1}_{q\times q})
\end{align*}
Therefore, Condition 1 holds.
\begin{equation*}
    \frac{1}{T}\sum_{t=1}^T\E[\psi_t\psi_t^\top| \mathcal{F}(H_t, A_t) ] = \frac{1}{T}\sum_{t=1}^T \tilde \sigma^4_t(S_t) \E[\xi_t^2 | \mathcal{F}(H_t, A_t) ] f_t(S_t)f_t(S_t)^\top \overset{p}{\rightarrow} \Gamma_\beta
\end{equation*}

To verify Condition 2, For any $e>0$ and unit vector $\mathbf{c}$:
\begin{align*}
    \frac{1}{T}\sum_{t=1}^T & \E[\Vert\mathbf{c}^\top \Gamma_\beta^{-1/2} \psi_t(\beta^\star,\boldsymbol{\eta}^\star,H_{t},A_{t}) \Vert^2 \mathbbm{1}_{\Vert \mathbf{c}^\top\Gamma_\beta^{-1/2} \psi_t\Vert>eT}|\mathcal{F}(H_{t},A_{t})]\\
    & = \frac{1}{T}\sum_{t=1}^T \E[\Vert\mathbf{c}^\top \Gamma_\beta^{-1/2} \psi_t(\beta^\star,\boldsymbol{\eta}^\star,H_{t},A_{t}) \Vert^2 \mathbbm{1}_{\Vert \mathbf{c}^\top\Gamma_\beta^{-1/2} \psi_t\Vert^{\delta}> (eT)^{\delta}}|\mathcal{F}(H_{t},A_{t})]\\
    & \leq \frac{1}{e^{\delta} T^{1+\delta}}\sum_{t=1}^T\E[\Vert\mathbf{c}^\top \Gamma_\beta^{-1/2} \psi_t(\beta^\star,\boldsymbol{\eta}^\star,H_{t},A_{t}) \Vert^{2+\delta} |\mathcal{F}(H_{t},A_{t})]\\
    & = \frac{1}{e^{\delta} T^{\delta}} \times \frac{1}{T}\sum_{t=1}^T\E[\Vert\mathbf{c}^\top \Gamma_\beta^{-1/2} \psi_t(\beta^\star,\boldsymbol{\eta}^\star,H_{t},A_{t}) \Vert^{2+\delta} |\mathcal{F}(H_{t},A_{t})]\\
    & = \frac{1}{e^{\delta} T^{\delta}} \times \frac{1}{T}\sum_{t=1}^T\E[\Vert\mathbf{c}^\top \Gamma_\beta^{-1/2} \tilde \sigma^2_t(S_t)\xi_t f_t(S_t) \Vert^{2+\delta} |\mathcal{F}(H_{t},A_{t})]\\
    & \leq \frac{1}{e^{\delta} T^{\delta}} \times \frac{1}{T}\sum_{t=1}^T\E[\Vert\mathbf{c}^\top \Gamma_\beta^{-1/2} \tilde \sigma^2_t(S_t) f_t(S_t) \Vert^{2+\delta} \cdot \xi_t ^{2+\delta} |\mathcal{F}(H_{t},A_{t})]\\
    & = \frac{1}{e^{\delta} T^{\delta}} \times \frac{1}{T}\sum_{t=1}^T\Vert\mathbf{c}^\top \Gamma_\beta^{-1/2} \tilde \sigma^2_t(S_t) f_t(S_t) \Vert^{2+\delta} \E[ \xi_t ^{2+\delta} |\mathcal{F}(H_{t},A_{t})]\\
    & = o_p(1)
\end{align*}
The last inequality holds because each summand is finite. Thus, with $T$ sufficiently large, the RHS converges to $0$ in probability. Therefore Condition 2 holds true. At this point, we can state the asymptotic normality property of $\hat\beta^{(DR)}$ when $T \rightarrow \infty$:
\begin{equation}
    \sqrt{T}(\hat\beta^{(DR)} - \beta^\star) \sim \mathcal{N}(0,B_{\beta}^{-1}\Gamma_\beta B_{\beta}^{-1})
\end{equation}

The bread $B_{\beta}$ can be consistently estimated by $\P_{n,T}[\tilde \sigma_t^2 (S_t)f_t(S_t)f_t(S_t)^\top]$. As for the meat term, define: 
\begin{align*}
    \frac{1}{T}\sum_{t=1}^T M_t & = \frac{1}{T}\sum_{t=1}^T \psi_t(\beta^\star,\boldsymbol{\hat\eta}, H_{t},A_{t})\psi_t(\beta^\star,\boldsymbol{\hat\eta},H_{t},A_{t})^\top \\
    &~~~~~~~~~~~~~~~~~~~~~~ - \E[\psi_t(\beta^\star,\boldsymbol{\eta}^\star, H_{t},A_{t})\psi_t(\beta^\star,\boldsymbol{\eta}^\star, H_{t},A_{t})^\top| \mathcal{F}(H_t, A_t) ]\\
    & = \frac{1}{T}\sum_{t=1}^T \hat\psi_{t}\hat\psi_{t}^\top - \E[\psi_{t}\psi_{t}^\top|\mathcal{F}(H_t, A_t)] \\
    & = \underbrace{\frac{1}{T}\sum_{t=1}^T \hat\psi_{t}\hat\psi_{t}^\top - \psi_{t}\psi_{t}^\top}_{I}+ \underbrace{\frac{1}{T}\sum_{t=1}^T\psi_{t}\psi_{t}^\top -\E[\psi_{t}\psi_{t}^\top|\mathcal{F}(H_t, A_t)]}_{II}.
\end{align*}
We can show that by Martingale WLLN, term II is $o_p(1)$. As for term I, based on the decomposition of $\hat\psi_t$ and Assumption \ref{ass:T-infinity} (4), we can also conclude that it is $o_p(1)$. To put it more concrete, we can write the decomposition of $\hat\psi_t = \psi_t(\beta^\star,\boldsymbol{\hat\eta}, H_{t},A_{t})$ as follows:
\begin{align}
    \hat\psi_t =&   \tilde \sigma^2_t(S_t) \big(\frac{ \tilde p_t (A_t|S_t) (A_t - \tilde p_t (1|S_t))(Y_{t+1}-  \hat g_t(H_t,A_t))}{\hat p_t(A_t|H_t) \tilde \sigma^2_t(S_t)} +  \hat \beta (t; H_t) - f_t(S_t)^\top \beta^\star\big) f_t(S_t) \nonumber  \\
    =&    \tilde \sigma^2_t(S_t) \Big(\frac{ \tilde p_t (A_t - \tilde p_t(1|S_t) )\big(Y_{t+1}- g^\star_t(H_t,A_t)+g^\star_t(H_t,A_t)- \hat g_t(H_t,A_t)\big)}{\tilde \sigma^2_t(S_t)}\Big(\frac{1}{\hat p_t }-\frac{1}{ p_t }+\frac{1}{ p_t }\Big)  \nonumber \\
    &~~~~~~~~~~~~~~~~~~~~~~~~~~~~~~ + \beta (t; H_t)-f_t(S_t)^\top \beta^\star  +\big(\hat \beta (t; H_t)-\beta (t; H_t)\big) \Big) f_t(S_t)  \nonumber\\
    =&    \tilde \sigma^2_t(S_t)\Big( \frac{ W_t (A_t - \tilde p_t(1|S_t) )(Y_{t+1}- g^\star_t(H_t,A_t))}{\tilde \sigma^2_t(S_t)}+\beta (t; H_t)-f_t(S_t)^\top \beta^\star\Big) f_t(S_t) \\
    &  +    \tilde p_t (A_t - \tilde p_t(1|S_t) )(Y_{t+1}- g^\star_t(H_t,A_t))\Big(\frac{1}{\hat p_t }-\frac{1}{ p_t }\Big)f_t(S_t)\\
    & +    \tilde p_t (A_t - \tilde p_t(1|S_t) )(g^\star_t(H_t,A_t)- \hat g_t(H_t,A_t))\Big(\frac{1}{\hat p_t }-\frac{1}{ p_t }\Big) f_t(S_t)\\
    &+    W_t (A_t - \tilde p_t (1|S_t))\big(g^\star(H_t,A_t)-\hat g_t(H_t,A_t)\big)f_t(S_t)\\
    & +     \tilde \sigma^2_t(S_t) \big(\hat \beta (t; H_t)-\beta (t; H_t)\big)f_t(S_t)
\end{align}
Here, \(\hat\psi_t\) is the sum of terms (22) to (26), where Term (22) is exactly \(\psi_t\). Using the inequality \((\sum_{i=1}^5 x_i)^2 \leq 5 \sum_{i=1}^5 x_i^2\), we can prove that Term I converges at the \(o_p(1)\) rate. To see this, the square of Term (22) cancels with the \(\psi_t\psi_t^\top\) in Term I. 
The mean sum of squares for Terms (23), (25), and (26) converges at \(o_p(1)\) based on Assumption \ref{ass:T-infinity} (4). The only remaining mean sum of squares is for Term (24), which we can prove as follows:
\begin{align*}
    & \frac{1}{T}\sum_{t=1}^T\tilde p^2_t (A_t - \tilde p_t(1|S_t) )^2(g^\star_t(H_t,A_t)- \hat g_t(H_t,A_t))^2\left(\frac{1}{\hat p_t }-\frac{1}{ p_t }\right)^2\\
    \lesssim &~ \frac{1}{T} \left( \sum_{t=1}^T (p_t^\star(1|H_t)-\hat p_t(1|H_t))^2 \right)  \times \left(\sum_{t=1}^T \big(g_t^\star(H_t,A_t)-\hat g_t(H_t,A_t)\big)^2\right) \\
    = &~ T  \times \Vert p_t^\star(1|H_t)-\hat p_t(1|H_t) \Vert^2_T \times \Vert g_t^\star(H_t,A_t)-\hat g_t(H_t,A_t) \Vert^2_T\\
    =& o_p(1).
\end{align*}

Overall, we have 
\begin{equation*}
    \frac{1}{T}\sum_{t=1}^T M_t \overset{P}{\rightarrow} 0
\end{equation*}
which is equivalent to:
\begin{align*}
    \frac{1}{T} \sum_{t=1}^T \hat\psi_t \hat\psi_t^\top  - \E[\psi_t \psi_t^\top|\mathcal{F}(H_t,A_t)]\overset{P}{\rightarrow} 0.
\end{align*}
Thus, we ensure that even with the nuisance function estimates as plug-ins, we can still consistently estimate $\Gamma_{\beta}$ .

\subsection{Connection to Previous Theoretical Results}

To clarify, this paper focuses on using micro-randomized trials (MRTs) as a study design to evaluate the causal effect of sequential stochastic interventions. In this setup, adaptiveness comes in through the randomization probability \( p_t(A_t | H_t) \), which is designed to balance optimizing participant outcomes during the study and ensuring valid post-study causal effect inference. 
Because of this, a sublinear rate of selecting a particular treatment arm is not supposed to happen, and we make the strong positivity assumption to reflect this point.

When data are collected under an adaptive intervention design, standard least squares estimators may fail to be asymptotically normal as $T \rightarrow \infty$ \citep{deshpande2018accurate, hadad2021confidence, zhang2020inference}. MRTs with long time horizons share similar probabilistic tools for asymptotic analysis with frameworks such as response-adaptive randomization (RAR, \cite{hu2006theory}) and bandit problems, but differ in goals and design. MRTs involve within-person randomization to improve participant outcomes during the study while enabling post-study causal inference and policy evaluation. Bandit algorithms, which aim to optimize cumulative reward, can be used to guide treatment allocation in MRTs, so long as the exploration–exploitation trade-off is carefully managed to preserve valid inference. In contrast, RAR adjusts treatment allocations across independent participants in clinical trials to balance overall patience welfare and statistical efficiency.

In \cite{zhang2021statistical}, the authors note that ``different forms of adaptive weights are used by existing methods for simple models \citep{deshpande2018accurate, hadad2021confidence, zhang2020inference}''. They show that M-estimators can provide valid statistical inference on adaptively collected data when adjusted with appropriately chosen adaptive weights
which represents a step toward developing a general framework for statistical inference on data collected with adaptive algorithms, including (contextual) bandit algorithms. 
Our discussion below focuses on drawing a connection between the asymptotic normality results we establish and those presented in \cite{zhang2021statistical}.

First, regarding estimator \textbf{consistency}, both our approach and that of \cite{zhang2021statistical} rely on a strong positivity assumption on the treatment randomization probability. While \cite{zhang2021statistical} assumes the outcome regression model is correctly specified, our method is based on a Z-estimator framework. In both cases, these conditions are sufficient to ensure consistency. To show \textbf{asymptotic normality}, \cite{zhang2021statistical} and we follow the same central limit theorem for Martingale Difference Sequences \citep{dvoretzky1972asymptotic}. The key difference lies in how the convergence of the conditional variance to a finite constant matrix is established.

\cite{zhang2021statistical} assume the environment of the contextual bandit is stationary, where $\{X_t, Y_t(a), a \in \mathcal{A}\} \overset{i.i.d}{\sim} \mathcal{P}$ for $t \in [1:T]$, and use a stabilizing weight to derive an explicit form of the limiting variance matrix. In contrast, we allow for general temporal dependence and establish conditional variance convergence by assuming that the unconditional variance of the estimator converges a finite constant matrix. Under an additional assumption that the correlation among conditional residuals decays over time, we show that the conditional variance also converges to this same limit. This, in turn, justifies the use of the empirical robust sandwich estimator for consistently estimating the asymptotic variance.

Similar assumptions to ours have been made in \cite{bojinov2019time}, \cite{ yu2023multiplicative} and \cite{liu2024incorporating}, where it is directly assumed that the conditional variance of the estimator converges to a finite constant matrix. Our set of assumptions can be viewed as sufficient conditions that guarantee the variance convergence assumed in the prior literature.

In summary, as noted by \cite{hadad2021confidence} in the section titled ``Asymptotically Normal Test Statistics'' of their paper,
 ``Formally, what is required is that the sum of conditional variances of each term in the sequence converges to the unconditional variance of the estimator'', and our Theorem 4 provided sufficient conditions for this purpose.

\subsection{Simulations results under a static treatment policy} 

We adopt the simulation setup described in Equation \eqref{eq:generativemodel} in Section \ref{sec:sim}, focusing on the DR-WCLS approach. The empirical results presented below correspond to $n = 3$ and $T = 150$. The randomization probability is $p_t(1|H_{t }) = \text{expit}(-0.8 A_{t-1,j})$; the state dynamics are given by $\mathbb{P}(S_{t,j}=1|A_{t-1 },H_{t-1 })=1/2$ with $A_0 = 0$, and the independent error term satisfies $e_{t,j} \sim \mathcal{N}(0,1)$ with $\text{Corr}(e_{u,j}, e_{t,j'}) = {\bf 1}(j=j') 0.1^{|u-t|/2}$. As in Section \ref{sec:sim}, we set $ \beta_{10}=-0.2$, and $\beta_{11} = \{0.2, 0.5, 0.8\}$. The marginal proximal effect is equal to $\beta_{10} + \beta_{11} \E [S_{t,j} ]=\beta_{10} = -0.2$. The marginal treatment effect is therefore constant over time and is given by $\beta_0^\star = \beta_{10} =  -0.2$. Nuisance functions are estimated using the time-wise block cross-fitting procedure detailed in Appendix E.2.

\begin{table}[htbp]
\caption{\centering Fully marginal causal effect estimation. The true value of the parameters is $\beta_{0}^\star = -0.2$.}
\label{tab:tabfour}
\begin{center}
\begin{tabular}{cccccccc}
\hline
Method & $\beta_{11}$ & Est & SE & RMSE & CP  \\\hline
\multirow{3}{*}{DR-WCLS} & 0.2 & -0.201 & 0.070 & 0.066 & 0.960  \\
& 0.5&-0.196 & 0.080 & 0.077 & 0.957   \\
&0.8& -0.200 & 0.085 & 0.081 & 0.956 \\
\hline
\end{tabular}
\end{center}
\end{table}

\subsection{Simulations results under a clipped bandit treatment policy} 

We adopt a similar setup to that in \cite{zhang2021statistical}, where we assume a stochastic contextual bandit environment in which $\{O_t(a), a \in \mathcal{A}\} \overset{i.i.d}{\sim} \mathcal{P}$ for $t \in [1:T]$. Even though the potential outcomes are i.i.d., the observed data are not because the treatment are selected using policies $p_t(A_t|H_t)$, which is a function of past data, $H_{t}$. The modification on treatment randomization we've made here is to use a clipped linear bandit algorithm, with binary treatment randomization probabilities constrained between 0.01 and 0.99. The clipped bandit algorithm randomizes treatment with probability $p_t(A_t = 1 \mid H_t) = \text{expit}(2(\hat\gamma_{1,t} + \hat\gamma_{3,t} S_t))$, where the coefficients $\boldsymbol{\hat\gamma}_t$ are updated online by regressing the outcome $Y_{t+1}$ on $A_t$, $S_t$, and their interaction $A_t S_t$, i.e., $Y_{t+1} \sim \gamma_0 + \gamma_1 A_t + \gamma_2 S_t + \gamma_3 A_t S_t$, allowing the algorithm to favor treatment based on context.

We use the data-generating process specified in Equation \eqref{eq:generativemodel}, with the independent error term satisfies $e_{t,j} \overset{i.i.d}{\sim} \mathcal{N}(0,1)$ and fixed coefficients shown as below:
\begin{equation*}
    Y_{t,j} =  g_t(H_t) + \big(A_{t,j} - p_t(1|H_t)\big)(-0.2 + 0.8 S_{t,j})+  e_{t,j}.
\end{equation*}

Below, we present empirical results for $n = 3$ and $T = 150$. Nuisance functions were estimated using the time-wise block cross-fitting approach described in Appendix E.2.

\begin{table}[htbp]
\caption{\centering Fully marginal causal effect estimation. The true value of the parameters is $\beta_{0}^\star = -0.2$.}
\label{tab:tabfive}
\begin{center}
\begin{tabular}{cccccccc}
\hline
Method  & Est & SE & RMSE & CP  \\\hline
DR-WCLS & -0.193 & 0.150 & 0.148  & 0.964  \\
\hline
\end{tabular}
\end{center}
\end{table}

\section{Cross-Fitting for Learning Nuisance Functions}
\label{app:samplesplit}

\subsection{Sample Split}

The estimation technique developed in this paper relies on K-fold cross-validation obtained by randomly partitioning the sample, i.e., estimate the nuisance models $\hat g_t(H_t, A_t)$ and $\hat p_t(t;H_t)$ on one part of the data (training data) and estimate the parameter of interest $\hat \beta$ on the other part of the data (test data). To partition the entire sample into K folds, we assume that the individuals are independently distributed. As a result, we can divide the entire population into K groups and perform cross-fitting.

Cross-fitting plays an important role here: the defined regression procedure estimates the pseudo-outcome on a separate sample, independent of the one used in the second-stage regression \citep{kennedy2020optimal}, which allows informative error analysis while being agnostic about the first-stage methods.




\subsection{Time-Wise Sample Split}
\label{app:time-samplesplit}

A straightforward approach to train the nuisance function is to use the history data $H_t$ at each time point. This ensures the consistency and asymptotic normality as proved in Appendix \ref{app:T-infinity}, because $H_t$ is conditionally independent of current observations after conditioning on itself. However, this method can lead to highly variable estimates for early time points due to the small training set size. Given that we are conducting an offline evaluation of the causal excursion effect, it is reasonable to incorporate some future data into the estimation of the current nuisance functions to improve precision.

Sample split is challenging due to time dependence, and it requires assumptions about local dependence to be valid. However, it can be done under additional assumptions:

\begin{assumption}
\label{ass:time-split}
    There exists a positive integer $r$ such that $Y_{t+1}$ is conditionally mean independent of the future at least $r>0$ steps ahead, i.e., $\E[Y_{t+1}|H_t,A_t,\tilde H_{t+r}] = \E[Y_{t+1}|H_t,A_t]$, where $\tilde H_{t+r}$ denotes the information accumulated after time $t+r$.
\end{assumption}

This assumption allows us to train the nuisance function $\hat g_t(H_t, A_t)$ using both historical and future data, $\{H_t,A_t,\tilde H_{t+r}\}$, without introducing bias. This condition can be met if the outcome sequence $\{Y_{t+1}\}_{t=1}^T$ exhibits only local dependence, thus conditioning on observations sufficiently distant does not change the current outcome's conditional expectation (the same applies to the treatment sequence $\{A_t\}_{t=1}^T$). Consider a scenario where the outcome sequence $\{Y_{t+1}\}_{t=1}^T$ follows a finite-order moving average process, MA($l$). In this case, setting $r = l+1$ ensures that Assumption \ref{ass:time-split} is satisfied. In contrast, if the sequence instead follows an autoregressive AR($l$) process, the equality posed by Assumption \ref{ass:time-split} generally fails to hold. Nevertheless, as $r$ increases, the discrepancy between $\mathbb{E}[Y_{t+1} \mid H_t, A_t, \tilde{H}_{t+r}]$ and $\mathbb{E}[Y_{t+1} \mid H_t, A_t]$ diminishes, becoming negligible for sufficiently large $r$.  

It is important to emphasize that Assumption \ref{ass:time-split} serves to eliminate bias introduced by time-wise sample splitting in the estimation procedure. This assumption is stricter than what is necessary for ensuring the consistency of causal parameter estimates. Consequently, minor deviations between $\mathbb{E}[Y_{t+1} \mid H_t, A_t, \tilde{H}_{t+r}]$ and $\mathbb{E}[Y_{t+1} \mid H_t, A_t]$ may be permissible without compromising consistency. A formal characterization of the allowable discrepancy and its implications for inference remains an open question, which we defer to future work.

Based on Assumption \ref{ass:time-split} and inspired by the Leave-One-Out and K-fold cross validation methods in the i.i.d. case, we outline two time-wise sample split procedures below.

\begin{figure}[htbp]
\begin{center}
\includegraphics[width=5in]{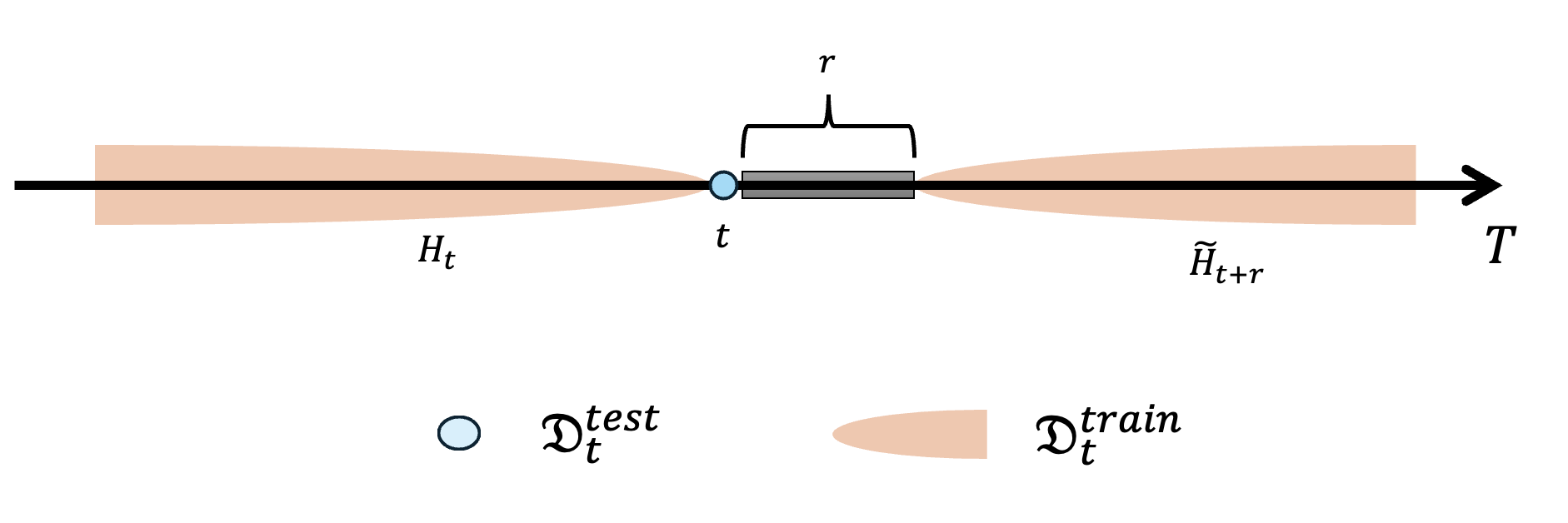}
\end{center}
\caption{Time-Wise Cross Fitting.} 
\label{fig:time-split}
\end{figure}

First, based on the principle of maximizing the time points in the training data, we propose using the training data defined as $\mathcal{D}^{train}_{t} = \{H_t, A_t, \tilde H_{t+r}\}$. For implementation, we train the nuisance function on $\mathcal{D}^{train}_{t}$ at each time point $t$ and apply it to the testing data, which is a single observation made at time point $t$. In this way, the training and testing data are conditionally independent given the history $H_t$. Appendix \ref{app:T-infinity} shows that this strategy guarantees the consistency and asymptotic normality of the causal parameter estimation as $T \rightarrow \infty$. A significant advantage of this procedure is that it allows us to take advantage of the largest possible training set to learn nuisance functions, which helps to improve their accuracy and ensures that Assumption \ref{ass:T-infinity} (4) holds. Below, we provide an illustrative figure in Figure \ref{fig:time-split}. 

\begin{figure}[htbp]
\begin{center}
\includegraphics[width=5in]{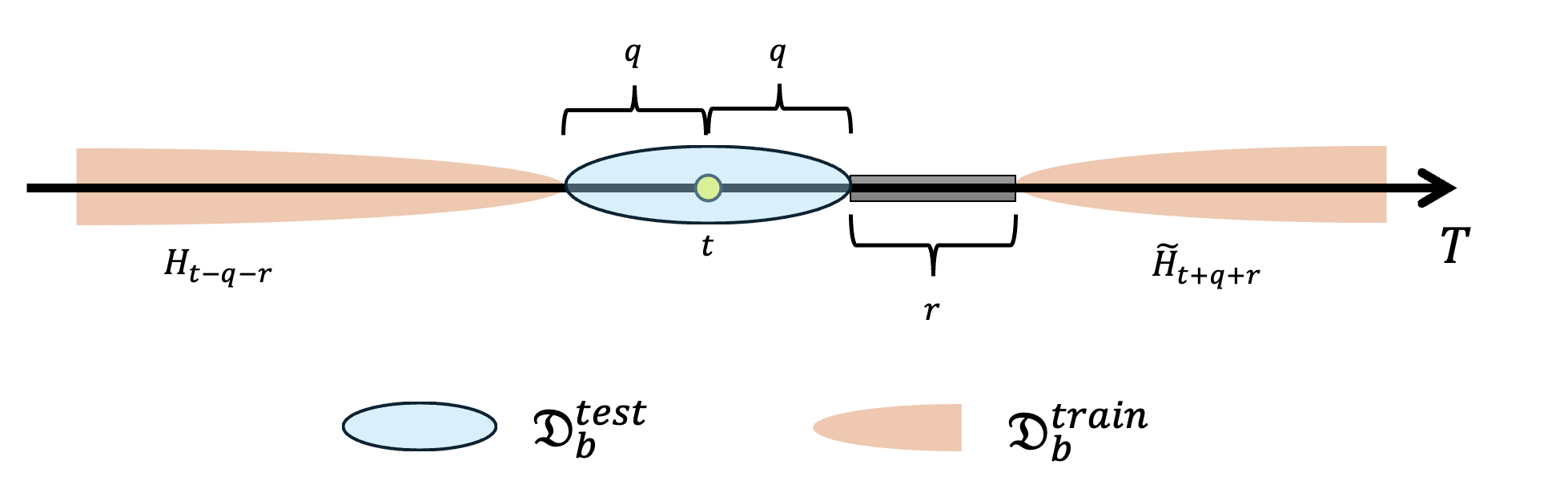}
\end{center}
\caption{Time-Wise Block Cross Fitting.} 
\label{fig:time-block}
\end{figure}

One drawback of the aforementioned procedure is its substantial computational burden, as it requires training the nuisance model at every time point. This becomes particularly challenging when the total number of time points $T$ is significantly large. To address this, we propose a time-block sample split method similar to the scheme introduced in \cite{gilbert2021causal}. Below, we provide an illustrative figure in Figure \ref{fig:time-block}. 

\textbf{Step 1}: Select a single observation at random and then subsample all observations within a distance of $q$ from that point. We denote this set of observations as $\mathcal{D}^{test}_{b}$.

\textbf{Step 2}: The training set, denoted as $\mathcal{D}^{train}_{b}$, consists of all units in the past, and future observations where the minimum distance from any time point in $\mathcal{D}^{test}_{b}$ is at least $r$. The choice of $r$ should refer to Assumption \ref{ass:time-split}.

\textbf{Step 3}: The nuisance functions are learned using the data in $\mathcal{D}^{train}_{b}$ and then applied to the data in $\mathcal{D}^{test}_{b}$ to obtain $\boldsymbol{\hat\eta}^b_t(H_t, A_t)$ for the time points between $t-q$ and $t+q$.

\textbf{Step 4}: Repeat Step 1-3 for $b= 1,2,\dots, B$ times and the resulting estimates averaged to obtain $\boldsymbol{\hat\eta}_t(H_t, A_t) = \sum_{b=1}^B \boldsymbol{\hat\eta}^b_t(H_t, A_t) \mathbf{1}(t \in \mathcal{D}^{test}_{b})/\sum_{b=1}^B \mathbf{1}(t \in \mathcal{D}^{test}_{b})$ for each time point.



Here, $q$ is chosen based on $T$ to approximately reach a target subsample size, such as $n/K$ for some fixed $K$ in K-fold sample split. As the total number of subsamples \( B \) approaches infinity, we can summarize that for time \( t \), the limit training set for \( \boldsymbol{\hat\eta}_t \) is \( \{H_{t}, \tilde H_{t+r}\} \) and the test set is always \( \{H_t, A_t\} \). Thus, \( \boldsymbol{\hat\eta}_t(H_t, A_t) \) is a function of \( \{H_t, A_t, \tilde H_{t+r}\} \), which is aligned with the first scenario. Therefore, the proof in Appendix \ref{app:T-infinity} remains valid after modifying the training set definition.

This block split approach saves the effort of training the nuisance function separately for each time point. However, in this case if $r$ appears to be significantly large, the drawback is that we choose to utilize less data to fit the nuisance function, which might cause the error term between the fitted function and the true conditional expectation to fail to meet Assumption \ref{ass:T-infinity}. This could jeopardize the consistency of causal parameter estimation. However, this is not the case for the first approach.

\section{Doubly Robust Estimation with Missing Observations}
\label{sec:missingdata}

In mHealth studies, it is common for both the proximal outcome~$Y_{t+1}$ and elements of the history~$H_t$ to be missing.  In the case study of a 6-month MRT on medical interns presented in Section~\ref{sec:casestudy}, for example, the proximal outcomes are self-reported mood score and step count. Self-reports are often missing due to non-response, while step count can be missing due to individuals not wearing the wrist sensors. Previous approaches are not equipped to address missing data in the context of MRTs and require complete observation data for their application  \citep{boruvka2018, dempsey2020, qian2020estimating}. Here, we extend the DR-WCLS criterion to be robust to missing data. 
 

Specifically, we focus on missing outcomes $\{Y_{t+1}\}_{t=1}^T$, assuming the moderator set $\{S_t\}_{t=1}^T$ is fully observed. Let $R_{t}$ be the binary indicator of whether the proximal outcome $Y_{t+1}$ is observed ($R_t = 1$) or not ($R_t = 0$) at the decision time $t$, and $R_t (\bar a_t)$ denotes the potential observation status.  Clearly, missingness is a post-treatment variable and therefore we require additional assumptions:

\begin{assumption}
\label{ass:md_drwcls}
We assume consistency, missing at random, and positivity:
\begin{enumerate}
 \item Consistency: For each $t \leq T$,
    $R_t (\bar A_t) = R_t$, i.e., the observed missing data indicator is equal to the corresponding potential outcome observation status; 
 \item Missing at random: For each $t \le T$, $R_t(\bar a_{t})$ is independent of $A_t$ conditional on the observed history $H_t$;
 \item Positivity: if the joint density $\{R_t = r_t, H_t = h_t, A_t = a_t\}$ is greater than zero, then $ \epsilon^R < p(R_t=1|H_t,A_t) = p(R_t|H_t) < 1-\epsilon^R$ for some constant $\epsilon^R >0$.
\end{enumerate}
\end{assumption}

Under Assumption \ref{ass:md_drwcls}, we can derive a doubly robust extension for missing data by modifying the DR-WCLS criterion as follows:
\begin{equation}
\label{eq:md_drwcls}
    \resizebox{0.99\textwidth}{!}{%
        $\mathbb{P}_n \Big[\sum_{t=1}^T \tilde \sigma^2_t(S_t) \Big(
       \frac{{\bf 1}(R_t = 1)}{p (R_t | H_t)} \frac{W_t (A_t - \tilde p_t (1|S_t))(Y_{t+1}- g_t(H_t,A_t))}{\tilde \sigma^2_t(S_t)} +\beta (t; H_t) - f_t(S_t)^\top \beta\Big) f_t(S_t) \Big]  =0.$%
        }
\end{equation}
\sloppy Equation \eqref{eq:md_drwcls} extends Equation \eqref{eq:drwcls} by weighting the first term with the inverse probability of missingness. Since the missingness mechanism is another complex nuisance component, it naturally fits within the meta-learning framework. Let $\boldsymbol{\eta}_t(H_t, A_t) = (g(H_t, A_t), p(R_t,A_t|H_t))$. The estimator $\hat \beta^M_n$ from Equation \eqref{eq:md_drwcls} is asymptotically normal as $n \to \infty$; see Appendix \ref{app:md_drwcls} for proofs and results for large $T$.

\begin{corollary}(Asymptotic property for the DR-WCLS estimator with missing data)
\label{cor:dr_wcls}
Under Assumptions \ref{ass:po}, \ref{ass:directeffect}, \ref{ass:nuisance-op1} in the main text, and Assumption \ref{ass:md_drwcls}, given invertibility and moment conditions, the estimator $\hat \beta^{M}_n$ that solves \eqref{eq:md_drwcls} is subject to an error term, which (up to a multiplicative constant) is bounded 
by:
\begin{equation}
    \mathbf{\hat B}^{M} = \sum_{t=1}^T  \sum_{a \in \{0,1 \}}\left\Vert \hat p_t(R_t=1,A_t = a|H_t)- p_t(R_t=1,A_t = a|H_t)\right\Vert \left\Vert \hat g_t(H_t,a)- g_t(H_t,a)\right\Vert .
\end{equation}
If we further assume $\mathbf{\hat B}^{M} =  o_p(n^{-1/2})$, $\hat \beta^{M}_n$ is consistent and asymptotically normal such that $\sqrt{n}(\hat \beta^{M}_n - \beta^\star) \rightarrow \mathcal{N}(0,\Sigma^M_{DR})$, where $\Sigma^M_{DR}$ is defined in Appendix \ref{app:md_drwcls}.
\end{corollary}

\subsection{Proof of Corollary \ref{cor:dr_wcls}}
\label{app:md_drwcls}

\subsubsection{Double robustness property}
To derive the DR-WCLS criterion \eqref{eq:md_drwcls} with the missing indicator $R_t$. Under Assumption \ref{ass:po} and \ref{ass:md_drwcls}, the pseudo outcome $\tilde Y_{t+1}^{(DR)}$ can be written as: 

\begin{align*}
    \tilde Y_{t+1}^{(DR)} =&\beta(t;H_t) + \frac{A_t R_t(Y_{t+1}-g(H_t,A_t))}{p_t(A_t, R_t|H_t)} - \frac{(1-A_t) R_t(Y_{t+1}-g(H_t,A_t))}{p_t(A_t, R_t|H_t)} \\
    =&\beta(t;H_t) + \frac{A_t R_t(Y_{t+1}-g(H_t,A_t))}{p(R_t|H_t)p(A_t|H_t)} - \frac{(1-A_t) R_t(Y_{t+1}-g(H_t,A_t))}{p(R_t|H_t)p(A_t|H_t)} \\
    =&\beta(t;H_t) + \frac{R_t}{p(R_t|H_t)}\Big[\frac{A_t (Y_{t+1}-g(H_t,A_t))}{p(A_t|H_t)} - \frac{(1-A_t)(Y_{t+1}-g(H_t,A_t))}{p(A_t|H_t)}\Big] \\
    =& \beta(t;H_t) + \frac{{\bf 1}(R_t = 1)}{p (R_t | H_t)}\frac{W_t (A_t - \tilde p_t (1|S_t))(Y_{t+1}- g_t(H_t,A_t))}{\tilde \sigma^2_t(S_t)}
\end{align*}
and the corresponding estimating equation is:
\begin{align*}
    \mathbb{P}_n \Big[\sum_{t=1}^T \tilde \sigma^2_t(S_t) \Big(\frac{{\bf 1}(R_t = 1)}{p (R_t | H_t)} \frac{W_t (A_t - \tilde p_t (1|S_t))(Y_{t+1}- g_t(H_t,A_t))}{\tilde \sigma^2_t(S_t)} + \beta (t; H_t) - f_t(S_t)^\top \beta\Big) f_t \Big]
\end{align*}

Furthermore, based on previous proofs, we can conclude that the $\hat\beta_n$ obtained by solving the above estimating equation is doubly robust. 

\subsubsection{Asymptotic normality}

We use the same notation as the previous section, assume $\hat \beta^{(DR)}_n$ setting the following equation to $0$:
\begin{align*}
    \mathbb{P}_n \Big[\sum_{t=1}^T \tilde \sigma^2_t(S_t) \Big(\frac{{\bf 1}(R_t = 1)}{p (R_t | H_t)} \frac{W_t (A_t - \tilde p_t (1|S_t))(Y_{t+1}- g_t(H_t,A_t))}{\tilde \sigma^2_t(S_t)} + \beta (t; H_t) - f_t(S_t)^\top \beta\Big) f_t(S_t) \Big]
\end{align*}

When the true randomization probability $p_t = p(A_t|H_t)$ and missing mechanism $p_t^R = p(R_t|H_t)$ are unknown, we have the weight $W_t$ estimated by $ \hat W_t = \tilde p_t(A_t|S_t)/ \hat p_t(A_t|H_t)$, and missing mechanism estimated by $\hat p(R_t|H_t)$. Then the estimating equation can be decomposed as:
\begin{align*}
    &\P_n \Big[\sum_{t=1}^T \tilde \sigma^2_t(S_t) \Big(\frac{R_t}{\hat p (R_t | H_t)}\frac{ \tilde p_t (A_t|S_t) (A_t - \tilde p_t (1|S_t))(Y_{t+1}-  \hat g_t(H_t,A_t))}{\hat p_t(A_t|H_t) \tilde \sigma^2_t(S_t)} +  \hat \beta (t; H_t) - f_t(S_t)^\top \beta\Big) f_t(S_t) \Big] \\
    =&\P_n \Big[\sum_{t=1}^T \tilde \sigma^2_t(S_t) \Big(\frac{R_t \tilde p_t (A_t - \tilde p_t(1|S_t) )(Y_{t+1}- g^\star_t(H_t,A_t)+g^\star_t(H_t,A_t)- \hat g_t(H_t,A_t))}{\tilde \sigma^2_t(S_t)}\Big(\frac{1}{\hat p_t \hat p_t^R }-\frac{1}{ p_t p_t^R }+\frac{1}{ p_t p_t^R}\Big) \\
    &  + \beta (t; H_t)-f_t(S_t)^\top \beta^\star  +(\hat \beta (t; H_t)-\beta (t; H_t)) - f_t(S_t)^\top (\beta-\beta^\star)\Big) f_t(S_t) \Big]\\
    =& \P_n \Big[\sum_{t=1}^T \tilde \sigma^2_t(S_t)\Big( \frac{ R_t W_t (A_t - \tilde p_t(1|S_t) )(Y_{t+1}- g^\star_t(H_t,A_t))}{p_t^R \tilde \sigma^2_t(S_t)}+\beta (t; H_t)-f_t(S_t)^\top \beta^\star\Big) f_t(S_t)\Big]\\
    &   + \P_n \Big[\sum_{t=1}^T R_t \tilde p_t (A_t - \tilde p_t(1|S_t) )(Y_{t+1}- g^\star_t(H_t,A_t))\Big(\frac{1}{\hat p_t^R \hat p_t }-\frac{1}{p_t^R p_t }\Big)f_t(S_t)\Big]\\
    & + \P_n \Big[\sum_{t=1}^T R_t\tilde p_t (A_t - \tilde p_t(1|S_t) )(g^\star_t(H_t,A_t)- \hat g_t(H_t,A_t))\Big(\frac{1}{\hat p_t^R \hat p_t }-\frac{1}{ p_t^R p_t }\Big) f_t(S_t)\Big] \\
    & + \P_n \Big[\sum_{t=1}^T \frac{R_t}{p_t^R}W_t (A_t - \tilde p_t (1|S_t))(g^\star(H_t,A_t)-\hat g_t(H_t,A_t))f_t(S_t)\Big] \\
    & + \P_n \Big[\sum_{t=1}^T  \tilde \sigma^2_t(S_t) (\hat \beta (t; H_t)-\beta (t; H_t))f_t(S_t)\Big]-\P_n \Big[\sum_{t=1}^T \tilde \sigma^2_t(S_t) f_t(S_t)f^\top_t(S_t)\Big](\beta_n-\beta^\star)
\end{align*}
By the WLLN, we have the following (element-wise) convergence results:
\begin{align*}
 & \P_n \Big[\sum_{t=1}^T R_t \tilde p_t (A_t - \tilde p_t(1|S_t) )(Y_{t+1}- g^\star_t(H_t,A_t))\Big(\frac{1}{\hat p_t^R \hat p_t }-\frac{1}{ p_t^R p_t }\Big)f_t(S_t)\Big] \overset{P}{\to} 0, \\
    &\P_n \Big[\sum_{t=1}^T \tilde \sigma^2_t(S_t) f_t(S_t)f^\top_t(S_t)\Big] \overset{P}{\to} \E\Big[\sum_{t=1}^T \tilde \sigma^2_t(S_t) f_t(S_t)f^\top_t(S_t)\Big] ,
\end{align*}
and
\begin{align*}
    & \P_n \Big[\sum_{t=1}^T \frac{R_t}{p_t^R} W_t (A_t - \tilde p_t (1|S_t))(g^\star(H_t,A_t)-\hat g_t(H_t,A_t))f_t(S_t)\Big] +\\
    &~~~~~~~~~~~~~~~~~~~~~~~~~~~~~\P_n \Big[\sum_{t=1}^T  \tilde \sigma^2_t(S_t) (\hat \beta (t; H_t)-\beta (t; H_t))f_t(S_t)\Big]\overset{P}{\to} 0 .
\end{align*}
Apart from the nicely-behaved terms above, the only term that might be problematic and causes bias is:
$$
\P_n \Big[\sum_{t=1}^T R_t \tilde p_t (A_t - \tilde p_t(1|S_t) )(g^\star_t(H_t,A_t)- \hat g_t(H_t,A_t))\Big(\frac{1}{\hat p_t^R \hat p_t }-\frac{1}{ p_t^R p_t }\Big) f_t(S_t)\Big],
$$
which equals:
\begin{align*}
    &\sum_{t=1}^T \sum_{a \in \{0,1 \}} \Big(\underbrace{\P_n\Big[  {\bf c}(a)(p_t^R(1|H_t) p_t(a|H_t)-\hat p_t^R(1|H_t) \hat p_t(a|H_t))(g_t^\star(H_t,a)-\hat g_t(H_t,a))}_{\text{(II)}}\Big) f_t(S_t)\Big],
\end{align*}
where ${\bf c}(a)= \frac{ \tilde \sigma^2_t(S_t)/\hat p_t^R(1|H_t)}{a\hat p_t(1|H_t)+ (a-1)(1-\hat p_t(1|H_t))}$. In our context, $T$ is finite and fixed. Therefore, by the fact that $\hat p_t(1|H_t))$ is bounded away from zero and one, along with the Cauchy–Schwarz inequality, we have that (up to a multiplicative constant) term (II) is bounded above by:
\begin{equation}
    \mathbf{\hat B}^M = \sum_{t=1}^T \sum_{a \in \{0,1 \}} \left\Vert p_t^R(1|H_t)p_t(a|H_t)-\hat p_t^R(1|H_t) \hat p_t(a|H_t)\right\Vert \cdot \left\Vert g_t^\star(H_t,a)-\hat g_t(H_t,a)\right\Vert .
\end{equation}
Same argument as in the previous section, if $\hat p(a|H_t)$ and $\hat p_t^R(1|H_t)$ are based on a correctly specified parametric model, so that $\left\Vert \hat p_t^R(1|H_t) \hat p_t(a|H_t) - p_t^R(1|H_t) p_t(a|H_t)\right\Vert = O_p(n^{-1/2})$, then we only need $\hat g_t(H_t,a)$ to be consistent, $\left\Vert g_t^\star(H_t,a)-\hat g_t(H_t,a)\right\Vert = o_p(1)$, to make $\mathbf{\hat B}^M$ asymptotically negligible. Thus if we know the treatment and data missingness mechanism, the outcome model can be very flexible. Another way to achieve efficiency is if we have both $\left\Vert \hat p_t^R(1|H_t) \hat p_t(a|H_t) - p_t^R(1|H_t) p_t(a|H_t)\right\Vert = o_p(n^{-1/4})$ and $\left\Vert g^\star(H_t,a)-\hat g(H_t,a)\right\Vert = o_p(n^{-1/4})$, so that their product term is $o_p(n^{-1/2})$ and asymptotically negligible \citep{kennedy2016semiparametric}. This of course occurs if both $\hat g_t(H_t,a)$ and $\hat p_t^R(1|H_t) \hat p_t(a|H_t)$ are based on correctly specified models, but it can also hold even for estimators that are very flexible and not based on parametric models.

Assuming we have nuisance estimates that can make $\mathbf{\hat B}^M$ asymptotically negligible. Along side with other terms converge at a $o_p(n^{-1/2})$ rate:
\begin{align*}
    & \Big(\P_n \Big[\sum_{t=1}^T 
    \frac{R_t}{p_t^R} W_t (A_t - \tilde p_t (1|S_t))(g^\star(H_t,A_t)-\hat g_t(H_t,A_t))f_t(S_t) \Big]\Big)^2\\
    &\lesssim \frac{1}{n}\P_{n} \Big[\sum_{t=1}^T \big(\frac{R_t}{p_t^R}  W_t (A_t - \tilde p_t (1|S_t))(g^\star(H_t,A_t)-\hat g_t(H_t,A_t))\big)^2 f_t(S_t)f_t(S_t)^\top \Big]\\
    & \lesssim \P_{n}\Big[ \sum_{t=1}^T (g^\star(H_t,A_t)-\hat g_t(H_t,A_t))^2\Big]\times \frac{1}{n}\pmb{1}_{q\times q}\\
    & = o_p(1) \times \frac{1}{n}\pmb{1}_{q\times q} \\
    & = o_p(n^{-1}) \times \pmb{1}_{q\times q},
\end{align*}
The first inequality follows from the assumption that $T$ is finite and fixed, while the second inequality holds because $p_t^R$, $R_t$, $W_t$ and $f_t(S_t)$ are all bounded. The $o_p(1)$ comes from the assumption that the nuisance functions should at least estimate the true outcome conditional expectation consistently. By an abuse of notation, we still use $\tilde Y_{t+1}^{(DR)} = \frac{ R_t W_t (A_t - \tilde p_t(1|S_t) )(Y_{t+1}- g^\star_t(H_t,A_t))}{p_t^R \tilde \sigma^2_t(S_t)}+\beta (t; H_t)$. Then the DR-WCLS estimator satisfies:
\begin{align*}
    n^{1/2}(\hat\beta^{(DR)}_n-\beta^{\star}) = n^{1/2}~ \P_n \Big[\sum_{t=1}^T ~ \E& \Big[\sum_{t=1}^T \tilde \sigma^2_t(S_t) f_t(S_t)f^\top_t(S_t)\Big]^{-1} \tilde \sigma^2_t(S_t) \big(\tilde Y_{t+1}^{(DR)} - \\
    & f_t(S_t)\beta^\star\big)f_t(S_t) \Big] + o_p(1),
\end{align*}
and it is efficient with influence function:
\begin{align*}
        &\sum_{t=1}^T ~  \E\Big[\sum_{t=1}^T \tilde \sigma^2_t(S_t) f_t (S_t)f_t (S_t)^\top  \Big]^{-1} \tilde \sigma^2_t(S_t) (\tilde Y_{t+1}^{(DR)} - f_t(S_t)\beta^\star)f_t(S_t).
\end{align*}
In conclusion, under moment conditions, we have asymptotic normality with variance given by $\Sigma^M_{DR}=Q^{-1} W Q^{-1}$, where
\begin{align*}
    Q & = \E\Big[\sum_{t=1}^T \tilde \sigma^2_t(S_t) f_t (S_t)f_t (S_t)^\top  \Big], \\
    W &= \E \Big[ \Big(\sum_{t=1}^T \tilde \sigma^2_t(S_t) (\tilde Y_{t+1}^{(DR)} - f_t(S_t)\beta^\star)f_t(S_t)\Big)^2\Big].
\end{align*}

\subsection{Algorithm}
\label{app:sec:md-algorithm}

\textbf{Step I} Let $K$ be a fixed integer. Form a K-fold random partition of $\{1,2,\dots,N\}$ by dividing it to equal parts, each of size $n := N/K$, assuming $N$ is a multiple of $K$. Form each set $I_k$, let $I_k^\complement$ denote the observation indices that are not in $I_k$.

\textbf{Step II} For each fold, use any supervised learning algorithm to estimate the appropriate working models.  Let~$\hat g_t^{(k)} (H_t, A_t)$, $\hat p_t^{(k)} (1 | H_t)$, $\hat p_t^{(k)} (R_t | H_t)$ and $\hat{\tilde p}_t^{(k)} (1 | S_t)$ denote the estimates for~$\E[Y_{t+1} | H_t, A_t]$, $\E[A_t | H_t]$,$\E[R_t | H_t]$, and $\E[A_t | S_t]$ respectively using individuals in $I_k^\complement$, i.e., estimates of the nuisance parameters the $k$th fold.  

\textbf{Step III} Construct the pseudo-outcomes and perform weighted regression estimation:
    $$
    \tilde Y^{(DR)}_{t+1} := \frac{{\bf 1}(R_t = 1) \hat W^{(k)}_t (A_t - \hat{\tilde p}^{(k)}_t (1|S_t)) (Y_{t+1} - \hat g_t^{(k)} (H_t, A_t))}{\tilde \sigma^2_t(S_t)\hat p_t^{(k)} (R_t | H_t)} + \left( \hat g_t^{(k)} (H_t, 1) - \hat g_t^{(k)} (H_t, 0)\right)
    $$
    where nuisance parameters are from the appropriate fold.  Then regress~$\tilde Y^{(DR)}_{t+1}$ on $f_t(S_t)^\top \beta$ with weights $\hat{\tilde p}^{(k)}_t (1|S_t) (1-\hat{\tilde p}^{(k)}_t (1|S_t))$ to obtain $\hat\beta^{M}_n$.









\subsection{Simulations}

We extend the simulation setup from Section \ref{sec:sim} to assess the empirical performance of our proposed estimators. The observation indicator $R_t$ follows a Bernoulli distribution with $p_t(R_t = 1 | H_t) = 0.9 \times {\bf 1}(S_t = -1) + 0.8\times{\bf 1}(S_t = 1)$. The marginal treatment effect remains constant over time, with $\beta_0^\star = \beta_{10} = -0.2$. The simulation results are reported below.

\begin{table}[htbp]
\caption{\centering Fully marginal causal effect estimation with missing outcomes. The true value of the parameters is $\beta_{0}^\star = -0.2$.}
\label{tab:tabsix}
\begin{center}
\begin{tabular}{cccccccc}
\hline
Method & $\beta_{11}$ & Est & SE & RMSE & CP  \\\hline
\multirow{3}{*}{DR-WCLS} & 0.2 & -0.199 & 0.026 & 0.026 & 0.950  \\
& 0.5&-0.200 & 0.027 & 0.027 & 0.963   \\
&0.8& -0.199 & 0.030 & 0.031 & 0.940 \\
\hline
\end{tabular}
\end{center}
\end{table}

\subsection{Time dimension asymptotic property}

Redefine $\boldsymbol{\eta}_t(H_t, A_t) = (g(H_t, A_t), p(R_t,A_t|H_t))$ for this section. First define:
\begin{align*}
    \psi^M_t (\beta^\star ;H_{t},& A_t,R_t) = \\
    & \tilde \sigma^2_t(S_t) \Big(\underbrace{\frac{{\bf 1}(R_t = 1)}{p (R_t | H_t)} \frac{W_t (A_t - \tilde p_t (1|S_t))(Y_{t+1}- g_t(H_t,A_t))}{\tilde \sigma^2_t(S_t)} + \beta (t; H_t)}_{\tilde Y^{(DR)}_{t+1}} - f_t(S_t)^\top \beta\Big) f_t
\end{align*}
And then reformulate the estimating equation as follows: 
\begin{equation}
\label{eq:md_drwcls-T}
    \P_n \Big[\frac{1}{T}\sum_{t=1}^T \psi^M_t(\beta^\star ;H_{t},A_t,R_t)  \Big] =0
\end{equation}

We need to further adjust Assumption \ref{ass:T-infinity} as follows. The psuedo-outcome $\tilde Y^{(DR)}_{t+1}$ is defined as in Appendix \ref{app:sec:md-algorithm} Step III.
\begin{assumption}
\label{app:ass:md-T-infty}
In the presence of missing data, we require the following to hold when $T \rightarrow \infty$:
    \begin{enumerate}
    \item There exists $\beta^\star$, such that $\lim_{T \rightarrow \infty}\frac{1}{T}  \sum_{t=1}^T\E[\psi^M_t(\beta^\star ;H_{t},A_t,R_t)] =0$.
    \item Denote the second-stage residual as \( \xi_t \coloneqq \tilde Y^{(DR)}_{t+1} - f_t(S_t)^\top \beta^\star \). There exists constants \( \delta > 0 \) and \( c_1 > 0 \) such that  $\sup_{t}\mathbb{E}[\xi_t^{2+\delta}| H_t, A_t] < c_1$. The correlation of the sequence \( \{\mathbb{E}[\xi_t^2| H_t, A_t]\}_{t=1}^T \) decreases as the time points \( t \) and \( t' \) move further apart, and there exists a constant positive definite matrix $\Gamma^M_{\beta}$, such that $\lim_{T \rightarrow \infty}\frac{1}{T}  \sum_{t=1}^T \E \big[ \psi^M_{t}(\beta;H_{t},A_t)\psi^M_{t}(\beta;H_{t},A_t)^\top\big] = \Gamma^M_{\beta} $.
    \item The Euclidean norm of the causal effect moderator $f_t(S_t) \in \mathbb{R}^q$ is bounded almost surely by some constant $c_2 >0$ for $\forall t$.
    \item $\Vert \boldsymbol{\hat\eta}_t - \boldsymbol{\eta}_t \Vert_T^2 = o_p(1)$
    and
    \begin{equation}
    \sum_{a \in \{0,1 \}}\left\Vert \hat p(R_t=1,a|H_t)- p(R_t=1,a|H_t)\right\Vert_T  \left\Vert \hat g_t(H_t,a)- g_t(H_t,a)\right\Vert_T =o_p(T^{-1/2}).
        \end{equation}
    \end{enumerate}
\end{assumption}
In addition to the key assumptions, we define the following quantity:
\begin{align*}
    B^M_{\beta} &= \lim_{T \rightarrow \infty} \frac{1}{T} \sum_{t=1}^T \E[\dot \psi^M_{t}(\beta;H_{t},A_t)].
\end{align*}
The first term $B^M_{\beta}$ matches the expression of $B_{\beta}$ in Theorem \ref{thm:T-infinity}, while $\Gamma^M_{\beta}$ is slightly adjusted to accommodate the missing indicator. With all the notation and assumptions set, we can now state the following corollary:
\begin{corollary}
\label{thm:T-infinity-md}
Assume that the sample size $n$ is finite and fixed and $\hat p_t(A_t|H_t)$ is bounded away from 0 and 1. Under Assumptions \ref{ass:po}, \ref{ass:directeffect}, \ref{ass:md_drwcls}, and \ref{app:ass:md-T-infty}, given invertibility and moment conditions, as $T \rightarrow \infty$, the estimator $\hat\beta^{M}$ that solves Equation \eqref{eq:md_drwcls-T} is consistent and asymptotically normal such that $\sqrt{T}(\hat\beta^{M} - \beta^\star)\rightarrow \mathcal{N}(0,(B_{\beta}^M)^{-1}\Gamma^M_{\beta} (B_{\beta}^M)^{-1})$.
\end{corollary}
The proof resembles closely that in Appendix \ref{app:T-infinity}. We omit the details here.

\section{Assessing Time-Lagged Effects}
\label{app:lagged}

Beyond the interest in proximal outcomes, additional attention has been paid to lagged outcomes defined at future decision points with a fixed window length $\Delta > 1$, denoted as $Y_{t,\Delta}$, which is a known function of the observed history and the latest treatment: $Y_{t,\Delta}=y(H_{t+\Delta-1}, A_{t+\Delta-1})$.  In practice, $\Delta$ is explicitly chosen to avoid the curse of the horizon problem~\citep{dempsey2020}. 
While this has been true to date, we acknowledge that larger $\Delta$  will be more common as MRT data sets grow in size and these longer-term outcomes become of primary interest.  Under Assumption~\ref{ass:po}, the causal estimand for lagged effect can be expressed in terms of observable data \citep{shi2022assessing}:
\begin{equation}
\label{eq:causalexursion_lag}
    \beta_{\bfp,\pi}(t+\Delta;s)= \E \left[ \E_{{\bfp}} \left[ W_{t,\Delta-1} Y_{t,\Delta} \mid A_t = 1, H_t \right] - \E_{{\bfp}} \left[ W_{t,\Delta-1} Y_{t,\Delta} \mid A_t = 0, H_t \right] \mid S_t = s \right], 
\end{equation}
where $W_{t, u} = \prod_{s=1}^{u} \pi_t (A_{t+s} | H_{t+s}) /  p_t (A_{t+s} | H_{t+s})$, with~$W_{t,0} =1$. Here, we assume the reference distribution for treatment assignments from $t+1$ to $t+\Delta-1$ ($\Delta>1$) is given by a randomization probability generically represented by $\{\pi_{u}(a_{u} | H_{u})\}_{u=t+1}^{t+\Delta-1}$. This generalization contains previous definitions such as lagged effects~\citep{boruvka2018} where $\pi_{u} = p_{u}$ and deterministic choices such as $a_{t+1:(t+\Delta-1)} = {\bf 0}$~\citep{dempsey2020,qian2020estimating}, where $\pi_{u} = {\bf 1}\{a_{u} = 0\}$ and ${\bf 1}\{\cdot\}$ is the indicator function. Furthermore, we assume the time-lagged effects defined in \eqref{eq:causalexursion_lag} takes a linear form $\beta_{{\bfp},\pi} (t +\Delta;s) = f_t(s)^\top \beta^\star$, where $f_t(s) \in \mathbb{R}^q$ is a feature vector depending only on state $s$ and decision point $t$. 

A brief discussion in \cite{shi2022assessing} presented an approach to improve the efficiency of the estimation of the lagged effect and alleviate the curse of the horizon \citep{liu2018breaking}. Specifically, it was shown that an optimal estimating function will be orthogonal to the score functions for the treatment selection probabilities~\citep{bickel1993efficient}. This implies that the estimator can be improved by replacing the estimating equation by itself minus its projection on the score functions for the treatment selection probabilities~\citep{murphy2001marginal}. This can be done in the case of the DR-WCLS estimating equation as follows:
\begin{equation}
\label{eq:dr-lagged}
\begin{split}
\P_n  \Big[\sum_{t=1}^{T-\Delta+1} \Big[ & W_{t}(A_{t} - \tilde p_t(1|S_{t}))\Big( W_{t,\Delta-1} \left( Y_{t,\Delta} - g_{t+\Delta-1} (H_{t+\Delta-1}, A_{t+\Delta-1}) \right) \\
&- \sum_{u=0}^{\Delta-2} W_{t,u} \big[ g_{t+u} (H_{t+u}, A_{t+u}) - \sum_{a_{t+u+1}} \pi (a_{t+u+1} | H_{t+u+1} ) g_{t+u+1} (H_{t+u+1}, a_{t+u+1}) \big]\Big) \\
&+ \tilde\sigma_t^2(S_t) \left( \beta(t+\Delta,H_t) - f_t(S_t)^\top \beta \right)\Big]  f_t(S_t)^\top\Big] =0,
\end{split}
\end{equation}
\sloppy where $g_{t+u} (H_{t+u}, A_{t+u})$ is a working model for $\E[ W_{t+u+1: t+\Delta-1} Y_{t,\Delta} | H_{t+u}, A_{t+u}]$.  Specifically, $g_{t+\Delta-1} (H_{t+\Delta-1}, A_{t+\Delta-1}) = \E [ Y_{t,\Delta} | H_{t+\Delta-1}, A_{t+\Delta-1} ] $, and $\E[g_{t+u-1} (H_{t+u-1}, A_{t+u-1})] = \E [ \sum_{a_{t+u}} \pi_{t+u} (a_{t+u} | H_{t+u}) g_{t+u} (H_{t+u}, a_{t+u})]$. The parameterized linear working model of the conditional expectation $g_{t+u}(H_{t+u}, A_{t+u})$ in \cite{murphy2001marginal} can be improved by using supervised learning algorithms to construct data-adaptive estimates. 
Therefore the $\hat\beta^{\Delta}_n$ obtained by solving Equation \eqref{eq:dr-lagged} has the following asymptotic property as $n$ approaches infinity. In addition, the asymptotic property of the time dimension is provided in Appendix \ref{app:lagged}.

\begin{corollary}
[Asymptotic property for the DR-WCLS estimator for lagged outcomes]
\label{corollary:lagged}
Under Assumptions \ref{ass:po}, \ref{ass:nuisance-op1}, assuming the time-lagged causal effect  $\beta_{{\bfp},\pi} (t +\Delta;s) = f_t(s)^\top \beta^\star$, and given invertibility and moment conditions, the $\hat\beta^{\Delta}_n$ obtained by solving Equation \eqref{eq:dr-lagged} is subject to an error term, which is  (up to a multiplicative constant) bounded above by $\sum_{u=0}^{\Delta-1}\mathbf{\hat B}_{u}$, where
\begin{equation}
\begin{split}
    \mathbf{\hat B}_{u} = \sum_{t=1}^{T-\Delta +1} \sum_{a_{t+u}} \left\Vert \hat p_{t+u}(a_{t+u}|H_{t+u}) - p_{t+u}(a_{t+u}|H_{t+u})\right\Vert \cdot \left\Vert \hat g_{t+u}(H_{t+u},a_{t+u})- g_{t+u}(H_{t+u},a_{t+u})\right\Vert.
\end{split}
\end{equation}
If we assume that $\mathbf{\hat B}_{u} = o_p (n^{-1/2})$, the estimator $\hat\beta^{\Delta}_n$ is consistent and asymptotically normal such that $\sqrt{n}(\hat\beta^{\Delta}_n - \beta^\star) \rightarrow \mathcal{N}(0,\Sigma^\Delta_{DR})$, where $\Sigma^\Delta_{DR}$ is defined in Appendix \ref{app:lagged}.
\end{corollary}

When $\Delta$ is large, correctly specifying the conditional expectation model $g_t(H_t,A_t)$ is particularly useful to avoid the variance estimation growing exponentially due to the weight $W_{t,\Delta}$, thus offering a remedy for the curse of the horizon \citep{liu2018breaking}.

\subsection{Proof of Corollary \ref{corollary:lagged}}

The estimating equation is written as the following:
\begin{equation*}
\begin{split}
\P_n  \Bigg[\sum_{t=1}^{T-\Delta+1} \bigg[ & W_{t,j}(A_{t,j} - \tilde p_t(1|S_{t}))\bigg( W_{t,\Delta-1, j} \left( Y_{t,\Delta,j} - g_{t+\Delta-1} (H_{t+\Delta-1}, A_{t+\Delta-1,j}) \right) \\
&- \sum_{u=0}^{\Delta-2} W_{t,u,j} \Big( g_{t+u} (H_{t+u}, A_{t+u,j}) - \sum_{a_{t+u+1}} \pi (a_{t+u+1} | H_{t+u+1} ) g_{t+u+1} (H_{t+u+1}, a_{t+u+1}) \Big)\bigg) \\
&+ \tilde\sigma_t^2 \left( \beta(t+\Delta,H_t) - f_t(S_t)^\top \beta \right)\bigg]  f_t(S_t)^\top\Bigg] =0,
\end{split}
\end{equation*}
where $W_{t,0,j}=1$. Using supervised learning estimates, we can get data-adaptive plug-ins for $g$'s and $p$'s. First, we prove this statement:
\begin{equation*}
    \E[g_{t+u-1} (H_{t+u-1}, A_{t+u-1})] = \E \Big[ \sum_{a_{t+u}} \pi_{t+u} (a_{t+u} | H_{t+u}) g_{t+u} (H_{t+u}, a_{t+u}) \Big],
\end{equation*}
which follows a simple iterative conditional expectation:
\begin{align*}
    \E\left[g_{t+u-1} (H_{t+u-1}, A_{t+u-1})\right]= &\E \left[W_{t+u}g_{t+u}(H_{t+u},A_{t+u})\right]\\
    = &\E \left[\E\left[W_{t+u}g_{t+u}(H_{t+u},A_{t+u})|H_{t+u}\right]\right]\\
    = & \E \Big[ \sum_{a_{t+u}} \pi_{t+u} (a_{t+u} | H_{t+u}) g_{t+u} (H_{t+u}, a_{t+u}) \Big]
\end{align*}

\subsubsection{Double Robustness}

To show the double robustness property of the estimator, we first assume the weight is correctly specified, we have the cancellation terms:
\begin{align*}
    & W_{t,u} g_{t+u} (H_{t+u}, A_{t+u})  =  \sum_{a_{t+u}}W_{t,u-1} \pi (a_{t+u} | H_{t+u} ) g_{t+u} (H_{t+u}, a_{t+u}), \\
    & \E[W_{t}(A_t - \tilde p_t(1|S_t))g_t(H_t,A_t)] = \tilde \sigma_t^2 \beta(t+\Delta,H_t),
\end{align*}
where $u \in \{1,2,\dots,\Delta-1\}$. Therefore, we are left with solving:
\begin{align*}
    &\E \Big[\sum_{t=1}^{T-\Delta+1} \big(W_{t}(A_t-\tilde p_t) W_{t,\Delta-1}Y_{t,\Delta} - \tilde \sigma_t^2 f_t(S_t)^\top\beta\big)f_t(S_t)^\top\Big]\\
    =& \E\Big[\sum_{t=1}^{T-\Delta+1} \tilde \sigma^2_{t}  \big(  \E[W_{t,\Delta-1}Y_{t,\Delta}|H_t,1]-\E[W_{t,\Delta-1}Y_{t,\Delta}|H_t,0] - f_t(S_t)^\top \beta \big) f_t(S_t)   \Big]\\
    =& 0
\end{align*}
Then when we assume the $g$'s are correctly specified, the following holds:
\begin{align*}
    &\E\left[ Y_{t,\Delta} - g_{t+\Delta-1} (H_{t+\Delta-1}, A_{t+\Delta-1})\right] = 0 \\
    &\E\left[g_{t+u} (H_{t+u}, A_{t+u})\right] - \sum_{a_{t+u+1}} \pi (a_{t+u+1} | H_{t+u+1} ) g_{t+u+1} (H_{t+u+1}, a_{t+u+1})  = 0,
\end{align*}
where $u \in \{1,2,\dots, \Delta-2\}$. As a result, we are left to solve:
\begin{align*}
    \E\Big[\sum_{t=1}^{T-\Delta+1} \tilde \sigma^2_{t}  \left( \beta(t+\Delta;H_t)  - f_t(S_t)^\top \beta \right) f_t(S_t)   \Big] =0.  
\end{align*}

When $\Delta$ is large and we have a fairly accurate understanding of the nuisance functions $\{g_{t+u}\}_{u=1}^{\Delta-1}$, our proposed method is especially useful. It helps to prevent the variance estimation from growing exponentially due to the weight $W_{t,\Delta}$, thus providing a solution to the curse of the horizon. Furthermore, under the assumption that the $g$'s are correctly specified, we have the DR-WCLS estimator satisfies:
\begin{align*}
    n^{1/2}(\hat\beta^{\Delta}_n-\beta^{\star}) = n^{1/2}~ \P_n & \Big[\sum_{t=1}^{T-\Delta+1} ~ \E\Big[\sum_{t=1}^{T-\Delta+1} \tilde \sigma^2_t(S_t) f_t(S_t)f^\top_t(S_t)\Big]^{-1} \times \\
    & ~~~~~~~~~~~~~~~~~~~~~~~ \tilde \sigma^2_t(S_t) (\beta(t+\Delta;H_t) - f_t(S_t)\beta^\star)f_t(S_t) \Big] + o_p(1),
\end{align*}
and it is efficient with influence function:
\begin{align*}
        &\sum_{t=1}^{T-\Delta+1} ~  \E\Big[\sum_{t=1}^{T-\Delta+1} \tilde \sigma^2_t(S_t) f_t (S_t)f_t (S_t)^\top  \Big]^{-1} \tilde \sigma^2_t(S_t) \big(\beta(t+\Delta;H_t) - f_t(S_t)\beta^\star \big)f_t(S_t).
\end{align*}
In conclusion, under moment conditions, we have asymptotic normality with variance given by $\Sigma^\Delta_{DR}=Q^{-1} W Q^{-1}$, where
\begin{align*}
    Q & = \E\Big[\sum_{t=1}^{T-\Delta+1} \tilde \sigma^2_t(S_t) f_t (S_t)f_t (S_t)^\top  \Big], \\
    W &= \E \Big[ \Big(\sum_{t=1}^{T-\Delta+1} \tilde \sigma^2_t(S_t) (\beta(t+ \Delta;H_t) - f_t(S_t)\beta^\star)f_t(S_t)\Big)^2\Big].
\end{align*}

\subsubsection{Asymptotic property}

We start the proof with the smallest lagged effect, i.e. setting $\Delta=2$. Thus, we can rewrite the \eqref{eq:dr-lagged} as:

\begin{equation*}
\begin{split}
\P_n &\bigg[\sum_{t=1}^{T-1} \Big[\tilde p_t\left(\frac{1}{\hat p_t}-\frac{1}{p_t}+\frac{1}{ p_t}\right) (A_t - \tilde p_t(1))\bigg(\pi_{t+1}\left(\frac{1}{\hat p_{t+1}}-\frac{1}{p_{t+1}}+\frac{1}{ p_{t+1}}\right) \left( Y_{t,2} -g^\star_{t+1} +g^\star_{t+1} -  \hat g_{t+1}  \right) \\
&-\Big( \hat g_{t}  - g^\star_{t} + g^\star_{t} - \sum_{a_{t+1}} \pi_{t+1} (\hat g_{t+1}(a_{t+1}) -g^\star_{t+1}(a_{t+1})+g^\star_{t+1}(a_{t+1}) ) \Big) \bigg)\\
&+ \tilde \sigma^2_t \left( \hat\beta(t+\Delta;H_t) - \beta(t+\Delta;H_t)+ \beta(t+\Delta;H_t) - f_t(S_t)^\top (\hat\beta -\beta^\star +\beta^\star) \right) \Big] f_t(S_t)^\top\bigg] =0
\end{split}
\end{equation*}

Recall the previous cancellation terms when $g$'s and $p_t$'s are correctly specified, we can simplify the equation above as:
\begin{equation*}
\begin{split}
0 &= \P_n \bigg[\sum_{t=1}^{T-1} \Big[\tilde p_t\Big(\frac{1}{\hat p_t}-\frac{1}{p_t}+\frac{1}{ p_t}\Big) (A_t - \tilde p_t(1))\Big(\pi_{t+1}\Big(\frac{1}{\hat p_{t+1}}-\frac{1}{p_{t+1}}\Big) \left( g^\star_{t+1} -  \hat g_{t+1}  \right) -\left( \hat g_{t}  - g^\star_{t}  \right) \Big)\\
&~~~~~~~~~~~+\tilde \sigma^2_t \left( \hat\beta(t+\Delta;H_t) - \beta(t+\Delta;H_t)+ \beta(t+\Delta;H_t) - f_t(S_t)^\top (\hat\beta -\beta^\star +\beta^\star) \right) \Big] f_t(S_t)^\top\bigg]\\
&= \P_n \bigg[\sum_{t=1}^{T-1} \Big[\tilde p_t(A_t - \tilde p_t(1))\Big(\frac{1}{\hat p_t}-\frac{1}{p_t}\Big) \Big(\pi_{t+1}\Big(\frac{1}{\hat p_{t+1}}-\frac{1}{p_{t+1}}\Big) \left( g^\star_{t+1} -  \hat g_{t+1}  \Big) -\left( \hat g_{t}  - g^\star_{t}  \right) \right)\\
&~~~~~~~~~~~+W_t(A_t - \tilde p_t(1))\Big(\pi_{t+1}\Big(\frac{1}{\hat p_{t+1}}-\frac{1}{p_{t+1}}\Big) \left( g^\star_{t+1} -  \hat g_{t+1}  \right)  \Big) \\
&~~~~~~~~~~~+\tilde \sigma^2_t \left(  \beta(t+\Delta;H_t) - f_t(S_t)^\top (\hat\beta -\beta^\star +\beta^\star) \right) \Big] f_t(S_t)^\top\bigg].
\end{split}
\end{equation*}
Then the deviation is:
\begin{equation*}
\begin{split}
&\P_n \bigg[\sum_{t=1}^{T-1} \Big[\tilde p_t(A_t - \tilde p_t(1))\Big(\frac{1}{\hat p_t}-\frac{1}{p_t}\Big) \Big(\pi_{t+1}\left(\frac{1}{\hat p_{t+1}}-\frac{1}{p_{t+1}}\right) \left( g^\star_{t+1} -  \hat g_{t+1}  \right) +\left(g^\star_{t} - \hat g_{t}  \right) \Big)\\
&~~~~~~~~~~~+W_t(A_t - \tilde p_t(1))\Big(\pi_{t+1}\left(\frac{1}{\hat p_{t+1}}-\frac{1}{p_{t+1}}\right) \left( g^\star_{t+1} -  \hat g_{t+1}  \right)  \Big)\Big] f_t(S_t)^\top\bigg],
\end{split}
\end{equation*}
which could be decomposed into two parts, first is inherited from the previous stage:
\begin{equation}
\label{app:eq-deviation_1}
\begin{split}
\P_n \bigg[\sum_{t=1}^{T-1} \Big[\tilde p_t(A_t - \tilde p_t(1))\Big(\frac{1}{\hat p_t}-\frac{1}{p_t}\Big) \left( \hat g_{t}  - g^\star_{t} \right)\Big] f_t(S_t)^\top\bigg],
\end{split}
\end{equation}
and the second part contains two more terms which are the deviation generated from the second stage:
\begin{equation}
\label{app:eq-deviation_2}
\begin{split}
&\P_n \bigg[\sum_{t=1}^{T-1} \Big[\tilde p_t(A_t - \tilde p_t(1))\pi_{t+1}\Big(\frac{1}{\hat p_t}-\frac{1}{p_t}\Big) \Big(\frac{1}{\hat p_{t+1}}-\frac{1}{p_{t+1}}\Big) \left( g^\star_{t+1} -  \hat g_{t+1}  \right) \\
&~~~~~~~~~~~+W_t(A_t - \tilde p_t(1))\pi_{t+1}\Big(\frac{1}{\hat p_{t+1}}-\frac{1}{p_{t+1}}\Big) \left( g^\star_{t+1} -  \hat g_{t+1}  \right)  \Big] f_t(S_t)^\top\bigg].
\end{split}
\end{equation}

We focus on analyzing the second term above. Because if the second term is asymptotically negligible, then the first term will naturally be asymptotically negligible. The second term can be written as:
\begin{align*}
    \P_n\bigg[\sum_{t=1}^{T-1} \Big(\sum_{a_t,a_{t+1} }  \mathbf{c}_{t+1} \big(\hat p_{t+1}(a_{t+1}|H_{t+1}(a_t))&- p_t(a_{t+1}|H_{t+1}(a_t))\big)\\
    &\times \big(\hat g(H_{t+1}(a_t),a_{t+1})- g(H_{t+1}(a_t),a_{t+1})\big)\Big) f_t(S_t)\bigg],
\end{align*}
where
\begin{align*}
    \mathbf{c}_{t+1} & = \frac{ \tilde \sigma^2_t}{a_{t+1}\hat p_{t+1}(1|H_{t+1}(a_t))+ (1-a_{t+1})(1-\hat p_{t+1}(1|H_{t+1}(a_t)))}.
\end{align*}

In our context, $T$ is finite and fixed. Therefore, by the fact that $\hat p_{t+1}(a_{t+1}|H_{t+1}(a_t))$ is bounded away from zero and one, along with the Cauchy-Schwarz inequality, we have that (up to a multiplicative constant) the term within the parentheses is bounded above by:
\begin{equation}
         \mathbf{\hat B}_{1} = \sum_{t=1}^{T-1} \sum_{a_t,a_{t+1}}\left\Vert \hat p_{t+1}(a_{t+1}|H_{t+1}(a_t))- p_t(a_{t+1}|H_{t+1}(a_t)) \right\Vert \left\Vert \hat g(H_{t+1}(a_t),a_{t+1})- g(H_{t+1}(a_t),a_{t+1})\right\Vert.
\end{equation}

Summarizing the deviation term above, the estimated $\hat\beta^{\Delta}_n$ ($\Delta$ =2) is subject to an error term, which is  (up to a multiplicative constant) bounded above by $\mathbf{\hat B} +\mathbf{\hat B}_{1}$, where $\mathbf{\hat B}$ is defined as in \eqref{app:eq-error}. To make $\mathbf{\hat B} +\mathbf{\hat B}_{1}$ asymptotically negligible, not only do we have the same requirement of the convergence rate of $\hat g_t(H_t,A_t)$ and $\hat p_t(A_t|H_t)$ as discussed before, but also require the lagged nuisance terms, $\hat g_{t+1}(H_{t+1},A_{t+1})$ and $\hat p_{t+1}(A_{t+1}|H_{t+1})$ to satisfy $\mathbf{\hat B}_{1}= o_p(n^{-1/2})$. 

Furthermore, when $\Delta = 3$, apart from the two deviation parts presented in \eqref{app:eq-deviation_1} and \eqref{app:eq-deviation_2}, we have a third part containing four more terms written as:
\begin{equation}
\label{app:eq-deviation_3}
\begin{split}
&\P_n \bigg[\sum_{t=1}^{T-2} \Big[\tilde p_t(A_t - \tilde p_t(1))\pi_{t+1}\pi_{t+2}\Big(\frac{1}{\hat p_t}-\frac{1}{p_t}\Big) \Big(\frac{1}{\hat p_{t+1}}-\frac{1}{p_{t+1}}\Big)\Big(\frac{1}{\hat p_{t+2}}-\frac{1}{p_{t+2}}\Big) \left( g^\star_{t+2} -  \hat g_{t+2}  \right) \\
&~~~~~~~~~~~+\tilde p_t(A_t - \tilde p_t(1))\pi_{t+1}\pi_{t+2}\Big(\frac{1}{\hat p_t}-\frac{1}{p_t}\Big) \frac{1}{p_{t+1}}\Big(\frac{1}{\hat p_{t+2}}-\frac{1}{p_{t+2}}\Big) \left( g^\star_{t+2} -  \hat g_{t+2}  \right)\\
&~~~~~~~~~~~+W_t(A_t - \tilde p_t(1))\pi_{t+1}\pi_{t+2}\Big(\frac{1}{\hat p_{t+1}}-\frac{1}{p_{t+1}}\Big) \Big(\frac{1}{\hat p_{t+2}}-\frac{1}{p_{t+2}}\Big) \left( g^\star_{t+2} -  \hat g_{t+2}  \right) \\
&~~~~~~~~~~~+W_t(A_t - \tilde p_t(1))\pi_{t+1}\pi_{t+2}\frac{1}{p_{t+1}}\Big(\frac{1}{\hat p_{t+2}}-\frac{1}{p_{t+2}}\Big) \left( g^\star_{t+2} -  \hat g_{t+2}  \right)  \Big] f_t(S_t)^\top\bigg].
\end{split}
\end{equation}

The same argument as above, our focus should be on analyzing the last term above. Because if the last term is asymptotically negligible, the first three terms will naturally be asymptotically negligible. This conclusion can be readily generalized to a bigger $\Delta$, and the proof follows the same expansion as shown for $\Delta=2$ and 3. For a specific value of $\Delta$, the estimation $\hat\beta^{\Delta}_n$ obtained by solving equation \eqref{eq:dr-lagged} is subject to an error containing in total $2^{\Delta}-1$ terms, which are  (up to a multiplicative constant) bounded above by $\sum_{u=0}^{\Delta-1}\mathbf{\hat B}_{u}$, where
\begin{equation*}
\begin{split}
    \mathbf{\hat B}_{u} = \sum_{t=1}^{T-u} \sum_{a_t, \dots,a_{t+u}} & \left\Vert \hat p_{t+u}(a_{t+u}|H_{t+u}(a_t, \dots,a_{t+u-1})) - p_{t+u}(a_{t+u}|H_{t+u}(a_t, \dots,a_{t+u-1}))\right\Vert\\
    &~~~~~~~~~~~~~~~~~\times \left\Vert \hat g(H_{t+u}(a_t, \dots,a_{t+u-1}),a_{t+u})- g(H_{t+u}(a_t, \dots,a_{t+u-1}),a_{t+u})\right\Vert,
\end{split}
\end{equation*}
which can be simplified as:
\begin{equation}
\begin{split}
    \mathbf{\hat B}_{u} = \sum_{t=1}^{T-u}\sum_{a_{t+u}} & \left\Vert \hat p_{t+u}(a_{t+u}|H_{t+u}) - p_{t+u}(a_{t+u}|H_{t+u})\right\Vert \left\Vert \hat g(H_{t+u},a_{t+u})- g(H_{t+u},a_{t+u})\right\Vert.
\end{split}
\end{equation}
If for each $u \in \{0,1,\dots, \Delta-1\}$, $\mathbf{\hat B}_{u}= o_p(n^{-1/2})$, then the summation $\sum_{u=0}^{\Delta-1}\mathbf{\hat B}_{u}= o_p(n^{-1/2})$.

\subsection{Time dimension asymptotic property}

First define:
\begin{align*}
    \psi^{\Delta}_t (\beta^\star ;H_{t+\Delta-1},& A_{t+\Delta-1}) = W_{t}(A_{t} - \tilde p_t(1|S_{t}))\Big( W_{t,\Delta-1} \left( Y_{t,\Delta} - g_{t+\Delta-1} (H_{t+\Delta-1}, A_{t+\Delta-1}) \right) \\
&- \sum_{u=0}^{\Delta-2} W_{t,u} \big[ g_{t+u} (H_{t+u}, A_{t+u}) - \sum_{a_{t+u+1}} \pi (a_{t+u+1} | H_{t+u+1} ) g_{t+u+1} (H_{t+u+1}, a_{t+u+1}) \big]\Big) \\
&+ \tilde\sigma_t^2(S_t) \left( \beta(t+\Delta,H_t) - f_t(S_t)^\top \beta \right)\Big]  f_t(S_t)^\top
\end{align*}
And then reformulate the estimating equation as follows: 
\begin{equation}
\label{eq:lagged_drwcls-T}
\P_n  \Big[\frac{1}{T-\Delta+1}\sum_{t=1}^{T-\Delta+1} \psi^{\Delta}_t (\beta^\star ;H_{t+\Delta-1}, A_{t+\Delta-1}) \Big] =0.
\end{equation}
To state the following assumption, we define the pseudo-outcome as:
\begin{equation}
\begin{split}
        \tilde Y^{(DR)}_{t,\Delta} &= \frac{W_{t}(A_{t} - \tilde p_t(1|S_{t}))}{\tilde\sigma_t^2(S_t)}\Big( W_{t,\Delta-1} \left( Y_{t,\Delta} - g_{t+\Delta-1} (H_{t+\Delta-1}, A_{t+\Delta-1}) \right) \\
&- \sum_{u=0}^{\Delta-2} W_{t,u} \big[ g_{t+u} (H_{t+u}, A_{t+u}) - \sum_{a_{t+u+1}} \pi (a_{t+u+1} | H_{t+u+1} ) g_{t+u+1} (H_{t+u+1}, a_{t+u+1}) \big]\Big)+  \beta(t+\Delta,H_t).
\end{split}
\end{equation}
Then we further adjust Assumption \ref{ass:T-infinity} as follows.
\begin{assumption}
\label{app:ass:lagged-T-infty}
In the presence of missing data, we require the following to hold when $T \rightarrow \infty$:
    \begin{enumerate}
    \item There exists $\beta^\star$, such that $\lim_{T \rightarrow \infty}\frac{1}{T-\Delta+1}  \sum_{t=1}^T\E[\psi^{\Delta}_t (\beta^\star ;H_{t+\Delta-1}, A_{t+\Delta-1})] =0$.
    \item Denote the second-stage residual as \( \xi_{t, \Delta} \coloneqq \tilde Y^{(DR)}_{t+\Delta} - f_t(S_t)^\top \beta^\star \). There exists constants \( \delta > 0 \) and \( c_1 > 0 \) such that  $\sup_{t}\mathbb{E}[\xi_{t, \Delta}^{2+\delta}| H_t, A_t] < c_1$. The correlation of the sequence \( \{\mathbb{E}[\xi_{t, \Delta}^2| H_t, A_t]\}_{t=1}^T \) decreases as the time points \( t \) and \( t' \) move further apart, and there exists a constant positive definite matrix $\Gamma^{\Delta}_{\beta}$, such that $\lim_{T \rightarrow \infty}\frac{1}{T -\Delta +1}  \sum_{t=1}^T \E \big[ \psi^{\Delta}_t (\beta^\star ;H_{t+\Delta-1}, A_{t+\Delta-1})\psi^{\Delta}_t (\beta^\star ;H_{t+\Delta-1}, A_{t+\Delta-1})^\top\big]=  \Gamma^{\Delta}_{\beta}$.
    \item The Euclidean norm of the causal effect moderator $f_t(S_t) \in \mathbb{R}^q$ is bounded almost surely by some constant $c_2 >0$ for $\forall t$.
    \item $\Vert \boldsymbol{\hat\eta}_t - \boldsymbol{\eta}_t \Vert_T^2 = o_p(1)$
    and
    \begin{equation}
\begin{split}
    \sum_{u=0}^{\Delta -1}  \sum_{a_{t+u} \in \{0,1\} } & \left\Vert \hat p_{t+u}(a_{t+u}|H_{t+u}) - p_{t+u}(a_{t+u}|H_{t+u})\right\Vert_T \times\\
    &~~~~~~~~~~~~~~~~~~~~~~\left\Vert \hat g(H_{t+u},a_{t+u})- g(H_{t+u},a_{t+u})\right\Vert_T = o_p(T^{-1/2}).
\end{split}
\end{equation}
\end{enumerate}
\end{assumption}
In addition to the key assumptions, we define the following quantity:
\begin{align*}
    B^{\Delta}_{\beta} &= \lim_{T \rightarrow \infty} \frac{1}{T} \sum_{t=1}^T \E[\dot \psi^{\Delta}_t (\beta^\star ;H_{t+\Delta-1}, A_{t+\Delta-1})].
\end{align*}
Again, the first term $B^{\Delta}_{\beta}$ matches the expression of $B_{\beta}$ in Theorem \ref{thm:T-infinity}. With all the notation and assumptions set, we can now state the following corollary:
\begin{corollary}
\label{thm:T-infinity-lagged}
Assume that $n$ is finite and fixed, and $\hat p_t(A_t|H_t)$ is bounded away from 0 and 1. Under Assumptions \ref{ass:po}, \ref{ass:directeffect} and \ref{app:ass:lagged-T-infty}, given invertibility and moment conditions, as $T \rightarrow \infty$, the estimator $\hat\beta^{\Delta}$ that solves Equation \eqref{eq:lagged_drwcls-T} is consistent and asymptotically normal such that $\sqrt{T}(\hat\beta^{\Delta} - \beta^\star)\rightarrow \mathcal{N}(0,(B^{\Delta})^{-1}\Gamma^{\Delta}_{\beta} (B^{\Delta})^{-1})$.
\end{corollary}
The proof resembles closely that in Appendix \ref{app:T-infinity}. We omit the details here.

\section{Binary Outcomes}
\label{app:binaryoutcomes}

\citet{qian2020estimating} proposed an estimator of the marginal excursion effect (EMEE) by adopting a log relative risk model to examine whether a particular time-varying intervention has an effect on a binary longitudinal outcome. The causal excursion effect is defined by:
\begin{align}
        \beta_{\mathbf{p}} (t;s) &= \log \frac{\E \left[ Y_{t+1}(\bar A_{t-1}, 1)\given S_t(\bar A_{t-1})=s \right]}{\E \left[ Y_{t+1}(\bar A_{t-1}, 0)\given S_t(\bar A_{t-1})=s \right]}\\
        & = \log \frac{\E \left[ \E \left[ Y_{t+1}\given{A_t=1,H_t}\right]\given S_t =s \right]}{\E \left[ \E \left[Y_{t+1}\given{A_t=0,H_t}\right]\given S_t=s\right]}.
\end{align}

Assuming $\beta_{{\bf p}} (t;s) = f_t(s)^\top \beta^\star$, where $f_t(s) \in \mathbb{R}^q$ is a feature vector of a $q$-dimension and only depends on state $s$ and decision point $t$, a consistent estimator for $\beta^*$ can be obtained by solving a set of weighted estimating equations:
\begin{equation}
\label{eq:mrtstandard_binary}
\mathbb{P}_n \left[ \sum_{t=1}^T W_t e^{-A_{t} f_t (S_t)^\top \beta} \left( Y_{t+1} - e^{g_t(H_t)^\top \alpha + A_t f_t (S_t)^\top \beta} \right) \left( \begin{array}{c} g_t(H_t) \\ (A_{t} - \tilde p_t (1 \mid S_t) ) f_t(S_t) \end{array} \right) \right] = 0.
\end{equation}
See \cite{qian2020estimating} for more details on the estimand formulation and consistency, asymptotic normality, and robustness properties of the estimation method EMEE.




Based on Equation \eqref{eq:mrtstandard_binary}, we propose a doubly robust alternative to EMEE, termed ``DR-EMEE''. A doubly robust estimator for the log-relative risk is constructed by solving the following set of estimating equations:
\begin{equation}
\label{eq:dr-emee}
\begin{split}
    \mathbb{P}_n \bigg[\sum_{t=1}^T \tilde \sigma^2_t(S_t) \bigg (&\frac{W_t e^{-A_{t} f_t (S_t)^\top \beta} (A_t - \tilde p_t (1|S_t))(Y_{t+1}- g_t(H_t,A_t))}{\tilde \sigma^2_t(S_t)}  \\
    &~~~~~~~~~~~~~~~~~~~~~~~~~~~~~~~~~~~~~~~+ e^{-f_t (S_t)^\top \beta} g_t(H_t, 1) - g_t (H_t, 0)\bigg) f_t \bigg] = 0.
\end{split}
\end{equation}



\begin{corollary}[Asymptotic property for DR-EMEE estimator]
\label{corollary:binary}
Upon correctly specifying either conditional expectation model $g_t(H_t,A_t)$ or treatment randomization probability $p_t(A_t|H_t)$, given invertibility and moment conditions, the estimator $\hat\beta_n$ obtained from solving Equation \eqref{eq:dr-emee} is consistent and asymptotically normal such that $\sqrt{n}(\hat\beta_n - \beta^\star) \rightarrow \mathcal{N}(0,\Sigma^b_{DR})$, where $\Sigma^b_{DR}$ is defined in Appendix \ref{app:dr-emee}.
\end{corollary}

\subsubsection{Doubly robust property}
Equation \eqref{eq:dr-emee} presented a doubly-robust alternative to estimating the causal excursion effect for a binary longitudinal outcome. Here we prove that the estimator is doubly robust.
If the conditional mean model $g(H_t,A_t)$ is specified correctly, then we have the following.
\begin{align*}
    &\E \Big[\sum_{t=1}^T \tilde \sigma^2_t(S_t) \Big(\frac{W_t e^{-A_{t} f_t (S_t)^\top \beta^\star} (A_t - \tilde p_t (1|S_t))(Y_{t+1}- g_t(H_t,A_t))}{\tilde \sigma^2_t(S_t)} + e^{-f_t (S_t)^\top \beta^\star} g_t(H_t, 1) - g_t (H_t, 0)\Big) f_t \Big] \\
    =& \E \Big[\sum_{t=1}^T \tilde \sigma^2_t(S_t) \left( e^{-f_t (S_t)^\top \beta^\star} g_t(H_t, 1) - g_t (H_t, 0)\right) f_t \Big]=0,
\end{align*}
which indicates the estimator satisfies the following at every time point:
\begin{equation*}
    f_t (S_t)^\top \beta^\star = \text{log}\frac{g_t(H_t, 1)}{g_t (H_t, 0)} = \text{log} \frac{\E[Y_{t+1}|H_t, A_t = 1]}{\E[Y_{t+1}|H_t, A_t = 0]}.
\end{equation*}
Under regularity conditions, the estimator 
$\hat\beta_n \overset{P}{\to} \beta^{\star}$; that is, $\hat\beta_n$ obtained from solving Equation \eqref{eq:dr-emee} is a consistent estimator of $\beta^{\star}$. On the other hand, if the treatment randomization probability $p(A_t|H_t)$ is correctly specified, then we have:
\begin{align*}
    &\E \Big[\sum_{t=1}^T \tilde \sigma^2_t(S_t) \Big(\frac{W_t e^{-A_{t} f_t (S_t)^\top \beta^\star} (A_t - \tilde p_t (1|S_t))(Y_{t+1}- g_t(H_t,A_t))}{\tilde \sigma^2_t(S_t)} + e^{-f_t (S_t)^\top \beta^\star} g_t(H_t, 1) - g_t (H_t, 0)\Big) f_t \Big] \\
    =&\E \Big[\sum_{t=1}^T  W_t e^{-A_{t} f_t (S_t)^\top \beta^\star} (A_t - \tilde p_t (1|S_t))(Y_{t+1}- g_t(H_t,A_t))f_t + \tilde \sigma^2_t(S_t) \left( e^{-f_t (S_t)^\top \beta^\star} g_t(H_t, 1) - g_t (H_t, 0)\right) f_t \Big]\\
    = & \E \Big[\sum_{t=1}^T \tilde \sigma^2_t(S_t) \left( e^{-f_t (S_t)^\top \beta^\star}\E[Y_{t+1}|H_t, A_t = 1]  - \E[Y_{t+1}|H_t, A_t = 0]\right) f_t \Big],
\end{align*}
which indicates the estimator satisfies that at every time point:
\begin{equation*}
    f_t (S_t)^\top \beta^\star  = \text{log} \frac{\E[Y_{t+1}|H_t, A_t = 1]}{\E[Y_{t+1}|H_t, A_t = 0]}.
\end{equation*}
Under regularity conditions, the estimator 
$\hat\beta_n \overset{P}{\to} \beta^{\star}$; that is, $\hat\beta_n$ obtained from solving Equation \eqref{eq:dr-emee} is a consistent estimator of $\beta^{\star}$. In conclusion, $\hat\beta_n$ obtained by solving Equation \eqref{eq:dr-emee} is doubly robust.

\subsubsection{Proof of Corollary \ref{corollary:binary}}
\label{app:dr-emee}

Denote the following estimating equation as:
\begin{equation}
\begin{split}
    \psi^b_t (\beta;H_t, A_t) &= W_t   e^{-A_t f_t(S_t)^\top \beta}  A_t \left(A_t-\tilde{p}_t(1|S_t)\right) \left( Y_{t+1} - g_t(H_t,A_t)\right)f_t(S_t)f_t(S_t)^\top +\\
&~~~~~~~~~~~~~~~~~~~~~~~~~~~~~~~~~~~~~~~~~~~~~~~~~~~~~~~~~~~~~~~~ \tilde \sigma^2_t(S_t) e^{-f_t(S_t)^\top \beta}  g(H_t,1)f_t(S_t)f_t(S_t)^\top. 
\end{split}
\end{equation}
Therefore, Equation \eqref{eq:dr-emee} can be written as:
\begin{align*}
    m_n(\beta) &= \sum_{t=1}^{T} \psi^b_t (\beta;H_t, A_t). 
\end{align*}

For the log-linear model, there is no closed-form solution; However, by Theorem 5.9 and Problem 5.27 of \cite{van2000asymptotic}. Given either nuisance model is correctly specified, $m_n(\beta)$ is continuously differentiable and hence Lipschitz continuous, Theorem 5.21 of \cite{van2000asymptotic} implies that $\sqrt{n}\{\hat\beta^{(DR)}_n-\beta^\star\}$ is asymptotically normal with mean zero and covariance matrix:
$$
\E\left[\dot{m}_n(\beta^\star)\right]^{-1}\E\left[m_n(\beta^\star)m_n(\beta^\star)^\top\right]\E\left[\dot{m}_n(\beta^\star)\right]^{-1^\top}.
$$

Since $e^{- f_t(S_t)^\top \beta^\star} \E \left(Y_{t,\Delta} \given H_t ,A_t=1 \right) =\E \left(Y_{t,\Delta} \given H_t ,A_t=0 \right)$, thus, we have the following:
\begin{align*}
    \E\left[\dot{m}_n(\beta^\star)\right] = \E \Big[\sum_{t=1}^{T}\tilde \sigma^2_t(S_t) g^\star_t(H_t,0)f_t(S_t)f_t(S_t)^\top   \Big]
\end{align*}
and, 
\begin{align*}
\E\left[m_n(\beta^\star)m_n(\beta^\star)^\top\right] = \E \Big[   \sum_{t=1}^{T} \tilde \sigma^2_t(S_t) \tilde \epsilon_{t,J} f_{t,J}(S_t) \times    \sum_{t=1}^{T} \tilde \sigma^2_t(S_t) \tilde \epsilon_{t,J^\prime} f_{t,J^\prime}(S_t)^\top \Big]
\end{align*}
where $\tilde \epsilon_{t,J} = e^{- f_t(S_{t,J})^\top \beta^\star}g^\star_t(H_t,1) - g^\star_t(H_t,0) $.

\section{More on simulation studies}
\label{app:simulation}

\subsection{The decision tree}

We generated ten time-varying continuous variables and ten time-varying discrete variables, and randomly picked two of each indicating as $X_{1,t}, \dots,X_{4,t}$. The cutoff values are selected to ensure that each outcome has a nonzero probability to be reached, and the outcome values are five random numbers generated from a uniform distribution on $(-1,1)$.

\begin{figure}[htbp]
\begin{center}
\includegraphics[width=3in]{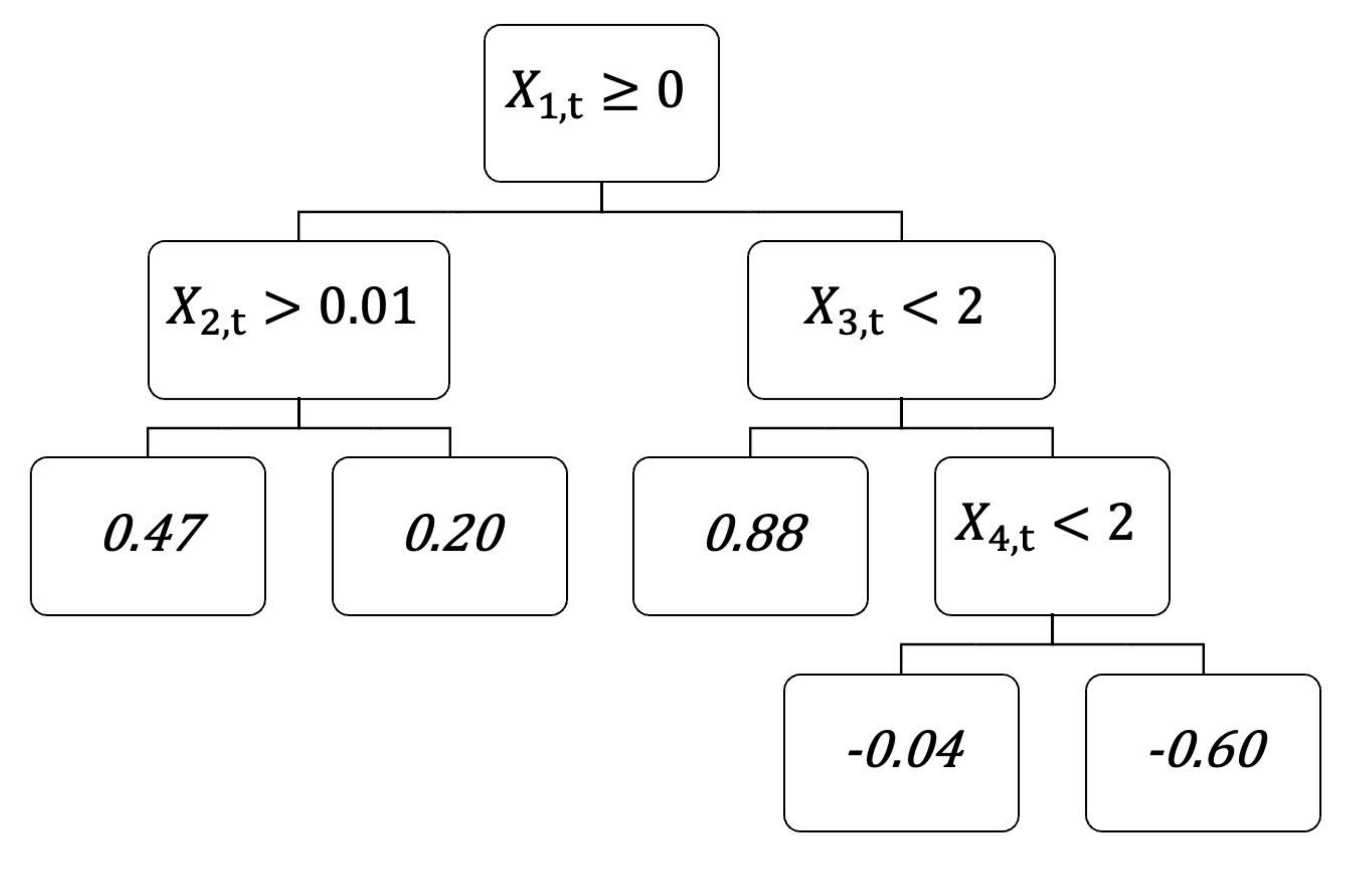}
\end{center}
\caption{The decision tree used to generate $g(H_t)$, where~$\{X_{1,t}, \dots,X_{4,t}\} \subset H_t$. }
\label{app:fig:generative_tree}
\end{figure}

\section{More on Case Study}
\label{app:casestudy}

\subsection{Control variable selection}

In Section \ref{sec:casestudy_1} and \ref{sec:casestudy_2}, we include in total 12 variables in the nuisance parameter estimation for R-WCLS and DR-WCLS methods, including the prior week's average step count, sleep time, and mood score, study week, sex, PHQ total score, depression at baseline, neuroticism at baseline, early family environment at baseline, pre-intern mood, sleep, and step count.  

In the WCLS model for mood outcome, we include the prior week's average mood score, depression at baseline, neuroticism at baseline, and study week as control variables; for step count outcome, we include the prior week's average step count, pre-intern step count and study week as control variables. 

In Section \ref{sec:casestudy_3}, we added two more variables to the estimation of nuisance parameters for the R-WCLS and DR-WCLS methods: the cumulative observation rate and the observation indicator from the previous week $R_{t-1,j}$. 

\subsection{Time-varying treatment effect on step count}

Estimated time-varying treatment moderation effects and their relative efficiency are shown in Figure \ref{fig:second} below.  We compare our proposed approach with the WCLS method. Similar to the mood outcome, a much narrower confidence band is observed when either R-WCLS or DR-WCLS method is used, indicating the estimation is more efficient at every time point.

\begin{figure}[htbp]
\begin{center}
\includegraphics[width=5in]{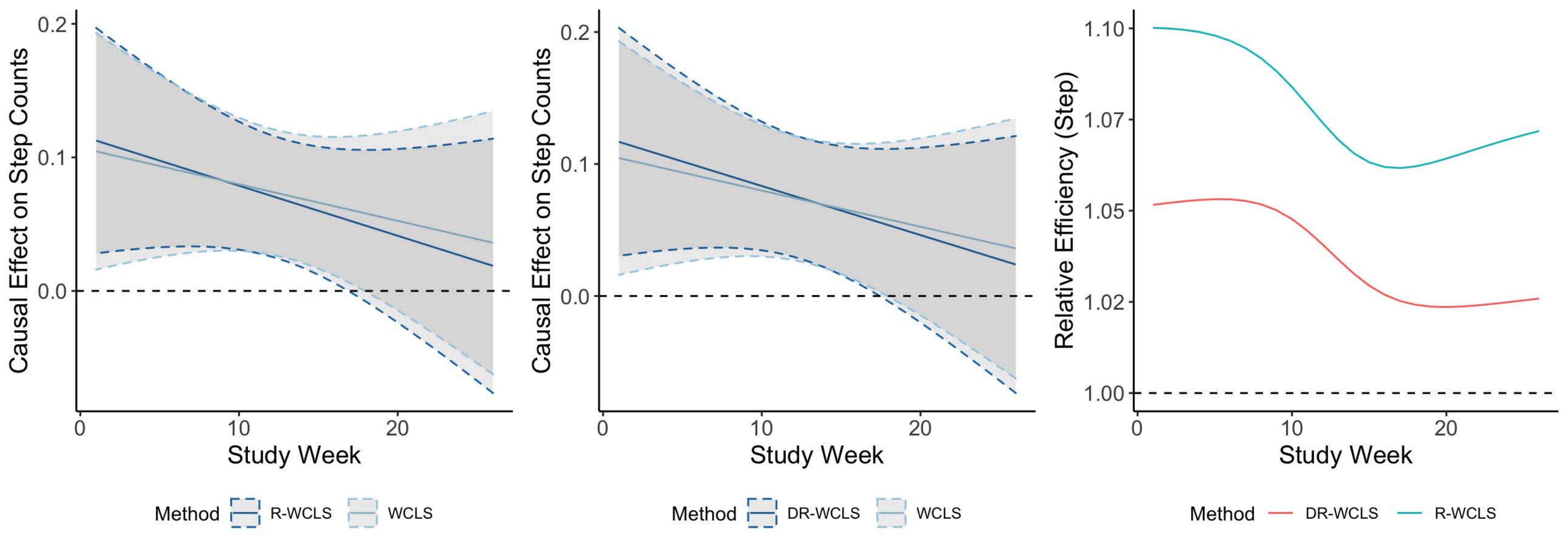}
\end{center}
\caption{Causal effects estimates with confidence intervals of R-WCLS (\textbf{left}) and DR-WCLS  (\textbf{middle}), and their relative efficiency in comparisons with WCLS (\textbf{right}). }
\label{fig:second}
\end{figure}

Furthermore, it is evident to see that the causal excursion effect of mobile prompts for step count change is positive in the first several weeks of the study, which means that sending targeted reminders is beneficial to increasing physical activity levels. In the later stages of the study, the effect fades away, possibly due to habituation to smartphone reminders. 

\subsection{Treatment Effect Estimation with Missing Data}
\label{sec:casestudy_3}

We apply our proposed methods to evaluate the treatment effect based on the raw observed data rather than the imputed dataset. To maintain consistency with previous analyses, we still use the weekly average mood score and step count (cubic root) as outcomes. Self-report mood scores and step counts are collected every day, so if no records were observed for the entire week, we indicate the weekly outcome as missing. Otherwise, the average mood score and step count (cubic root) are calculated as outcomes. For mood outcome, there is a total of 31.3\% person/week missing, and for step count outcome, 48.1\% person/week is missing. 

We carried out the same analysis as above for marginal treatment effects. Inverse probability weighting is used when implementing estimation using WCLS and R-WCLS criteria. Estimated treatment effects and their relative efficiency are shown in Table \ref{tab:fullymarginal_missing}. 
It is no longer evident that mood notifications have a significant overall impact on participants' moods, but the step count analysis still indicates a positive effect of sending activity notifications on participants' physical activity levels.

\begin{table}
\caption{IHS Study: Fully marginal treatment effect estimation with missing outcomes.}
\label{tab:fullymarginal_missing}
\begin{center}
\begin{tabular}{ccccc}
\hline
Outcome & Method & Estimation & Std.err & P-value  \\\hline
\multirow{3}{*}{Mood} & WCLS & $7.71 \times 10^{-3}$ & $1.73 \times 10^{-2}$  & 0.655  \\
& R-WCLS& $1.81 \times 10^{-3}$ & $1.62 \times 10^{-2}$  & 0.911  \\
&DR-WCLS & $3.00 \times 10^{-3}$ & $1.68 \times 10^{-2}$  & 0.858  \\
\hline 
\multirow{3}{*}{Steps} & WCLS & $6.71 \times 10^{-2}$ & $3.94 \times 10^{-2}$  & 0.088 \\
& R-WCLS& $7.43 \times 10^{-2}$ & $4.05 \times 10^{-2}$  & 0.067  \\
& DR-WCLS & 0.104 & $4.09 \times 10^{-2}$  & \textbf{0.011}  \\
\hline
\end{tabular}
\end{center}
\end{table}



\end{document}